\documentclass{amsbook}

\usepackage{amssymb}
\usepackage{amsfonts}
\usepackage{amsmath}
\usepackage{lastpage}
\usepackage[legalpaper,bookmarks=true,colorlinks=true,linkcolor=blue,citecolor=blue]{hyperref}
\usepackage{graphicx}%
\usepackage{fancyhdr}
\usepackage{color}
\usepackage[mathlines]{lineno}
\usepackage{lscape}
\usepackage{epsfig}

\newtheorem{theorem}{Theorem}
\theoremstyle{plain}

\newtheorem{lemma}{Lemma}

\numberwithin{equation}{section}

\begin{document}
\Large
\pagenumbering{roman}

\begin{center}

\huge \textbf{Gane Samb LO, Pape Djiby Mergane,\\
Thilabola Atozou Kpanzou, Mohamed Cheikh Haidara}\\
\vskip 6cm
\Huge \textbf{Weak Convergence (IIIA)} \\
\bigskip \textbf{Asymptotic Representations of Statistics in the Functional Empirical process : A portal and some applications}
\vskip 6cm

\huge \textit{\textbf{Statistics and Probability African Society (SPAS) Research Monographs Series}.\\
\textbf{Calgary, Alberta. 2018}}.\\

\end{center}

\newpage
\huge \textbf{SPAS Research Monographs Series}\\
\bigskip \bigskip

\Large \textbf{Advisers}\\

\bigskip \bigskip

\Large \textbf{List of published books}\\

\newpage
\noindent \textbf{Library of Congress Cataloging-in-Publication Data}\\

\noindent Main Author : Gane Samb LO, 1958-\\

\noindent Weak Convergence (IIIB). Asymptotic Representations of Statistics in the Functional Empirical process : A portal and some applications.\\

\noindent SPAS Research Monographs Series, 2018.\\

\newpage
\noindent \textbf{Author : Gane Samb LO}\\
\bigskip

\bigskip
\noindent \textbf{Emails}:\\
\noindent gane-samb.lo@ugb.edu.sn, ganesamblo@ganesamblo.net.\\

\bigskip
\noindent \textbf{Url's}:\\
\noindent www.ganesamblo@ganesamblo.net\\
\noindent www.statpas.net/cva.php?email.ganesamblo@yahoo.com.\\

\bigskip \noindent \textbf{Affiliations}.\\
Main affiliation : University Gaston Berger, UGB, SENEGAL.\\
African University of Sciences and Technology, AUST, ABuja, Nigeria.\\
Affiliated as a researcher to : LSTA, Pierre et Marie Curie University, Paris VI, France.\\

\noindent \textbf{Teaches or has taught} at the graduate level in the following universities:\\
Saint-Louis, Senegal (UGB)\\
African University of Sciences and Technology (AUST), Abuja, Nigeria\\
Banjul, Gambia (TUG)\\
Bamako, Mali (USTTB)\\
Ouagadougou - Burkina Faso (UJK)\\
African Institute of Mathematical Sciences, Mbour, SENEGAL, AIMS.\\
Franceville, Gabon\\

\bigskip \noindent \textbf{Dedication}.\\

\noindent \textbf{To my mother (1927-2011)}

\bigskip \noindent \textbf{Acknowledgment of Funding}.\\

\noindent The author acknowledges continuous support of the World Bank Excellence Center in Mathematics, Computer Sciences and Intelligence Technology, CEA-MITIC. His research projects in 2014, 2015 and 2016 are funded by the University of Gaston Berger in different forms and by CEA-MITIC.

\newpage
\noindent \textbf{Author : Tchilabalo A. KPANZOU}\\
\bigskip

\noindent Dr Tchilabalo holds a PhD from the University of Stellenbosch, South Africa (2011).\\
 
\noindent \textbf{Emails}:\\
\noindent kpanzout@gmail.com, kpanzout@yahoo.fr\\

\noindent \textbf{Url's}:\\
\noindent https://sites.google.com/a/aims.ac.za/tchilabalo\\
\noindent http://univi.net/spas/cvf.php?email=kpanzout@yahoo.fr\\
 
\bigskip \noindent \textbf{Affiliations}.\\
Main affiliation: University of Kara, Kara, TOGO.\\ 

\noindent \textbf{Teaches or has taught} at the graduate level in the following universities:\\
University of Kara (UK), TOGO\\
University of Lom\'e (UL), TOGO\\
Ecole Normale Sup\'erieure (ENS), TOGO\\
University of Abomey-Calavi (UAC), BENIN\\

\newpage
\noindent \textbf{Author : Pape Djiby Mergane}\\
\bigskip

\noindent Dr Pape Djiby Mergane holds a PhD from Gaston Berger University of Saint-Louis (2014).\\
\bigskip
\noindent \textbf{Emails}:\\
\noindent merganedjiby@gmail.com.\\

\bigskip
\noindent \textbf{Url's}:\\
\noindent https://arxiv.org/find/all/1/all:+AND+djiby+mergane/0/1/0/all/0/1\\

\bigskip \noindent \textbf{Affiliations}.\\
LERSTAD, Gaston Berger University (UGB), Saint-Louis, SENEGAL.\\
Alioune Diop University of Bambey (UADB), Bambey (SENEGAL)\\

\noindent \textbf{Teaches or has taught} at :\\
Alioune Diop University of Bambey (UADB), Bambey (SENEGAL)\\
Gaston Berger University (UGB), Saint-Louis, SENEGAL.\\
National Institute of Statistics and Demography, Dakar, SENEGAL.\\

\newpage
\noindent \textbf{Author : Mohamed Cheikh Haidara}\\
\bigskip
\noindent Dr Mohamed Cheikh Haidara holds a PhD from Gaston Berger University of Saint-Louis (2012).\\

\bigskip
\noindent \textbf{Emails}:\\
\noindent chheikhh@yahoo.fr, mcheikhhaidara@gmail.com\\

\bigskip
\noindent \textbf{Url's}:\\
\noindent \\http://univi.net/spas/cva.php?email=chheikhh@yahoo.fr\\
\noindent https://arxiv.org/find/all/1/all:+AND+haidara+AND+mohamed+cheikh/0/1/0/all/0/1\\

\bigskip \noindent \textbf{Affiliations}.\\
LERSTAD, Gaston Berger University (UGB), Saint-Louis, SENEGAL.\\
 
\noindent \textbf{Teaches or has taught} at :\\
Cheikh Anta Diop University (UCAD), Dakar, SENEGAL\\

\title[Weak Convergence and Functional Empirical Process]{Weak Convergence (IIIA). Asymptotic Representations of Statistics in the Functional Empirical process : A portal and some applications}

\begin{abstract} (Short Abstract) In this research monograph, we deal with a very general asymptotic representation for statistics named GRI expressed in the functional empirical process, both one-dimensional and multidimensional, and another call residual empirical process. Most of statistics in form of combination of L-statistics are covered by the asymptotic theory dealt here. We also treat three important exaples as show cases.\\


\noindent \textbf{AMS 2010 Classification Subjects :} 60XXX; 62G30
\end{abstract}

\maketitle

\newpage
\noindent \textbf{ABSTRACT} (\textbf{English}) In this research monograph, we deal with a very general asymptotic representation for statistics named GRI expressed in the functional empirical process, both one-dimensional and multidimensional, and another call residual empirical process. Most of statistics in form of combination of L-statistics are covered by the asymptotic theory dealt here. This treatise is conceived to be a kind of \textbf{spaceship} on which modules are hanged. The  spaceship is a functional Gaussian process and each module is the asymptotic representation of one statistic in terms of that Gaussian process. In that way, it is possible to navigate from one module to another, that is, to find the joint distribution of any pair of statistics, to compare them with respect to the areas and the times. In order to be able to do so, we should have a broad conception at the beginning. Within the constructed frame, the asymptotic joint law of any finite number of other statistics is automatically given as well as the joint distribution of its spatial variation or temporal variation, in absolute or relative values. We also deal with the general problem of decomposability of statistics by comparing statistical decomposability, a new view we introduce, versus functional decomposability. A general result only based on the GRI is provided.\\

\noindent This monograph is also the portal of a handbook of GRI that will cover the largest number possible of statistics. In prevision of that,  we treat three important examples as show cases.\\

\noindent It is expected that this portal and the handbook will attract the attention of researchers working in the asymptotic area and will furnish useful tools to scientists who are interested in application of asymptotic tests, completed by computer packages.\\

\noindent \noindent \textbf{RESUM\'E} (\textbf{Fran\c{c}ais}) Dans cette monographie de recherche, nous traitons d'une repr\'esentation g\'en\'erale asymptotique pour des statistiques exprim\'ee par rapport au processus empirique fonctionnel, \`a la fois unidimensionnel et multidimensionnel, et un autre processus empirique appel\'e r\'esiduel. La plupart des statistiques sous forme de combinaison de L-statistiques sont couvertes par la th\'eorie asymptotique trait\'ee ici. Ce trait\'e est conçu pour être une sorte de \textbf{vaisseau spatial} sur lequel les modules sont accroch\'es. Le vaisseau spatial est un processus gaussien fonctionnel et chaque module est la repr\'esentation asymptotique d'une statistique en fonction de ce processus gaussien. De cette mani\`ere, il est possible de naviguer d'un module \`a un autre, c'est-\`a-dire de trouver la distribution conjointe de n'importe quelle paire de statistiques, de les comparer par rapport spatialement et temporellement. Pour pouvoir le faire, nous devrions avoir une conception large au d\'ebut. \`a l'int\'erieur du cadre construit, la loi conjointe asymptotique d'un nouvel \'element avec un nombre fini d'autres statistiques est automatiquement donn\'ee ainsi que la distribution conjointe de sa variation spatiale ou variation temporelle, en valeurs absolues ou relatives. Nous traitons \'egalement du probl\`eme g\'en\'eral de la d\'ecomposabilit\'e des statistiques en comparant la d\'ecomposabilit\'e statistique, une nouvelle notion que nous introduisons, par rapport \`a la d\'ecomposabilit\'e fonctionnelle. Un r\'esultat g\'en\'eral bas\'e uniquement sur la repr\'esentation GRI est fourni.

\noindent Cette monographie est \'egalement annonciatrice d'un recueil de repr\'esentations GRI qui couvrira le plus grand nombre possible de statistiques. En pr\'evision de cela, nous traitons aussi de trois cas spécifiques importants.\\ 

\noindent Nous \'esp\'erons que ce portail et le recueil attireront l'attention de tous ceux qui travaillent dans le domaine des lois asymptotiques et fourniront aux sp\'ecialistes des domaines appliqu\'es des outils de travail qui seront compl\'e\'es par des programmes informatiques.\\

\frontmatter
\tableofcontents
\mainmatter
\Large

\chapter*{General Preface}

\noindent \textbf{This textbook} is the first of series whose ambition is to cover broad part of Probability Theory and Statistics .  These textbooks are intended to help learners and readers, both of of all levels, to train themselves.\\

\noindent As well, they may constitute helpful documents for professors and teachers for both courses and exercises.  For more ambitious  people, they are only starting points towards more advanced and personalized books. So, these texts are kindly put at the disposal of professors and learners.

\bigskip \noindent \textbf{Our textbooks are classified into categories}.\\

\noindent \textbf{A series of introductory  books for beginners}. Books of this series are usually accessible to student of first year in 
universities. They do not require advanced mathematics.  Books on elementary probability theory and descriptive statistics are to be put in that category. Books of that kind are usually introductions to more advanced and mathematical versions of the same theory. The first prepare the applications of the second.\\

\noindent \textbf{A series of books oriented to applications}. Students or researchers in very related disciplines  such as Health studies, Hydrology, Finance, Economics, etc.  may be in need of Probability Theory or Statistics. They are not interested by these disciplines  by themselves.  Rather, the need to apply their findings as tools to solve their specific problems. So adapted books on Probability Theory and Statistics may be composed to on the applications of such fields. A perfect example concerns the need of mathematical statistics for economists who do not necessarily have a good background in Measure Theory.\\

\noindent \textbf{A series of specialized books on Probability theory and Statistics of high level}. This series begin with a book on Measure Theory, its counterpart of probability theory, and an introductory book on topology. On that basis, we will have, as much as possible,  a coherent presentation of branches of Probability theory and Statistics. We will try  to have a self-contained, as much as possible, so that anything we need will be in the series.\\

\noindent Finally, \textbf{research monographs} close this architecture. The architecture should be so large and deep that the readers of monographs booklets will find all needed theories and inputs in it.\\

\bigskip \noindent We conclude by saying that, with  only an undergraduate level, the reader will  open the door of anything in Probability theory and statistics with \textbf{Measure Theory and integration}. Once this course validated, eventually combined with two solid courses on topology and functional analysis, he will have all the means to get specialized in any branch in these disciplines.\\

\bigskip \noindent Our collaborators and former students are invited to make live this trend and to develop it so  that  the center of Saint-Louis becomes or continues to be a renown mathematical school, especially in Probability Theory and Statistics.

\chapter*{General Preface of Our Series of Weak Convergence}

\noindent \textbf{The series Weak convergence} is an open project with three categories.\\

\noindent \textbf{The special series  Weak convergence I} consists of texts devoted to the core theory of weak convergence, each of them concentrated on the handling of one specific class of objects. The texts will have labels $A$, $B$, etc. Here are some examples.\\

\noindent (1) Weak convergence of Random Vectors (IA).\\

\noindent (2) Weak convergence of stochastic processes and empirical processes (IB).\\

\noindent (3) Weak convergence of random measures (IC).\\

\noindent (4) Weak convergence of random measures (ID).\\

\noindent (5) etc.\\

\bigskip \noindent \textbf{The special series Weak convergence II} consists of texts related to the theory of weak convergence, each of them concentrated on one specialized field using weak convergence. Usually, these subfields are treated apart in the literature. Here, we want to put them in our general frame as continuations of the Weak Convergence Series I. Some examples are the following.\\

\noindent (1) Weak laws of sums on independent randoms variables.\\

\noindent (2) Weak laws of sums on associated randoms variables.\\

\noindent (3) Univariate Extreme values Theory.\\

\noindent (4) Multivariate Extreme values Theory.\\

\noindent (5) Etc.\\

\bigskip \noindent \textbf{The special series Weak convergence III} consists of texts focusing on statistical applications of Parts of the Weak Convergence Series I and Weak Convergence Series II. Examples :\\

\noindent The present book falls in the category III of our series devoted to weak convergence. It constitutes a portal to a handbook of Gaussian Asymptotic Distributions Using the Functional Empirical Process as defined and introduced here.\\

\noindent here, we establish a general representation for a large class of statistics and indexes. Since these type of indexes are very recurrent in a significant number of disciplines, it seemed important to us to gather their asymptotic treatment in a unified approach and specifically deal with important issues in the same Gaussian field (a frame we lay out in the monograph) like :\\

\noindent (1) A general asymptotic representation for individual statistics.\\

\noindent (2) Asymptotic representations for temporal absolute or relative variation of statistics.\\

\noindent (3) Spatial Asymptotic representations for statistics.\\

\noindent (4) Estimation of decomposability default for statistics.\\

\noindent These points are important for any statistics and pay important roles in Applications. In the field of socio-economic studies, the important of last point quite significant for example.\\

\noindent The importance of this monograph resides in the fact that, virtually, the asymptotic theory of a significant number of statistics is implicitly done in this monograph even if they do not exist yet. Better than that, their asymptotic theory are placed in an already existing Gaussian field that allow to see get at one their interaction with other statistics whose representations are already available. A none less important feature is that the frame allows to make the interaction possible for statistics with different dimensions.\\

\noindent Once this portal settled, the monograph ay be extended by hanging on it a list of individual representations to form a handbook.\\

\chapter*{General Introduction} \label{hfep_intro}

\noindent  Some of my students, my collaborators and myself have spent more that one decade to contribute on the asymptotic theory of welfare indices. A list of the papers we wrote is at the appendix of this introduction. Some papers are published in indexed papers, other in 
non-indexed ones, others are posted in Arxiv (arxiv.org).\\

\noindent  The main reason which justifies such a monograph is two-fold. \\

\noindent  (a) One one side, we concluded that using the function empirical process \textit{fep} to achieve the results is powerful and efficient.\\

\noindent  At the beginning, we tried to use the real empirical process and the non less powerful tools of Hungarian constructions (\cite{kmt}, , \cite{cscshm}). When passing to the functional approach, everything became almost easy. However, the price has been paid for acquiring the technology of this wonderful theory of \textit{fep}, which has been popularized by \cite{vaart}, and based on the developments of many authors, for example \cite{dudley}, \cite{pollard}, Gaenssler \cite{gaenssler}, \cite{billingsley}, \cite{pollard}, etc.\\

\noindent  (b) On the other side, we discovered that behavior that asymptotic behavior of the indices, and by the way a large number of statistics, depend on two functions $h$ and $\ell$ in the following general asymptotic representation

$$
\mathbb{G}_{n,(1)}(h) + \int_{0}^{1} \mathbb{G}_{n,(1)}(\tilde{f}_s) \ell(s) \ ds, \ (GRI)
$$  

\bigskip \noindent  where $\tilde{f}_s$ is a function of $s \in (0,1)$ that will be precised later and $\mathbb{G}_{n,(1)}$ is the \textit{fep} in dimension one, based on a sample of size $n\geq 1$.\\

\noindent  From there comes the idea to share our experience in using the \textit{fep} and, by this, to devote one single broad study on the origin, the properties and the application of the representation (GRI), in which the main notation and terminology would be precised.\\

\noindent  Once this frame fixed, we open a king of spaceship on which we may attach modules, each module being the (GRI) formula of new statistics. An open handbook containing that spaceship and modules, will be the next step of this monograph.\\

\noindent  The monograph deals with the \textit{fep} which does not make differences between dimensions of the space since only the metrical topology is used. This allows the treatment of multivariate statistics.\\

\noindent  Actually, the \textit{fep} treats one-dimensional and multidimensional statistics in the same way. This allows to have a unique conception of the study. In that conception, for any statistic which is added to the vessel, its asymptotic joint law of any finite number of other statistics in the vessel is automatically known. As well, even the joint distribution of its spatial variation or temporal variation (in absolute or relative values) with other statistics is already established.\\

\noindent  To allow passing from one dimension to higher dimensions, we adopt notation in form of subscripts that clear indicated the dimension associated with the use of the natural projections. At first sight, this may be an over-notation. But at the end, it allows to keep the constructions and its use clear and unequivocal.\\

\noindent  We introduce and justified the notion of Gaussian field within the strict scope of the study.\\

\noindent  While the theoretical aspects are pretty well surrounded, the variances and covariances, might seem complicated.
But nowadays, computers take care of such questions, and there is nothing to worry about. We already have a number of own packages that work well. May be, my collaborators will be able to design an \textit{R} project in that sense.\\

\noindent  Before we announce the organization of the book, we wish to point out that researchers outside of Mathematics circles, will not find unavoidable difficulties to understand and to use the tools presented. The main reason is that most of the techniques are based on convergence of multivariate random variables, for with the book \cite{ips-wcia-ang} is enough.\\
 
 \noindent  Here is how is organized the monograph.\\
 
 \noindent  The first part, the gateway, concerns the intrinsic results. It includes four Chapters.\\
 
 \noindent  In Chapter \ref{hfep_gateway_intro}, we give the main notation on the \textit{fep} and its properties. Next, we explain the General Representation of Indices (GRI), its origin, its conditions and its potential applications. Three approaches are studied : Fixed-time, patial and time evolution.\\
 
 \noindent  In Chapter \ref{hfep_decomp}, we address the general problem of decomposability of statistics. We introduce the notion of statistical decomposability versus functional decomposability. The results are also general and may be applied to any statistic for which the (GRI) is admissible.\\

\noindent  In Chapter \ref{hfep_gateway_variation}, we show how to find the asymptotic laws of the variation (absolute and relative) of an index for a time to another, and the joint distribution of variations of two indices.\\

\noindent  In Chapter \ref{hfep_gateway_influence}, the joint law of two statistics admitting the GRI is given, having in mind potential applications to the pro-poor and anti-poor growth in Welfare analysis.\\

\noindent  In the second part, we provide first constituents of the announced handbook. We applied our techniques to important Welfare indices, as show-cases on how they work.\\

\noindent  \textbf{What next}? Computational resources will be gathered under an independent release. Also, a handbook of the applications of the method to as many as possible statistics is open.\\
 
\newpage \noindent  \textbf{List of papers of the authors of the monograph and co-authors}.\\

\noindent  1 - The asymptotic theory of the poverty intensity in view of Extreme value theory for two simple cases,(2007), Afrika Statistika, 41-55, (2). (With Serigne Touba Sall)\\

\noindent 2 - Estimation Asymptotique des Indices de Pauvreté : Modélisation Continue et Analyse spatio-temporelle de la pauvreté au Sénégal (Asymptotic estimation of poverty indices : continuous modelling and, time and space analysis of poverty in Senegal), (2009), Journal Africain des Sciences de la Communication et des Technologies, 341-377, (3).\\ 

\noindent 3 - The asymptotic theory of the Kakwani class of poverty measures, (2009), African Diaspora Journal of Mathematics, 54-67, 1. (WIth Serigne Touba Sall)\\

\noindent 4 - Une théorie Générale Asymptotique des Mesures de Pauvreté (A general theory of the asymptotics poverty measures) , (2009), C. R. Math. Rep. Acad. Sci. Canada, 45-52,  31 (2). (Withe Serigne Touba Sall and Cheikh Tidiane Seck)\\

\noindent 5 - Uniform Convergence of the Non-Weighted Poverty Measures, (2009), Commun. Stat., Theory Methods 38, No. 20, 3697-3704 (2009). (With Cheikh Tidiane Seck). (Zbl pre05648823).\\ 

\noindent 6 - Uniform weak convergence of the time-dependent poverty measures for continuous longitudinal data, Brazilian Journal of Probability and Statistics, 2010, Vol. 24, No. 3, 457–467 (avec Serigne Touba Sall)\\

\noindent 7 – A Simple Note on some Empirical Stochastic Process as a Tool in Uniform L-Statistics Weak Laws.  Afrika Statistika, Special volume (5)   : Proceedings of the International Workshop on Multiple Risks and Copula, Biskra 2010, pp. 245-251. Ed. Abdelhakim Necir.\\

\noindent 8 - Asymptotic Representation Theorems for Poverty Indices.  Afrika Statistika, Special Volume (5) : Proceedings of the International Workshop on Multiple Risks and Copula, Biskra 2010, pp. 238-244. Ed. Abdelhakim Necir.  (With serigne Touba Sall)\\

\noindent 9  -  On the General Poverty Index. (2013). Far East Journal of Theoretical Statistics. Volume 42. (1), 1-22\\

\noindent 10 -  On the influence of the Theil-like inequality measure on the growth (2013). arXiv:1210.3190. Applied Mathematics, 2013, 4, 986-1000 doi:10.4236/am.2013.47136. (With Pape Djiby Mergane)\\

\noindent 11 - Functional Weak Laws for the Weighted Mean Losses or Gains and Applications Applied Mathematics Vol.6 No.5. (with Serigne Touba Sall, Pape Djiby Mergane)\\

\noindent 12 - Asymptotic Confidence Bands for Copulas Based on the Local Linear Kernel Estimator Applied Mathematics. 2015. 6 (12), 2077-2095 (with Diam Ba, Cheikh Tidiane Seck) http://dx.doi.org/10.4236/am.2015.612183\\

\noindent 14 - Robust ordering of two income distributions by means of poverty indices. Fast East Journal of Theoretical Statistics. 50 (3), 2015, pages 203-230. $http://dx.doi.org/10.1765/FJTSMay2015_203_230$ (With Cheikh Tidiane Seck)\\

\noindent 15. Asymptotic inference in poverty indices: An empirical processes approach. Communications in Statistics - Theory and Methods, 46:12, 6192-6212, DOI: 10.1080/03610926.2015.1122060 (with Cheikh Tidiane Seck and J. Ngatchou).\\

\noindent 16. Asymptotic inference in poverty indices: an empirical processes approach, Asymptotic Theory and Statistical Decomposability gap Estimation for Takayama's Index. arXiv:1701.04735 (With Pape Djiby Mergane, Cheikh Mohamed Haidara, Cheikh Tidiane Seck).\\

\noindent 17. Sur la d\'ecomposabilité empirique des indicateurs de pauvret\textbf{}. arXiv:1701.02649. (With Cheikh Mohamed Haidara)\\

\part{The Gateway}
\chapter{Introduction and Notation} \label{hfep_gateway_intro}

This chapter opens the gateway and may be considered as a portal of all the parts of on angoing \textit{A handbook of Asymptotic Representations of Statistics in the Functional Empirical process and Applications}, as we explained earlier. Its gives the main aspects of the functional empirical process (\textit{fep}) which is the tool on which depend all the results in the remainder of the book and the quoted handbook.\\

\noindent As mentioned in the introduction, the monograph deals with asymptotic normality results and their applications. But, as we know, there are so many of such results, which may be combined in a great number of ways. But how many times did we have, for example, two asymptotic normality  results of two different statistics based on the same data, or such that one of them is based on some sub-data of the other, and we cannot see how to combine them to have the the joint asymptotic laws. The same situation may occur with one statistic which is observed in different areas or over different times. To find the joint asymptotic distributions of two or more statistics, combined with areas or periods of time, we are frequently obliged to do the work anew. The famous delta method, even if it is very powerful, requires new computations each time we have new situations.\\

\noindent In many fields, we already have working and existing statistics. New ones are regularly found. It would be better to have a kind of \textbf{spaceship} on which modules are hanged. In our situations, the  spaceship is a functional Gaussian process and each module is the asymptotic representation of one statistic in terms of that Gaussian process. We may call that spaceship a Gaussian field in which the asymptotic laws of the statistics are expressed. In that way, it is possible to navigate from one module to another, that is, to find the joint distribution of any pair of statistics, to compare them with respect to the areas and the times. In order to be able to do so, we should have a broad conception at the beginning. This chapter constitutes that construction.\\

\noindent We begin by some general facts on the empirical process, in its real and functional forms. Next, we present in details the functional form.\\

\noindent  It is amazing that we will not need all the sophisticated and extremely complicated aspects of uniform convergence and tightness we necessarily have to deal with when working on weak convergences in the space of bounded functions on some space $T$ ($T=\mathbb{R}^k$, here). There are some circumstances where they are useful and handy. But for the needs of our study, the finite-distributional convergence will be enough and then the multivariate central limit theorem is just needed. The readers who are interested in detailed results in the theory of empirical processes are directed to \cite{billingsley}, \cite{gaenssler}, \cite{pollard}, \cite{vaart}, etc. For the needs for the finite-distributions scheme are, we will back on \cite{ips-wcia-ang}.\\

\noindent Before we proceed, we point out that a similar enterprise has been done in \cite{barrettdonald2000}, but using real empirical processes. As we will see latter, a huge part of the limitations due to the use of real valued empirical processes are lifted by the functional empirical process to, the most important one of them being non-linearity.

\newpage

\section{The empirical Process} \label{hfep_gateway_sec1}

\noindent \text{I - The real empirical process}.\\

\noindent Let $X,X_{1},\ X_{2},...$ be a sequence of independent and identically distributed following a real-valued cumulative distribution
function $F$ and all defined on the same probability space $(\Omega ,\mathcal{A},\mathbb{P})$. For each $n\geq 1$, we may define the empirical
distribution function associated with $X_{1}$, $X_{2}$,...,$X_{n}$ : 
\begin{equation*}
\begin{array}{ccc}
\mathbb{R}\ni x & \mapsto  & \mathbb{F}_{n}(x)=\frac{1}{n}Card\{j,\text{ }1\leq j\leq
n\text{, \ }X_{j}\leq x\}%
\end{array}
\end{equation*}

\noindent The empirical process associated with $X_{1}$, $X_{2}$,...,$X_{n}$
is defined as follows 
\begin{equation*}
\alpha _{n}(x)=\sqrt{n}(\mathbb{F}_{n}(x)-F(x)),\text{ }x\in \mathbb{R}.
\end{equation*}

\bigskip \noindent In the real case, we have the two following keys results. :\\

\noindent \textbf{The Glivenko-Cantelli Law}  :\\

\begin{equation*}
\left\Vert \mathbb{F}_{n}-F\right\Vert_{\infty} =\sup_{x\in \mathbb{R}}\left\vert
F_{n}(x)-F(x)\right\vert \rightarrow 0\text{ }a.s.\text{ }as\text{ }%
n\rightarrow +\infty .
\end{equation*}

\noindent \textbf{The Donsker Law }. The sequence of stochastic processes $(\alpha _{n}(x),$ $x\in \mathbb{R})$ weakly
convergences on $\ell ^{\infty }(\mathbb{R})$ - the space of bounded real-valued function defined on R - to a re-scaled Brownian brigde $(B(F(x)),$ $x\in \mathbb{R})$, denoted as

\bigskip 
\begin{equation*}
(\alpha _{n}(x),x\in \mathbb{R})\rightsquigarrow (B(F(x)),x\in \mathbb{R})\text{ in }\ell
^{\infty }(R)\text{ as }n\rightarrow +\infty,
\end{equation*}

\bigskip \noindent where $(B(t),t\in \lbrack 0,1])$ \ is by definition the Brownian bridge, which is a centered Gaussian
process of variance-covariance function%
\begin{equation*}
\Gamma (s,t)=\min (s,t)-st,\text{ }(s,t)\in \lbrack 0,1]^{2}.
\end{equation*}

\bigskip \noindent In many occasions, we do not need the full version of the Donsker Theorem as
we will see in the sequel. We usually only need the finite-distributional
version, which is readily proved by using multinomial probabilities, and
which is stated as below.\\

\bigskip \noindent \textbf{The finite-distribution weak law of the empirical process}. For any finite number $k\geq 1,$ and for any real numbers $x_{1}<...<x_{k},$
we have the following weak convergence on $\mathbb{R}^{k}$%
\begin{equation*}
(\alpha _{n}(x_{1}),...,\alpha _{n}(x_{k}))^{t}\rightsquigarrow
(B(F(x_{1})),...,B(F(x_{1})))^t.
\end{equation*}

\bigskip \noindent where, throughout the monograph, $x^t$ stands for the transpose of a matrix, columun or line and we consider elements of $\mathbb{R}^d$, $d\geq 1$, as columns.\\

\noindent The real empirical process has been deeply investigated, mainly in the Skorohod topology in $D(0,1)$, the space of real-valued functions defined on $[0,1]$  with at most a countable number of discontinuity points which are all of the first kind (see \cite{billingsley}, as a main reference). But for a long time, the direct approach, which by the way a counting one, had hidden the linearity of this fundamental object. And linearity brings more powerful tools from the functional analysis prospective. Define for any
$x\in \mathbb{R}$

\begin{equation*}
f_{x}=1_{]-\infty ,x]},
\end{equation*}%

\noindent we get for any fixed $n\geq 1,$%
\begin{equation*}
\alpha _{n}(x)=G_{n}(f_{x})=\frac{1}{\sqrt{n}}\sum_{j=1}^{n}\left\{
f_{x}(X_{j})-Ef_{x}(X_{j})\right\} ,
\end{equation*}

\noindent and any real numbers $x_{1},..,x_{k}$ and any $a_{1},..,a_{k}$, we have for
any fixed $n\geq 1$%
\begin{equation*}
G_{n}\left( \sum_{h=1}^{k}a_{h}f_{x_{h}}\right)
=\sum_{h=1}^{k}a_{h}G_{n}\left( f_{x_{h}}\right) .
\end{equation*}

\bigskip \noindent This properties renders much easier the study of the empirical process. This
leads to the functional approach.\\

\noindent \textbf{II - The Functional Empirical Process}.\\

\noindent Let $Z_{1}$, $Z_{2}$, ... be a sequence of independent
copies of a random variable $Z$ defined on the same probability space with $(\Omega, \mathcal{A}, \mathbb{P})$
values on some metric space $(S,d)$. The mathematical expection symbol with respect to $\mathbb{P}$ is denoted by $\mathbb{E}$ and $\mathbb{P}_{Z}=\mathbb{P} \circ Z^{-1}$ is the probability measure image of $\mathbb{P}$ by a measurable mapping $Z$. Define for each $n\geq 1,$ the
functional empirical process by 
\begin{equation*}
\mathbb{G}_{n}(f)=\frac{1}{\sqrt{n}}\sum_{j=1}^{n}(f(Z_{j})-\mathbb{E}f(Z_{j})),
\end{equation*}

\bigskip \noindent where $f$ is a real and measurable function defined on $\mathbb{R}$ such that

\begin{equation}
\mathbb{V}_{Z}(f)=\int \left( f(x)-\mathbb{P}_{Z}(f)\right)
^{2}dP_{Z}(x)<\infty ,  \label{var}
\end{equation}

\bigskip \noindent which entails

\begin{equation}
\mathbb{P}_{Z}(\left\vert f\right\vert )=\int \left\vert f(x)\right\vert
dP_{Z}(x)<\infty \text{.}  \label{esp}
\end{equation}

\bigskip \noindent Let us denote by $\mathcal{F}(S)$ - $\mathcal{F}$ for short -
the class of real-valued measurable functions that are defined on S such
that (\ref{var}) holds. The space $\mathcal{F}$ , when endowed with the
addition and the external multiplication by real scalars, is a linear space.
Next, it is remarkable that $\mathbb{G}_{n}$ is linear on $\mathcal{F}$, that
is for $f$ and $g$ in $\mathcal{F}$ and for $(a,b)\in \mathbb{R}{^{2}}$, we
have

\begin{equation*}
a\mathbb{G}_{n}(f)+b\mathbb{G}_{n}(g)=\mathbb{G}_{n}(af+bg).
\end{equation*}

\bigskip \noindent We have this result

\begin{lemma}
\label{lemma.tool.1} \bigskip Given the notation above, then for any finite
number of elements $f_{1},...,f_{k}$ of $\mathcal{S},k\geq 1,$ we have

\begin{equation*}
(\mathbb{G}_{n}(f_{1}),...,\mathbb{G}_{n}(f_{k}))^{t} \rightsquigarrow 
\mathcal{N}_{k}(0,\Gamma (f_{i},f_{j})_{1\leq i,j\leq k}),
\end{equation*}

\bigskip \noindent where 
\begin{equation*}
\Gamma (f_{i},f_{j})=\int \left( f_{i}-\mathbb{P}_{Z}(f_{i})\right) \left(
f_{j}-\mathbb{P}_{Z}(f_{j})\right) d\mathbb{P}_{Z}(x),1\leq ,j\leq k.
\end{equation*}
\end{lemma}

\bigskip \noindent \textbf{PROOF}. It is enough to use the Cram\'{e}r-Wold
Criterion (see for example \cite{billingsley}, page 45, or \cite{ips-wcia-ang}, Chapter one), that
is to show that for any $a=^{t}(a_{1},...,a_{k})\in \mathbb{R}^{k},$ by
denoting $T_{n}=^{t}(\mathbb{G}_{n}(f_{1}),...,\mathbb{G}_{n}(f_{k})),$ we
have $<a,T_{n}>\rightsquigarrow <a,T>$ where $T$ follows the $\mathcal{N}%
_{k}(0,\Gamma (f_{i},f_{j})_{1\leq i,j\leq k})$\ law and $<\circ ,\circ >$
stands for the usual product scalar in $\mathbb{R}^{k}.$ But, by the
standard central limit theorem in $\mathbb{R}$, we have%
\begin{equation*}
<a,T_{n}>=\mathbb{G}_{n}\left( \sum\limits_{i=1}^{k}a_{i}f_{i}\right)
\rightsquigarrow \mathcal{N}(0,\sigma _{\infty }^{2}),
\end{equation*}

\bigskip \noindent where, for $g=\sum_{1\leq i\leq k}a_{i}f_{i},$%
\begin{equation*}
\sigma _{\infty }^{2}=\int \left( g(x)-\mathbb{P}_{Z}(g)\right) ^{2}dP_{Z}(x)
\end{equation*}

\bigskip \noindent and this easily gives%
\begin{equation*}
\sigma _{\infty }^{2}=\sum\limits_{1\leq i,j\leq k}a_{i}a_{j}\Gamma
(f_{i},f_{j}),
\end{equation*}

\bigskip \noindent so that $\mathcal{N}(0,\sigma _{\infty }^{2})$ is the law of $<a,T>.$ The proof is
finish.

\bigskip \noindent This functional approach leads to an almost universal method for
finding the asymptotic laws of multidimensional statistics.\\

\noindent We first give, as an application of the delta method, an easy way to find simple asymptotic laws.
 
\newpage

\section[Simple Asymptotic laws deriving by the \textit{fep}]{The General and Simple Method of Using the \textit{fep} for Asymptotic laws deriving} \label{hfep_gateway_sec2}

\noindent We usually work with usual asymptotic statistics on $\mathbb{R}%
^{k}.$ Once we have our sample $Z_{1},Z_{2},...$ as random variables defined
in the same probability space with values in $\mathbb{R}^{k},$ the studied
statistics, say $T_{n},$ is usually a combinations of expressions of the form%
\begin{equation*}
H_{n}=\frac{1}{n}\sum\limits_{j=1}^{k}H(Z_{j})
\end{equation*}

\bigskip \noindent for $H\in \mathcal{F}.$ We use this simple expansion, for 
$\mu (H)=\mathbb{E}H(Z),$ 
\begin{equation}
H_{n}=\mu (H)+n^{-1/2}\mathbb{G}_{n}(H).  \label{expan}
\end{equation}

\bigskip \noindent We have that $\mathbb{G}_{n}(H)$ is asymptotically
bounded in probability since $\mathbb{G}_{n}(H)$ weakly converges to, say $%
M(H)$ and then by the continuous mapping theorem $\left\Vert \mathbb{G}%
_{n}(H)\right\Vert \rightsquigarrow \left\Vert M(H)\right\Vert .$ Since all
the $\mathbb{G}_{n}(H)$ are defined on the same probability space, we get
for all $\lambda >0,$ by the assertion of the Portmanteau Theorem for
concerning open sets,%
\begin{equation*}
\limsup_{n\rightarrow \infty }P(\left\Vert \mathbb{G}_{n}(H)\right\Vert
>\lambda )\leq P(\left\Vert M(H)\right\Vert >\lambda )
\end{equation*}

\bigskip \noindent and then 
\begin{equation*}
\liminf_{\lambda \rightarrow \infty }\lim \sup_{n\rightarrow \infty
}P(\left\Vert \mathbb{G}_{n}(H)\right\Vert >\lambda )\leq \lim \sup
P(\left\Vert M(H)\right\Vert >\lambda )=0.
\end{equation*}

\bigskip \noindent From this, we use the big $O_{\mathbb{P}}$ notation, that
is $\mathbb{G}_{n}(H)=O_{\mathbb{P}}(1).$ Formula (\ref{expan}) becomes%
\begin{equation*}
H_{n}=\mu (H)+n^{-1/2}\mathbb{G}_{n}(H)=\mu (H)+O_{\mathbb{P}}(n^{-1/2})
\end{equation*}

\bigskip \noindent and we will be able to use the delta method. Indeed, let $g:\mathbb{R}\longmapsto \mathbb{R}$ be continuously differentiable on a
neighborhood of $\mu (H).$ The mean value theorem leads to \begin{equation}
g(H_{n})=g(\mu (H))+g^{\prime }(\mu _{n}(H))\text{ }n^{-1/2}\mathbb{G}_{n}(H)
\label{expan01}
\end{equation}

\bigskip \noindent where 
\begin{equation*}
\mu _{n}(H)\in \lbrack (\mu (H)+n^{-1/2}\mathbb{G}_{n}(H))\wedge \mu
(H),(\mu (H)+n^{-1/2}\mathbb{G}_{n}(H))\vee \mu (H)]
\end{equation*}

\bigskip \noindent so that 
\begin{equation*}
\left\vert \mu _{n}(H)-\mu (H)\right\vert \leq n^{-1/2}\mathbb{G}_{n}(H)=O_{%
\mathbb{P}}(n^{-1/2}).
\end{equation*}

\bigskip \noindent Then $\mu _{n}(H)$ converges to $\mu _{n}(H)$ in
probability (denoted $\mu _{n}(H)$ $\rightarrow _{\mathbb{P}}\mu (H)).$ But
the convergence in probability to a constant is equivalent to the weak
convergence. Then $\mu _{n}(H)$ $\rightsquigarrow \mu (H).$ Using again the
continuous mapping theorem, $g^{\prime }(\mu _{n}(H))\rightsquigarrow
g^{\prime }(\mu (H))$ which in tern yields $g^{\prime }(\mu
_{n}(H))\rightarrow _{\mathbb{P}}g^{\prime }(\mu (H))$ by the
characterization of the weak convergence to a constant. Now (\ref{expan01})
becomes
\begin{eqnarray*}
g(H_{n}) &=&g(\mu (H))+(g^{\prime }(\mu (H)+o_{P}(1))\text{ }n^{-1/2}\mathbb{%
G}_{n}(H) \\
&=&g(\mu (H))+g^{\prime }(\mu (H)\times \text{ }n^{-1/2}\mathbb{G}%
_{n}(H)+o_{P}(1))\text{ }n^{-1/2}\mathbb{G}_{n}(H) \\
&=&g(\mu (H))+\text{ }n^{-1/2}\mathbb{G}_{n}(g^{\prime }(\mu
(H)H)+o_{P}(n^{-1/2})
\end{eqnarray*}

\bigskip \noindent We arrive at the final expansion%
\begin{equation}
g(H_{n})=g(\mu (H))+\text{ }n^{-1/2}\mathbb{G}_{n}(g^{\prime }(\mu
(H)H)+o_{P}(n^{-1/2}).  \label{expanFinal}
\end{equation}

\bigskip \noindent The method consists in using the expansion (\ref%
{expanFinal}) as many times as needed and next to do some algebra on these
expansions. By using the same techniques as above, we have the following three
formulas

\begin{lemma}
\label{lemma.tool.2} Let ($A_{n})$ and ($B_{n})$ be two sequences of real
valued random variables defined on the same probability space holding the
sequence $Z_{1}$, $Z_{2}$, .. Let A and B be two real numbers and let $L(z)$
and $H(z)$ be two real-valued functions$\ of$ $z\in S.$ Suppose that $%
A_{n}=A+n^{-1/2}\mathbb{G}_{n}(L)+o_{P}(n^{-1/2})$ and $A_{n}=B+n^{-1/2}%
\mathbb{G}_{n}(H)+o_{P}(n^{-1/2}).$ Then

\begin{equation*}
A_{n}+B_{n}=A+B+n^{-1/2}\mathbb{G}_{n}(L+H)+o_{P}(n^{-1/2}),
\end{equation*}%
\begin{equation*}
A_{n}B_{n}=AB+n^{-1/2}\mathbb{G}_{n}(BL+AH)
\end{equation*}%
and if $B\neq 0,$%
\begin{equation*}
\frac{A_{n}}{B_{n}}=\frac{A}{B}+n^{-1/2}\mathbb{G}_{n}(\frac{1}{B}L-\frac{A}{%
B^{2}}H)+o_{P}(n^{-1/2})
\end{equation*}
\end{lemma}

\bigskip \noindent By putting together all the described steps in a smart way, the methodology will lead us to a final result of the form%
\begin{equation*}
T_{n}=T+n^{-1/2}\mathbb{G}_{n}(h)+o_{P}(n^{-1/2}),
\end{equation*}

\bigskip \noindent where

$$
h=\frac{1}{B}L-\frac{A}{B^2}H,
$$

\bigskip \noindent which entails the following weak convergence

\begin{equation} \label{repSimple}
\sqrt{n}(T_{n}-T)=\mathbb{G}_{n}(h)+o_{P}(1)\rightsquigarrow N(0,\Gamma
(h,h)).
\end{equation}

\bigskip 

\noindent We are now in position to apply right here the methodology in the
welfare environment. But, we need for once a broad constriction in which we
may achieve anything we want to have. So we need :\\

\noindent a) to have the essential key of the Bahadur representation laws which allows
to deal with L-Statistics.\\

\noindent b) to combine both the real and the functional approaches.\\

\noindent c) to integrate the copula methodology through the Sklar's theorem from the
computation prospective.\\

\bigskip \noindent The notation used in the paper may be seen as complicated, but
knowing the following simple facts may help in making them very
comprehensive. The subscript $(1)$ means that we are working in one dimension, where the randoms variables do not have a superscript. In
dimension 2, we always have the subscript (2) to main functions : \textit{cdf%
}'s, copulas, empirical process,etc. When followed by $i$, like $F_{(2),i}$,
it refers to a margin. For example $F_{(2),1}$ is the first marginal \textit{%
cdf} of $F_{(2)}$. Still in dimension 2, any superscript $i=1,2$ refers to
the first coordinate of a couple.\newline

\section{Notations and Probability Space} \label{hfep_gateway_sec3} 

\noindent In this Subsection, we complete the notations we already gave and
precise our probability space.\newline

\noindent \textbf{Univariate frame}. \noindent We are going to describe the
general Gaussian field in which we present our results. Indeed, we use a
unified approach when dealing with the asymptotic theories of the welfare
statistics. It is based on the Functional Empirical Process (\textit{fep})
and its Functional Brownian Bridge (\textit{fbb}) limit. It is laid out as
follows.\newline

\noindent When we deal with the asymptotic properties of one statistic or
index at a fixed time, we suppose that we have a non-negative random
variable of interest which may be the income or the expense $X$ whose
probability law on $(\mathbb{R},\mathcal{B}(\mathbb{R}))$, the Borel
measurable space on $\mathbb{R}$, is denoted by $\mathbb{P}_{X}.$ We
consider the space $\mathcal{F}_{(1)}$ of measurable real-valued functions $%
f $ defined on $\mathbb{R}$\ such that 
\begin{equation*}
V_{X}(f)=\int (f-\mathbb{E}_{X}(f))^{2}d\mathbb{P}_{X}=\mathbb{E}(f(X)-%
\mathbb{E}(f(X))^{2}<+\infty ,
\end{equation*}

\noindent where 
\begin{equation*}
\mathbb{E}_{X}(f)=\mathbb{E}f(X).
\end{equation*}

\bigskip \noindent On this functional space $\mathcal{F}_{(1)},$\ which is
endowed with the $L_{2} $-norm 
\begin{equation*}
\left\Vert f\right\Vert _{2}=\left( \int f^{2}d\mathbb{P}_{X}\right) ^{1/2},
\end{equation*}

\noindent we define the Gaussian process $\{\mathbb{G}_{(1)}(f),f\in 
\mathcal{F}_{(1)}\},$ which is characterized by its variance-covariance
function

\begin{equation}
\Gamma_{(1)}(f,g)=\int^{2}(f-\mathbb{E}_{X}(f))(g-\mathbb{E}_{X}(g))d\mathbb{P}_{X},(f,g)\in \mathcal{F}_{(1)}^{2}. \label{Gamma1}
\end{equation}

\noindent This Gaussian process is the asymptotic weak limit of the sequence of functional empirical processes (fep) defined as follows. Let $X_{1},X_{2},...$ be a sequence of independent copies of $X$. For each $n\geq1$, we define the functional empirical process associated with $X$ by 
\begin{equation*}
\mathbb{G}_{n,(1)}(f)=\frac{1}{\sqrt{n}}\sum_{j=1}^{n}(f(X_{j})-\mathbb{E}f(X_{j})),f\in \mathcal{F}_{(1)},
\end{equation*}

\noindent and denote the integration with respect to the empirical measure by

\begin{equation*}
\mathbb{P}_{n,(1)}(f)=\frac{1}{n} \sum_{i=1}^{n}f(X_{i}), \ f\in \mathcal{F}%
_{(1)},
\end{equation*}

\noindent Let us denote by $\ell^{\infty }(T)$ the space of real-valued bounded
functions defined on $T=\mathbb{R}$ equipped with its uniform topology. In
the terminology of the weak convergence theory, the sequence of objects $%
\mathbb{G}_{n,(1)}$ weakly converges to $\mathbb{G}_{(1)}$ in $\ell^{\infty}(%
\mathbb{R})$, as stochastic processes indexed by $\mathcal{F}_{(1)}$,
whenever it is a Donsker class. The details of this highly elaborated theory
may be found in \cite{billingsley}, \cite{pollard},  \cite{vaart} and similar sources.\newline

\noindent We only need the convergence in finite distributions which is a
simple consequence of the multivariate central limit theorem, as described
in Chapter 3 in \cite{ips-wcia-ang}.\newline

\noindent We will use the Renyi's representation of the random variable $X_i$%
's of interest by means (\textit{cdf}) $F_{(1)}$ as follows 
\begin{equation*}
X=_{d}F_{(1)}^{-1}(U),
\end{equation*}

\noindent where $U$ is a uniform random variable on $(0,1)$, $=_{d}$ stands
for the equality in distribution and $F^{-1}_{(1)}$ is the generalized
inverse of $F_{(1)}$, defined by

\begin{equation*}
F_{(1)}^{-1}(s)=\inf \{x, F_{(1)}(x)\geq s\}, \ s \in (0,1).
\end{equation*}

\bigskip \noindent Based on these representations, we may and do assume that we are
on a probability space $(\Omega,\mathcal{A},\mathbb{P})$ holding a sequence
of independent $(0,1)$-uniform random variables $U_1$, $U_2$, ..., and the
sequence of independent observations of $X$ are given by

\begin{equation}
X_{1}=F_{(1)}^{-1}(U_1), \ \ X_{2}=F_{(1)}^{-1}(U_2), \ \ etc.
\label{repRenyi}
\end{equation}

\bigskip  \noindent For each $n\geq 1$, the order statistics of $U_1,...,U_n$
and of $X_1,...,X_n$ are denoted respectively by $0 \equiv U_{0,n} < U_{1,n}\leq \cdots \leq
U_{n,n} <U_{n+1,n}=\equiv 1$ and $X_{1,n}\leq \cdots \leq X_{n,n}$.\newline

\noindent \noindent To the sequences of $(U_n)_{n\geq 1}$, we also associate
the sequence of real empirical functions

\begin{eqnarray}
\mathbb{U}_{n,(1)}(s)&=&\frac{1}{n} \#\{j,1\leq j \leq n, \ U_j \leq s\}, \
s\in(0,1) \ n\geq 1  \label{empiricalFunctionU}\\
&=& \sum_{j=1}^{n} \frac{j}{n} 1_{({U_{j,n}\leq s < U_{j+1,n}})},
\end{eqnarray}

\bigskip \noindent and the sequence of real uniform quantile functions 
\begin{equation}
\mathbb{V}_{n,(1)}(s)=U_{1,n}1_{(s=0)}+\sum_{j=1}^{n}U_{j,n}1_{((j-1)/n < s \leq (j/n))}, \ s\in(0,1), \
n\geq 1  \label{quantileFunctionU}
\end{equation}

\bigskip \noindent and next, the sequence of real uniform empirical processes 
\begin{equation}
\alpha_{n,(1)}(s)=\sqrt{n}(\mathbb{U}_{n,(1)}-s), 
\label{empiricalProcessU}
\end{equation}

\bigskip \noindent for $s\in (0,1)$ and $\geq 1$, and the sequence of real uniform quantile processes

\begin{equation}
\gamma_{n,(1)}(s)=\sqrt{n}(s-\mathbb{V}_{n,(1)}), \ s\in(0,1) \ n\geq 1.
\label{quantileProcessU}
\end{equation}

\bigskip \noindent The same can be done for the sequence $(X_n)_{n\geq 1}$,
and we obtain the associated sequence of real empirical processes a

\begin{equation}
\mathbb{G}_{n,r,(1)}(x)=\sqrt{n} \left( \mathbb{F}_{n,(1)}(x)-F_{(1)}(x)%
\right), \ x\in \mathbb{R}, \ n\geq 1  \label{empiricalProcess}
\end{equation}

\bigskip \noindent where

\begin{equation}
\mathbb{F}_{n,(1)}(x)=\frac{1}{n} \#\{j,1\leq j \leq n, \ X_j \leq x\}, \ x\in \mathbb{R} \ n\geq 1  \label{empiricalFunction}
\end{equation}

\noindent is the associated sequence of empirical functions. We also have
the associated sequence of quantile processes

\begin{equation}
\mathbb{Q}_{n,(1)}(x)=\sqrt{n} \left( \mathbb{F}^{-1}_{(n),(1)}(s) -
F^{-1}(s) \right), \ s\in (0,1), \ n\geq 1  \label{quantileProcess}
\end{equation}

\noindent where, for $n\geq 1$,

\begin{equation}
\mathbb{F}^{-1}_{n,(1)}(s)=X_{1,n}1_{(0\leq s \leq 1/n)}+\sum_{j=1}^{n}X_{j,n}1_{((j-1)/n\leq s \leq (j/n))}, \ s\in(0,1),
\label{quantileFunction}
\end{equation}

\noindent is the associated sequence of quantile processes.\newline

\noindent By passing, we recall that $\mathbb{F}^{-1}_{n,(1)}$ is actually
the generalized inverse of $\mathbb{F}_{(n),(1)}$ and for the uniform
sequence, we have

\begin{equation}
\mathbb{V}_{n,(1)}=\mathbb{U}^{-1}_{n,(1)}  \label{invCDFEMP}
\end{equation}

\noindent In virtue of Representation (\ref{repRenyi}), we have the
following remarkable relations

\begin{equation}
\mathbb{G}_{n,r,(1)}(x)=\alpha_{n,(1)}(F_{(1)}(x)), \ x\in \mathbb{R} \label{EmpEmpprocess}
\end{equation}

\bigskip \noindent and

\begin{equation}
\mathbb{Q}_{n,(1)}(x)=\sqrt{n}\left( F^{-1}_{(1)}(\mathbb{V}_{n,(1)}(s))-
F^{-1}_{(1)}(s)\right) \ s\in (0,1), \ n\geq 1,  \label{QQprocess}
\end{equation}

\bigskip \noindent We also have the following relations between the
empirical functions and quantile functions

\begin{equation}
\mathbb{F}_{n,(1)}(x)=\mathbb{U}_{n,(1)}(F_{(1)}(x)), \ x\in \mathbb{R}
\label{EEFunction}
\end{equation}

\bigskip \noindent and

\begin{equation}
\mathbb{F}^{-1}_{n,(1)}(s)=F^{-1}_{(1)}(\mathbb{V}_{(n),(1)}(s)), \ s\in
(0,1), \ n\geq 1.  \label{QQFunction}
\end{equation}

\noindent As well, the real and functional empirical processes are related
as follows : for $n\geq 1$,

\begin{equation}
\mathbb{G}_{n,r,(1)}(x)=\mathbb{G}_{n,(1)}(f_{x}^{\ast}), \
\alpha_{n,(1)}(s)=\mathbb{G}_{n,(1)}(\tilde{f}_s), \ s \in (0,1), \ x \in \mathbb{R},
\label{empiricalProcessRealFonct}
\end{equation}

\bigskip \noindent where for any $x \in \mathbb{R}$, $f_{x}^{\ast}=1_{]-%
\infty,x]}$ is the indicator function of $]-\infty,x]$ and for $s \in (0,1)$%
, $f_s=1_{[0,s]}$ and $\tilde{f}_s=1_{]-\infty,F^{-1}_{(1)}(s)]}$.\newline

\bigskip \noindent To finish the description, a result of Kiefer-Bahadur
(See \cite{bahadur66}) that says that the addition of the sequences of
uniform empirical processes and quantiles processes (\ref{empiricalProcessU}%
) and (\ref{quantileProcessU}) is asymptotically, and uniformly on $[0,1]$,
zero in probability, that is

\begin{equation}
\sup_{s\in [0,1]} \left\vert \alpha_{n,(1)}(s)+\gamma_{n,(1)}(s) \right\vert
=o_{\mathbb{P}}(1) \text{ as } n\rightarrow +\infty.  \label{bahadurRep}
\end{equation}

\noindent This result is a powerful tool to handle the rank statistics when
our studied statistics are $L$-statistics.\newline

\bigskip \noindent \textbf{Bivariate frame}. As to the bivariate case, we use the Sklar's theorem (See \cite{sklar}). We can also refer to \cite{losklar} for a quick proof of Sklar's Theorem. Let us begin to define a copula in $\mathbb{R}^2$ as bivariate probability distribution function $C(u,v)$, $(u,v)\in \mathbb{R}^2$ with support $[0,1]^2$ and with $[0,1]$-uniform margins, that is

\begin{equation*}
C(u,v)=0 \text{ for } (u,v)\in]-\infty,0[\times \mathbb{R}.
\end{equation*}

\noindent Let us denote by $F_{(2)}$ the bivariate distribution function of our random couple $Y=(X^{(1)}, X^{(2)})$ and by $F_{(21)}$ and $F_{(22)}$ its margins, which are the \textit{cdf} of $X^{(1)}$ and $X^{(2)}$ respectively. The Sklar's theorem (\cite{sklar}) says that there exists a copula $C_{(2)}$ such that we have

\begin{equation}
F_{(2)}(x,y)=C_{(2)}(F_{(21)}(x), F_{(22)}(y)), \text{ for any } (x,y)\in 
\mathbb{R}^2.  \label{theoSklar}
\end{equation}

\noindent This copula is unique if the marginal \textit{cdf}'s are continuous. In this paper, we will suppose that the marginal \textit{cdf}'s are continuous and then $C_{(2)}$ is unique and fixed for once. By the Kolmogorov Theorem, there exists a probability space $(\Omega,\mathcal{A},\mathbb{P})$ holding a sequence of independent random couples $(U^{(1)}_n,U^{(2)}_n)$, $n\geq 1$, of common bivariate distribution function $C_{(2)}$. On that space the random couples $(F_{(21)}^{-1}(U^{(1)}_n), \ F_{(22)}^{-1}(U^{(2)}_n))$ are independent and have a common bivariate
distribution function equal to $C_{(2)}$, since

\begin{eqnarray*}
&&\mathbb{P}(F_{(21)}^{-1}(U^{(1)}_i)\leq x_{1}, \
F_{(22)}^{-1}(U^{(2)}_i)\leq x_{2}) \\
&=&\mathbb{P}(U^{(1)}_i \leq F_{(21)}(x_{1}), \ U^{(2)}_i\leq
F_{(22)}(x_{2})) \\
&=&C_{(2)}(F_{(21)}(x_{1}), \ F_{(22)}(x_{2})) \\
&=& F_{(2)}(x_{1}, \ x_{2}),
\end{eqnarray*}

\noindent by (\ref{theoSklar}), and where we applied the general formula for
generalized inverses functions for a \textit{cdf} :

\begin{equation*}
F^{-1}(s) \leq y \Leftrightarrow s \leq F(x), \text{ for } (s,x) \in
[0,1]\times \mathbb{R}.
\end{equation*}

\bigskip \noindent For more on interesting properties of generalized inverses of
monotone functions, see \cite{ips-wcia-ang}, Chapter 4.\newline

\noindent Based on this remark, we place ourselves on the probability space
holding the sequence of independent random couples $(U^{(1)},U^{(2)})$, $%
(U^{(1)}_n,U^{(2)}_n)$, $n \geq 2$, with common distribution function $%
C_{(2)}$, and the observations from $Y=(X^{(1)},
X^{(2)})=(F_{(2),1}^{-1}(U^{(1)}),F_{(2),2}^{-1}(U^{(2)}))$, are generated
as follows :

\begin{equation}  \label{renyidim2}
Y_n=(F_{(21)}^{-1}(U^{(1)}_n),F_{(22)}^{-1}(U^{(2)}_n)), \ n\geq 1.
\end{equation}

\noindent We may directly study the empirical process

\begin{equation}
\mathbb{G}_{n,(2)}(h) = \frac{1}{\sqrt{n}}\,\sum_{j=1}^{n} \left( h(X_{j}^{(1)},X_{j}^{2}) - \mathbb{P}_{(X^{(1)},X^{(2)})}(h) \right).  \label{empProcXY}
\end{equation}

\noindent where $h \in L_2(\mathbb{R}^2, \ \mathcal{B}(\mathbb{R}^2), \mathbb{P}_{(X^{(1)},  X^{(2)})})$.\\

\noindent In this setting, we rather use the bidimensional functional empirical process based on $\left\{\left((U_{i}^{(1)},U_{i}^{(2)})\right) \right\}_{i=1,\ldots,n}$ and defined by

\begin{equation}
\mathbb{G}_{n,u,(2)}\left(\widetilde{h}\right) = \frac{1}{\sqrt{n}}\,\sum_{j=1}^{n}\,%
\left( \widetilde{h}  \left(U_{j}^{(1)},U_{j}^{(2)})\right) - \mathbb{P}%
_{\left((U^{(1)},U^{(2)})\right)}\left( \widetilde{h}\right)\right),  \label{empProcUV}
\end{equation}

\noindent whenever $\widetilde{h}$ is a function of $(u,v)\in [0,1]^2$ such that $%
\mathbb{E}(\widetilde{h}(U^{(1)},U^{(2)})^2)$ is finite.\newline

\noindent For any Donsker class $\mathcal{F}_{(2)}([0,1]^2)$, the stochastic
process $\mathbb{G}_{n,u,(2)}$ converges to a Gaussian process $\mathbb{T}$
with variance-covariance function, for $(f,g) \in L_{(2)}^{2}\biggr([0,1]^2,\mathbb{P}_{\left(U^{(1)},U^{(2)}\right)}\biggr)$, denoted by $\widetilde{\Gamma}_{(2)}\left(f,g\right)$, is given the following Formula we name (GammaStar) \label{GammaStar}

\begin{equation*}
\int_{[0,1]^2}\left(f(u,v) - \mathbb{P}_{\left(U^{(1)},U^{(2)}\right)}\left(f\right)\right)\left(g(u,v) - \mathbb{P}_{\left(U^{(1)},U^{(2)}\right)}\left(g\right)\right)\,dC(u,v)
\end{equation*}

\bigskip \noindent with

\begin{equation*}
\mathbb{P}_{\left(U^{(1)},U^{(2)}\right)}\left(f\right)=\mathbb{E}\left(f\left(U^{(1)},U^{(2)}\right) \right)=\int_{[0,1]^2}\,f(u,v)\,dC(u,v)
\end{equation*}

\noindent and the same is true for $g$. So, by using the transform

\begin{equation}
\widetilde{h}(s,t)=h\left(F^{-1}_{(2),1}(s),F^{-1}_{(2),2}(t)\right), \ (s,t) \in [0,1]^2,   \label{empProcXYUV}
\end{equation}

\noindent and the representation (\ref{renyidim2}), we get the remarkable following relation for any $h$, whenever one of the members makes sense,

\begin{equation}
\mathbb{G}_{n,(2)}\left(h\right) = \mathbb{G}_{n,u,(2)}\left(\widetilde{h}\right).  \label{empProcXY}
\end{equation}

\bigskip \noindent All the needed notation are now complete and will allow the expression of the asymptotic theory we undertake here.\newline

\section{The residual empirical process} \label{hfep_gateway_sec4}

\noindent \textbf{(A) - The origin}.\\

\noindent There is a considerable class of statistics which are combinations of one dimensional statistics of the form

$$
L_n=d_n \sum_{1\leq j \leq n} c(j,n) q_0(X_{j,n}), \ n\geq 1
$$

\bigskip \noindent where $q_0$ is some measurable mapping, $c(\circ,n)$ a function of $j \in \{1,\cdots,n\}$ and $(d_n)_{n\geq 1}$ is a sequence of real numbers. If $F_{(1)}$ is continuous, we may use the rank statistics $(R_{1,n}, \cdots, R_{n,n})$ defined by

$$
\forall 1\leq i \leq n, \ \forall 1\leq j \leq n, \ R_{j,n}=i \Leftrightarrow X_{i,n}=X_{j}.
$$

\bigskip \noindent Thus, for $n\geq 1$, $L_n$ becomes

$$
L_n=\sum_{1\leq j \leq n} \mathbb{F}_{n,(1)} q_0(X_{j}).
$$ 

\bigskip \noindent But it happens that for any $n\geq 1$, for any $1\leq j \leq n$,

$$
\frac{R_{j,n}}{n}=\mathbb{F}_{n,(1)}(X_j),
$$

\bigskip \noindent and this leads to

$$
L_n=\frac{1}{n} \sum_{1\leq j \leq n} \biggr(n d_n c\left(n\mathbb{F}_{n,(1)}(X_j)\right)\biggr) q_0(X_{j}), \ n\geq 1.
$$ 

\bigskip \noindent Fortunately, in many cases, there exists a measurable mapping $g$ such that $\mathbb{E}(|X|)<+\infty$ and

\begin{eqnarray*}
L_n&=&\frac{1}{n} \sum_{1\leq j \leq n} q_1(F_{(1)}(X_j)) q_0(X_{j})\\
&+&\frac{1}{n} \sum_{1\leq j \leq n} \biggr(n d_n c\left(n\mathbb{F}_{n,(1)}(X_j)\right)- q_1(F_{(1)}(X_j))\biggr) q_0(X_{j})\\
\end{eqnarray*}

\bigskip \noindent and that, by means of the mean value theorem,

\begin{eqnarray*}
&&\frac{1}{n} \sum_{1\leq j \leq n} \biggr(n d_n c\left(n\mathbb{F}_{n,(1)}(X_j)\right)- q_1(F_{(1)}(X_j))\biggr) h(X_{j})\\
&=&\frac{1}{n} \sum_{1\leq j \leq n} (\mathbb{F}_{n,(1)}(X_j)-F_{(1)}(X_j)) q_3(X_j) q_1(X_{j})\\
&+& o_{\mathbb{P}}(n^{-1/2}).
\end{eqnarray*}

\bigskip \noindent Upon specific conditions to be checked, we arrive at the form

$$
L_n=\frac{1}{n} \sum_{1\leq j \leq n} h(X_{j}) + \frac{1}{n} \sum_{1\leq j \leq n} (\mathbb{F}_{n,(1)}(X_j)-F_{(1)}(X_j)) q(X_j) + o_{\mathbb{P}}(n^{-1/2}).
$$  

\bigskip \noindent We conclude that, in our effort to asymptotically represent $L_n$ as an application of the empirical measure to some function $h$, that is $\mathbb{P}_n(h)$, we still have a residual term in the form of

$$
Re_n(\ell)=  \frac{1}{n}\sum_{j=1}^{n}\left(\mathbb{F}_{n,(1)}(X_j) - F_{(1)}(X_j) \right) q(X_j).
$$ 

\bigskip \noindent This made \cite{logs} to name it a residual empirical process and proceeded to its independent study.\\

\noindent Now let us describe deeper this stochastic process.\\

\bigskip \noindent \textbf{(B) - Residual empirical processes}.\\

\noindent A residual empirical process is any stochastic process of the form

\begin{equation*}
Re_n(\ell)=  \frac{1}{n}\sum_{j=1}^{n}\left(\mathbb{F}_{n,(1)}(X_j) - F_{(1)}(X_j) \right) q(X_j).
\end{equation*} 

\bigskip \noindent where $q$ is a measurable function from $[0,1]$ to $\mathbb{R}$ and 

$$
\ell(s)=q(F^{-1}_{(1)}(s)), \ s \in (0,1)
$$

\noindent and

$$
\Delta_n(s)=\biggr(\ell\left(\mathbb{V}_{n,(1)}(s)\right)-\ell(s)\biggr), \ s\in (0,1). 
$$

\bigskip \noindent \textbf{We stress} that the function $\ell$ depends of the \textit{cdf} $F_{(1)}$ and should bave been denoted $\ell(\circ)=\ell(F_{(1)},\circ)$. This warning is important in the situation of spatial analysis, as we will see it.\\

\subsection{General result} \label{hfep_gateway_ssec41}

\begin{theorem} \label{theoGenRes} If the following two assertions :\\

\noindent (1) (CRe1) $\mathbb{E}q(X)<+\infty$ \label{cre1}\\

\noindent and,\\

\noindent (2) and, as $n \rightarrow +\infty$, \label{cre2}
$$
\int_{0}^{1} \sqrt{n} \left( s - \mathbb{V}_{n,(1)}(s)\right) \Delta_n(s)\, ds \rightarrow 0 \ \ (CRe2)
$$ 

\bigskip \noindent holds,  we have the representation

$$
\sqrt{n} Re_n(\ell) = \int_{0}^{1} \mathbb{G}_{n,(1)}(\tilde{f}_s)\,\ell(s)\,ds + o_p(1),
$$
\end{theorem}

\noindent \textbf{Proof}. By using Formulas (\ref{empiricalFunctionU}) and (\ref{quantileFunctionU}), we get

\begin{equation*}
Re_n = \sum_{j=1}^{n} \int_{\frac{j-1}{n}}^{\frac{j}{n}}\left\{
\mathbb{F}_{n,(1)}(\mathbb{F}_{n,(1)}^{-1}(s)) - F_{(1)}(\mathbb{F}_{n,(1)}^{-1}(s))\right\}\, q\left(\mathbb{F}_{n,(1)}^{-1}(s)\right) \,ds,
\end{equation*}

\noindent and hence

\begin{equation}  \label{Rn}
Re_n = \int_{0}^{1}\left\{ F_{n,(1)}(F_{n,(1)}^{-1}(s)) -F_{(1)}(F_{n,(1)}^{-1}(s))\right\}\, q\left(F_{n,(1)}^{-1}(s)\right) \,ds.
\end{equation}

\bigskip

\noindent By using Formulas (\ref{invCDFEMP}), (\ref{EmpEmpprocess}) and (\ref{QQprocess}), we get
\begin{eqnarray*}
\sqrt{n} Re_n &=& - \int_{0}^{1} \sqrt{n} \left\{\mathbb{U}_{n,(1)}\left( \mathbb{V}_{n,(1)}(s) \right) - \mathbb{V}_{n,(1)}(s)\right\}\, q\left(F_{(1)}^{-1}\left(\mathbb{V}_{n,(1)}(s)\right)\right) \, ds\\ 
&=& - \int_{0}^{1} \sqrt{n} \left( s - \mathbb{V}_{n,(1)}(s)\right) q\left(F_{(1)}^{-1}\left(\mathbb{V}_{n,(1)}(s)\right)\right) \, ds\\
&-& \int_{0}^{1} \sqrt{n} \left( \mathbb{U}_{n,(1)}\left(\mathbb{V}_{n,(1)}(s)\right) -s\right)\, q\left(F_{(1)}^{-1}\left(\mathbb{V}_{n,(1)}(s)\right)\right)\, ds\\
&= :& Re_n(1) + Re_n(2).
\end{eqnarray*}

\bigskip

\noindent From \cite{shwell} (page 585), we have

\begin{equation*}
\sup_{0\leq s\leq 1} \left| \mathbb{U}_{n,(1)}\left(\mathbb{V}_{n,(1)}(s) \right)
-s \right| \leq \frac{1}{n}.
\end{equation*}

\bigskip \noindent We get

\begin{eqnarray*}
\left| Re_n(2) \right| &\leq&  \frac{1}{\sqrt{n}} \int_{0}^{1}  q\left(F_{(1)}^{-1}\left(\mathbb{V}_{n,(1)}(s)\right)\right)\, ds\\
&=& \frac{1}{\sqrt{n}} \left( \frac{1}{n} \sum_{j=1}^{n} q(X_j)\right), \ \ (CRe0)
\end{eqnarray*}

\bigskip \noindent which is an $o_{\mathbb{P}}(n^{-1/2})$ whenever $\mathbb{E}q(X)$ is finite. Under the Assumption $(Re2)$, we may replace $q\left(F_{(1)}^{-1}\left(\mathbb{V}_{n,(1)}(s)\right)\right)$ by 
$q\left(F_{(1)}^{-1}\left(s\right)\right)$, to get

\begin{eqnarray}
\sqrt{n} Re_n &=& - \int_{0}^{1} \sqrt{n} \left( s - \mathbb{V}_{n,(1)}(s)\right)F_{(1)}^{-1}\left(\mathbb{V}_{n,(1)}(s)\right)\, ds + o_p(1) \notag \\
&=& - \int_{0}^{1}\gamma_{n,(1)}(s) q\left(F_{(1)}^{-1}(s)\right) \,ds + o_p(1), \label{bahadur1}
\end{eqnarray}

\bigskip \noindent and by using the Bahadur's representation (See Formula \ref{bahadurRep}) and by applying Formula \ref{empiricalProcessRealFonct}, we arrive at

\begin{equation*}
\sqrt{n} Re_n = \int_{0}^{1} \mathbb{G}_{n,(1)}(\tilde{f}_s)\,\ell(s)\,ds + o_p(1),
\end{equation*}

\bigskip \noindent whenever 

$$
\mathbb{E}(\ell(X))=\int_{0}^{1} q(F_{(1)}^{-1}(s)) \,ds<+\infty.
$$

\bigskip \noindent This concludes the proof.\\

\subsection{Checking the Conditions (Re1) and (Re2)} \label{hfep_gateway_ssec42}

We preferred to state Theorem \ref{theoGenRes} with general the condition $(Re2)$ and not to enter in detailed forms based on convergence theorems. Instead, in each case, we will check whether or not they hold. Let us give here some general more specific conditions based on properties of the empirical process. Let us go back to the place  where we apply $(Re2)$ in the proof, that is, in Formula 
(\ref{bahadur1}). First, we replace $\gamma_{n,(1)}$ by the uniform empirical process $\mathbb{G}_{n,(1),r}$ to have

\begin{eqnarray*}
\sqrt{n} Re_n &=& \int_{0}^{1} \mathbb{G}_{n,(1),r}(s) \ell(\mathbb{V}_{n,(1)}(s)) \ ds \\
&-& \int_{0}^{1} \left(\gamma_{n,(1)}(s)+\mathbb{G}_{n,(1),r}(s)\right) \ell(\mathbb{V}_{n,(1)}(s)) \ ds + o_p(1).\\
&=:&Re_n(3)+Re_n(4)+ o_p(1).
\end{eqnarray*}

\bigskip \noindent The exact rate of convergence in the Bahadur-Kiefer Theorem (See \cite{shwell}, p.620) is $a_n=n^{-1/2}(\log \log n)^{1/4}$, $n>e$, that is

$$
\limsup_{n\rightarrow +\infty} \sup_{0\leq s \leq 1} a_n |\gamma_{n,(1)}(s)+\mathbb{G}_{n,(1),r}(s)|/a_n=1/2, \ a.s.
$$
 
\bigskip
\noindent A condition that $Re_n(3)=o_P(1)$ is

$$
\limsup_{n\rightarrow +\infty} a_n \int_{0}^{1}  \ell(\mathbb{V}_{n,(1)}(s)) \ ds.
$$

\bigskip
\noindent This is obviously true if $q$ is bounded, which will be the case in many situation. Next, we may write

$$
Re_n(4) = \int_{0}^{1} \mathbb{G}_{n,(1),r}(s) \ell(s) \ ds+ Re_n(5),
$$

\bigskip \noindent with, for $\nu$ fixed such that $0<\nu<1$,

\begin{eqnarray*}
Re_n(5)&=&\int_{0}^{1} \mathbb{G}_{n,(1),r}(s) \Delta_n(s)  ds\\
&\leq& \int_{0}^{1} (s(1-s))^{1-\nu} \sup_{0\leq s \leq 1} \biggr|\frac{\mathbb{G}_{n,(1),r}(s}{(s(1-s))^{1-\nu}}\biggr| |\Delta_n(s)|  ds\\
\end{eqnarray*}

\noindent But we have

$$
\Delta_n=\sup_{0\leq s \leq 1} \biggr|\frac{\mathbb{G}_{n,(1),r}(s}{(s(1-s))^{1-\nu}}\biggr|=O_P(1), \ as \ n\rightarrow +\infty.
$$

\bigskip \noindent (See for instance \cite{cscshm}, Formulas 2.7, 2.8, 4.2.18, third and fourth formulas in page 69, first formula in page 70). Now, a condition that $Re_n(5)=o_P(1)$ is

$$
\int_{0}^{1}  (s(1-s))^{1-\nu} \ell(\mathbb{V}_{n,(1)}(s)) \ ds=O_P(1), \ as \ n\rightarrow +\infty,
$$

\bigskip \noindent which, by $(Re1)$, is obviously obtained whenever

$$
\int_{0}^{1}  (s(1-s))^{1-\nu} \delta_n(s)=o_P(1), \ as \ n\rightarrow +\infty, \ \ (CRe4).
$$

\bigskip \noindent which is obtained if $\ell$, for instance, $\ell$ is continuous, and hence uniformly, on $(0,1)$.\\

\noindent Remind that $(Cre1)$ was used first in Formula $(CRe0)$ above. In reality, the conclusion was obtained if
$$
\frac{1}{\sqrt{n}} \left( \frac{1}{n} \sum_{j=1}^{n} q(X_j)\right)=o_P(1), \ as \ n\rightarrow +\infty, \ \ (CRe3).
$$

\noindent In conclusion, the result in Theorem \ref{theoGenRes} is still valid if the latter formulas $(CRe3)$ and (CRe4) hold.

\section{General handling} \label{hfep_gateway_sec5}

\noindent Let us show how works the methodology.\\

\noindent \textbf{Part A : Fixed time scheme}.\\

\noindent For a number of statistics, the representation of the form (\ref{repSimple}) is possible by directly applying the method of Section \ref{hfep_gateway_sec2}.\\

\noindent Unfortunately, most of the statistics, used in Welfare analysis, use the rank statistics so that the statistics is sum of terms that are products of a function of the ordered statistic 
$X_{j,n}$ by a function of the rank $j$. In such a case, it is usually possible, as we described in the lines above and as we will see it in the examples, to express the current statistic $I_n$ into a sum of two terms such that : \\

\noindent (a) the first is a functional empirical probability $\mathbb{P}_{n,(1)}(h)$,\\

\noindent (b) the second of the form :

\begin{equation*}
Re_n(\ell)=  \frac{1}{n}\sum_{j=1}^{n}\left(\mathbb{F}_{n,(1)}(X_j) - F_{(1)}(X_j) \right) q(X_j).
\end{equation*} 

\bigskip \noindent where $\ell(s)=q\left(F^{-1}_{(1)}(s)\right)$, $s\in [0,1]$. \cite{losall} called this process as a residual one. Among results, it is shown in the cited paper that, under smooth assumptions on $q$ (see the cited reference), the Bahadur representation exploitation leads to

$$
\sqrt{n}R_n(\ell)=\int_{0}^{1} \mathbb{G}_{n,(1)}(\tilde{f}_s)\,\ell(s)\,ds+o_{\mathbb{P}}(1),
$$   

\bigskip \noindent which in turn, leads to

$$
\sqrt{n}(I_n - \mathbb{E}h(X))=\mathbb{G}_{n,(1)}(h)+\beta_{n,(1)}(\ell)+o_{\mathbb{P}}(1) \ \ (GRI)
$$

\bigskip \noindent The ordered pair $(\mathbb{G}_{n,(1)}(h), \beta_{n,(1)}(\ell))$ is constructed such that it inherits the weak convergence $\mathbb{G}_{n,(1)}$ to $\mathbb{G}_{(1)}$, which entails the convergence of that couple to a Gaussian bivariate random variable $(\mathbb{G}_{(1)}(h), \beta_{(1)}(\ell))$. With the proper handling, as we will do in Section \ref{hfep_gateway_variation_sec1} in Chapter \ref{hfep_gateway_variation}, we will have no difficulty to have the general law :

\begin{equation} \label{genLawD1}
\sqrt{n}(I_n - \mathbb{E}h(X)) \rightsquigarrow \mathcal{N}(0,\sigma^{2}_I),
\end{equation}

\bigskip \noindent where $\Gamma =\gamma_1 + \gamma_2 +2 \gamma_3$, with

$$
\Gamma_{(1)}(h,h)=\int (h(x)-\mathbb{E}(h(X)))^2 dF_{(1)}(x)
$$

\bigskip \noindent and

$$
\gamma_1=\Gamma_{(1)}(h,h), \ \gamma_2=\int_{0}^{1} \int_{0}^{1} \Gamma_{(1)}(f_s,f_t)dsdt \ and \ \gamma_3=\int_{0}^{1} \Gamma_{(1)}(h,f_s)ds. 
$$

\bigskip \noindent We will come back to the computational aspects. For now, we have this summary : \\

\bigskip \noindent When dealing, in a fixed time, with a family $(I_{n}(\lambda), \ \lambda \in \Lambda)$ of welfare indices based on the real-valued variables $X>0$, we may represent them by the family of their representations

$$
\mathbb{G}_{n,(1)}(h_{\lambda})+\beta_{n,(1)}(\ell_{\lambda}), \ \lambda \in \Lambda.
$$

\newpage

\noindent \textbf{Part B : Spatial scheme}.\\

\noindent Suppose that we are monitoring the same index $I$ over a population divided on $K$ subgroups or areas and the particular value of the index in the $i$-th area, denoted $S_i$, is named as $I^{(i)}$, $i=1,...,K$. Let $X$ be the random variable which composes $I$ and let $F^{(i)}_{(1)}$ be the $cdf$ of $X$ on $S_i$, denoted  $X^{(i)}$, and $F$ be the $cdf$ of $X$ on the global population. Suppose that we perform independent studies on each area $S_i$ with respective samples of sizes $n_i$ for $X^{(i)}$. We get :

\noindent (a) For each $i$, a representation of the empirical index $I^{(i)}_n$ in the form

\begin{equation} \label{schemeSpat}
\mathbb{G}^{(i)}_{n_i,(1)}(h)+\beta^{(i)}_{n_j,(1)}(\ell_i)+ o_{\mathbb{P}}(n_{i}^{-1/2}).
\end{equation}

\bigskip \noindent where $\ell_i(.)=q(F^{(i)^{-1}}(\circ))$ and $\mathbb{G}^{(i)}_{n_i,(1)}$ is the \textit{fep} based on the sample sample $X^{(i)}$ with common $cdf$ $F^{(i)}$.

\noindent (b) It is important to see that the function $h$ may depend on the $cdf$. Thus, the function $h$ may vary with $i$.\\

\noindent From these two points, finding the laws of aggregated indices from the $I^{(i)}$'s are readily obtained. Interesting questions may also be treated if the sub-samples are not independent. For examples, the decomposability gap may be estimated in a purely random drawing in the whole population (See Chapter \ref{hfep_decomp} below). It is remarkable that, in Formula (\ref{schemeSpat}), the function $h$ is constant for over the areas since if depends on the mathematical form of the index.\\

\noindent If more than one index is monitored with respect to areas, we still may label them with $\lambda$ and use the results of Part A.\\

\newpage
\noindent \textbf{Part C : Time Evolution Scheme}.\\

\noindent To be simple, suppose that we monitor the same index $I$ over two periods $t=1$ and $s=2$ and we name as $I^{(j)}$ at the period $i=1,2$, and by $I^{(i)}_n$ their empirical counterparts. Let $X=(X^{(1)},X^{(2)})$ be the vector of the two incomes from time 1 to time 2. How do we set the frame in which the evolution of the index $I$ is easily handled, at least in the theoretical way?\\

\noindent It will be enough to use the joint \textit{fep} and next to use projections in the notations introduced in Section \ref{hfep_gateway_sec3}. Suppose that

$$
\sqrt{n}(I_n - \mathbb{E}h(X))=\mathbb{G}_n(h)+\beta_{n,(1)}(\ell)+o_{\mathbb{P}}(1)
$$

\bigskip \noindent is the general representation of $I$ at a fixed time. It is important to see that is form depends only on the mathematical form of $I$ and on the $cdf$ through $\ell$. As a reminder,H
$\mathbb{G}_{n,(2)}$ is the \textit{fep} based on the observations $(X^{(1)}_1,X^{(2)}_1)$, ..., $(X^{(1)}_n,X^{(2)}_n)$. Denote :

$$
h^{(1)}(x,y)=h_1(x), \ h^{(2)}(x,y)=h_2(y), \ (x,y) \in \mathbb{R}^2,
$$

$$
\tilde{f}^{(1)}_s(x,y)=1_{(x \leq F^{-1}_{(1)}(s))}, \tilde{f}^{(2)}_s(x,y)=1_{(y \leq F^{-1}_{(2)}(s))}, \ s \in [0,1] \ and \ (x,y)\in \mathbb{R}^2.
$$

\bigskip  \noindent and

$$
\ell^{(i)}(s)=q(F^{-1}_{(i)}(s)), \ i=1,2
$$

\bigskip\noindent We have

\begin{equation} \label{genLawD2}
I^{(i)}_n=I^{(i)} +n^{-1/2} \left( \mathbb{G}_{n,(2)}(h^{(i)}) + \beta_{n,(2)}(\ell^{(i)})  \right)+o_{\mathbb{P}}(n^{-1/2}), \ i=1,2.
\end{equation}

\bigskip \noindent where

\begin{equation}
\beta_{n,(2)}(\ell)=\int_{0}^{1} \mathbb{G}_{n,(2)}(\tilde{f}^{(i)}_s)\,\ell^{(i)}(s)\,ds+o_{\mathbb{P}}(1)
\end{equation}

\bigskip \noindent Here again, we conclude as follows :\\

\noindent The asymptotic probability law of $\biggr(I^{(1)}_n,I^{(2)}_n\biggr)$ is readily obtained through Formula (\ref{genLawD2}), allowing any kind of comparison or evolution study.\\

\noindent The frame we have set allows to express all needed variances or covariances.\\

\noindent The generalization to $k$ times and then to behavior of $k$ $(I^{(1)}_n,..., I^{(k)}_n$ is straightforward, even if the notation become heavier.\\

\noindent We will only describe it below.

\chapter{Statistical decomposability of indices} \label{hfep_decomp} 

\section{Introduction}

\noindent One of the most desired axiom of a welfare measure is the decomposability one.  Let us begin explain that concept.\\

\noindent Suppose that we are monitoring some index $I$ over a given population of size $N$. When $I$ is applied to the whole population, we may use the notation $I=I_{N}$. In a large population subjected to a number of  inequalities between areas and in which there are groups with specific features at the exclusion of the others, public policy efficiency usually requires to target disadvantaged areas or groups and to implement therein strong strategies aimed at improving the status of this group in relation to a given pattern (for example poverty, health covering, education level, etc.), monitored by the index $I$. In such a case, the population is divided into sensitive $K$ subgroups of interest $S_{1},...,S_{K}$ of respective sizes $N_i$, $i\in \{1,...,K\}$, and the studied behavior is followed up by an index, say $I$, taking the values $I^{(i)}=I_{N_{i}}^{(i)}$ in each subgroup $S_i$, $i\in \{1,...,K\}$.\\

\noindent The index $I$ is said to be decomposable if we may express the \textit{global} index on the whole population with respect to the partial indices at the subgroup level as follows, that is

\begin{equation}
I_N= \sum_{1\leq i \leq K} \frac{N_i}{N} I_{N_i}(i). \label{formule.decom}
\end{equation}

\noindent Formula (\ref{formule.decom}) offers the practical and comfortable latitude to work at the local level with the possibility to recompose the global index at the global level. This explains why decomposable indices are so preferred, in particular the Foster-Greer-Thorbecke (\cite{fgt84}) index of parameter $\alpha \geq 0$,

$$
FGT_n(\alpha)=\frac{1}{n} \sum_{1\leq j \leq n} \max\left(\frac{Z-X_j}{Z}, \ 0\right)^{\alpha}, \ \ \alpha\geq 0.
$$

\noindent The problem is that some the most interesting measures are not decomposable, in particular the weighted ones. Indeed, successful policies require to target disadvantaged or vulnerable groups. For example, suppose that we are dealing with poverty. A measure that counts all poor individuals with the same weight is less interesting than another that puts bigger weights to poorer individuals. A variation of such an index in the good direction tends to be negligible if the less poor individual behave better, and to be noticeable if the poorer individuals among the poor become better off.\\

\noindent Our problematic is to keep using weighted measures like the ones of \cite{sen}, \cite{kakwani}, \cite{shorrocks}, \cite{takayama}, to cite a few, and yet, to have a quick approach to report the global situation.\\

\noindent The solution resides certainly in the estimation of the decomposability gap : 

\begin{equation}
g_N=I_N- \sum_{1\leq i \leq K} \frac{N_i}{N} I_{N_i}. \label{formule.decom}
\end{equation}

\bigskip \noindent We will see later that we will be able to estimate this gap. Then we will be able to work at a local level and to report the global index in accurate confidence interval.\\

\noindent Recently, \cite{haidara-lo} motivated the estimation of decomposability gap of non-decomposable measures in the sense described above. Their results seem to be the first of that kind. The original work of Haidara and Lo concerned the general poverty index GPI \cite{loGPI2013}. But, these results implicitly include their extensions to any indice admitting the indice's general representation (GRI) in Section \ref{hfep_gateway_sec5}, Chapter \ref{hfep_gateway_intro}.\\

\noindent In the sequel, we suppose that we are working with indices satisfying the (GRI) representation. Let us precise the statistical problem.\\

\noindent We already described the decomposability in a non-random context. We are going to describe it in  the random frame.\\

\noindent Suppose that the population is divided into $K$ subgroups $S_{1},...,S_{K}$ and for each $i\in \{1,...,K\}$, let us denote the subset of the random sample $\{X_{1},...,X_{n}\}$ coming from $S_{i}$ by $\mathcal{E}_{i}=\{X_{i,1},...,X_{i,n_{i}^{\ast}}\}$ and then put $I_{n_{i}^{\ast}}^{(i)}=I(X_{i,1},...,X_{i,n_{i}^{\ast}})$ the random value of the index $I$ under study on the $i^{th}$ subgroup. We denote by $F_{i,(1)}$ the $cdf$ of $X$ on $S_i$. Let $I_n=I(X_1,...,X_n)$ be the observed index on the whole sample. The empirical decomposability gap is defined by

\begin{equation*}
gd_{n}=I_{n}-\frac{1}{n}\sum_{i=1}^{K}n_{i}^{\ast} I_{n_{i}^{\ast}}^{(i)}.
\end{equation*}

\bigskip \noindent At this step, we have to precise our random drawing. We are going to use a probability space in the form ($\Omega _{1}\times \Omega _{2},\mathcal{P}(\Omega_1) \otimes \mathcal{A}_{2}, 
\mathbb{P}^{(1)}\otimes \mathbb{P}^{(2)})$, with $\Omega_1=\{1,2,...,K\}$,  $\mathcal{P}(\Omega_1)$ is the power set of $\Omega_1$ and $\mathbb{P}^{(1)}$ is the discrete uniform probability on $\Omega_1$ such that $\mathbb{P}^{(1)}(\{i\})=p_i$, $1\leq i \leq K$. We draw the observations in the following way. In each trial $j$, we draw a subgroup according to $\mathbb{P}^{(1)}$.  We define  

\begin{equation*}
\pi _{i,j}(\omega _{1})=1_{(\text{the }i^{th}\text{ subgroup is drawn at the }j^{th}\text{ trial})}(\omega _{1}),
\end{equation*}

\bigskip \noindent where, $1\leq i\leq K,1\leq j\leq n$. Now, given that the $i^{th}$ subgroup is drawn at the $j^{th}$ trial, we pick one individual in this subgroup, according to $\mathbb{P}^{(2)}$, and observe its income $X_{j}(\omega _{1},\omega _{2}).$ We then have the observations 
\begin{equation*}
\{X_{j}(\omega _{1},\omega _{2}),\text{ }1\leq j\leq n\}.
\end{equation*}

\noindent Here, $\mathbb{P}^{(2)}$ is the probability in Section \ref{hfep_gateway_sec3} of Chapter \ref{hfep_gateway_intro}. We denote $\mathbb{P}=\mathbb{P}^{(1)}\otimes \mathbb{P}^{(2)}$, while keeping in mind that the representations in (GRI) in Section \ref{hfep_gateway_sec4} of Chapter \ref{hfep_gateway_intro} are valid with respect to  $\mathbb{P}^{(2)}$.\\ 

\noindent We have these simple facts. First, \ for $1\leq i\leq K$. 
\begin{equation}
n_{i}^{\ast }=\sum_{j=1}^{n}\pi _{i,j}.
\end{equation}

\noindent Let us denote the distribution of $X_{j}$ given $(\pi _{i,j}=1)$, by  $F_{i,(1)}$ that is 
\begin{equation*}
\mathbb{P}(X_{j}\leq y\text{ }\diagup \pi _{i,j}=1)=F_{i,(1)}(x).
\end{equation*}

\bigskip \noindent We simply put, in some places, $F_{i,(1}(x)=F_{i,(1)}$, $y\in \mathbb{R}$, to keep the notation simple. Then we have  
\begin{eqnarray*}
\forall (x\in \mathbb{R)},\mathbb{P}(X_{j}\leq y\text{ })&=&\sum_{i=1}^{K}\mathbb{P}(\pi _{i,j}=1)\mathbb{P}(X_{j}\leq y\text{ }\diagup \pi _{i,j}=1)\\
&=&\sum_{i=1}^{K}p_{i}F_{i,(1)}(x).\\
\end{eqnarray*}

\noindent We conclude that $\{X_{1},...,X_{n}\}$ is an independent sample drawn from $F_{(1)}(x)$ $=\sum_{i=1}^{K}p_{i}F_{i,(1)}(x)$, which is the mixture of the distribution functions of the subgroups incomes.\\

\noindent The formula above ensures that for any real-valued function $h$ such that the $h(X^{(i)})$'s are integrable, we have

\begin{equation}
\mathbb{E}h(X)=\sum_{1\leq i \leq K} p_i F_{(1)}^{(i)}.  \label{Exp}
\end{equation}

\noindent Finally, we readily see that conditionally on $n^{\ast }\equiv (n_{1}^{\ast },n_{2}^{\ast
},...,n_{K}^{\ast })=(n_{1},n_{2},...,n_{K})\equiv \overline{n}$ with $%
n_{1}+n_{2}+...+n_{K}=n,$ $\{X_{i,j},$ $1\leq j\leq n_{i}\}$ are
independent random variables with distribution function $F_{i,(1)}$.

\section{General Statistical Decomposition Theorem}

We suppose that the indice's general representation (GRI) in Section \ref{hfep_gateway_sec4}, Chapter \ref{hfep_gateway_intro}, with

$$
\ell(s)=q\left(F_{(1)}^{-1}(s)\right), \ \ s \in (0,1).
$$

\bigskip \noindent We already knew that the function $h$ in the (GRI) formula may depend on the $cdf$ $F_{i,(1)}$ on each subgroup to become $h_i$ and denote accordingly

$$
\ell_i(s)=q_i\left(F_{i,(1)}^{-1}(s)\right), \ \ s \in (0,1).
$$

\bigskip \noindent Let us introduce the constants :

\begin{equation*}
A_{1}=\sum_{i=1}^{K}p_{i}\left\{
\int_{0}^{1}(h-h_{i})^{2}(F_{i,(1)}^{-1}(t))dt-\left(
\int_{0}^{1}(h-h_{i})(F_{i,(1)}^{-1}(t))dt\right) ^{2}\right\} ,
\end{equation*}

\begin{equation*}
A_{2}=\sum_{i=1}^{K}p_{i}\int_{0}^{1}\int_{0}^{1}(s\wedge t-st)(p_{i}\ell-\ell _{i})(s)(p_{i}\ell -\ell _{i})(s)dsdt,
\end{equation*}

\begin{equation*}
A_{31}=\sum_{i=1}^{K}p_{i}^{2}\sum_{h\neq
i}^{K}p_{h}\int_{0}^{1}\int_{0}^{1} \left[ {F_{h}(F_{i,(1)}^{-1}(s))\wedge
F_{h}(F_{i,(1)}^{-1}(t))}\right.
\end{equation*}

\begin{equation*}
\left. {-F_{h}(F_{i,(1)}^{-1}(s))F_{h}(F_{i,(1)}^{-1}(t))}\right] \ell (s) q(F_{i,(1)}^{-1}(t))ds \ dt,
\end{equation*}

\begin{equation*}
A_{32}=\sum_{i=1}^{K}p_{i}\sum_{j\neq i}^{K}p_{j}\sum_{h\notin
\{i,j\}}^{K}p_{h}\int_{0}^{1}\int_{0}^{1}\left[ {F_{h}(F_{i,(1)}^{-1}(s))\wedge
F_{h}(F_{j,(1)}^{-1}(t))}\right.
\end{equation*}

\begin{equation*}
\left. {-F_{h}(F_{i,(1)}^{-1}(s))F_{h}(F_{j,(1)}^{-1}(t))}\right] \ell(s)\ell(s)ds \ dt,
\end{equation*}

\begin{equation*}
B_{1}=\sum_{i=1}^{K}p_{i}\int_{0}^{1}\left\{ {%
\int_{0}^{s}(h-h_{i})(F_{i,(1)}^{-1}(t))dt}\right.
\end{equation*}

\begin{equation*}
\left. {-s\int_{0}^{1}(h-h_{i})(F_{i,(1)}^{-1}(t))dt}\right\} (p_{i}\ell -\ell_{i})(s)ds,
\end{equation*}

\begin{equation*}
B_{2}=\sum_{j=1}^{K}p_{j}\sum_{i\neq
j}^{K}p_{i}\int_{0}^{1}\int_{0}^{1}[s\wedge
F_{i,(1)}(F_{j,(1)}^{-1}(t))-sF_{i,(1)}(F_{j,(1)}^{-1}(t))],
\end{equation*}

\begin{equation*}
\times (p_{i}\ell -\ell _{i})(s)\ell(s)ds \ dt,
\end{equation*}

\begin{equation*}
B_{3}=\sum_{j=1}^{K}p_{j}\sum_{i\neq j}^{K}p_{i}\int_{0}^{1}\newline
\left\{ {\int_{0}^{1}(h-h_{i})(F_{i,(1)}^{-1}(t))dt}\right.
\end{equation*}

\begin{equation*}
\left. {-F_{i,(1)}(F_{j,(1)}^{-1}(s))\times \int_{0}^{1}(h-h_{i})(F_{i,(1)}^{-1}(t))dt}
\right\} \ell(s) \ ds,
\end{equation*}

\begin{equation*}
gd=I-\sum_{i=1}^{K}p_{i} I^{(i)}
\end{equation*}

\noindent and, finally,
$$
gd_{0,n}=I-\sum_{i=1}^{K}(n_{i}^{\ast}/n)I^{(i)}.
$$

\bigskip \noindent We will need the following components of our variances. First, define for $i=1,...,K$

$$
L_{i}=\mathbb{E}h(X^{i})-I_{i}+\sum_{\alpha=1}^{K}p_{\alpha}\mathbb{E}F_{i,(1)}(X^{(\alpha)})q(X^{(\alpha)}),
$$

\noindent and

$$
M_{i}=\mathbb{E}h(X^{i})+\sum_{\alpha=1}^{K}p_{\alpha}\mathbb{E}F_{i,(1)}(X^{(\alpha)}) q(X^{(\alpha)}).
$$

\noindent Next, define

\begin{equation*}
\vartheta _{1}^{2}=A_{1}+A_{2}+A_{3}+2(B_{1}+B_{2}+B_{3})
\end{equation*}

\bigskip \noindent and 

\begin{equation*}
\vartheta _{2}^{2}=\sum_{i=1}^{K} L_{i}{}^{2}p_{i}-\left(\sum_{i=1}^{K}L_{i}p_{i}\right) ^{2}
\end{equation*}

\bigskip \noindent and

\begin{equation*}
\vartheta _{3}^{2}=\sum_{\alpha=1}^{K}M_{\alpha}^{2}p_{\alpha}-\left(\sum_{\alpha=1}^{K}M_{\alpha}p_{\alpha}\right) ^{2}
\end{equation*}

\bigskip \noindent Here is the general decomposability result.

\subsection{The theoretical result}

We have the following result.

\begin{theorem} \label{theo2} Let $\mathbb{E}X^{2}<\infty$, $\mathbb{E}(X^{(i)})^{2}<\infty$. Let us suppose also that$F_{(1)}$ and each $F_{i,(1)}$, $1\leq i \leq K$ are increasing so that they are invertible. Let assume also the the conditions (FHEP1) for the validity of the (GRI) representations of the indices holds on each subgroup and at the whole area.\\

\noindent Then we have
$$
gd_{n,0}^{\ast }=\sqrt{n}(gd_{n}-gd_{0,n})\leadsto \mathcal{N}(0,\vartheta _{1}^{2}+\vartheta_{3}^{2})
$$

\bigskip \noindent and \\

$$
gd_{n}^{\ast }=\sqrt{n}(gd_{n}-gd)\leadsto \mathcal{N}(0,\vartheta _{1}^{2}+\vartheta _{2}^{2})
$$ 
\end{theorem}

\bigskip \noindent A particular version of this theorem has already been proved in \cite{haidara-lo}, for specific welfare indices. A more general proof based only on the GRI is proposed below.\\

\section{Poof of the Theorem}.

\noindent From the assumptions, we write

$$
\sqrt{n}(I_n-I)=\mathbb{G}_{n,(1)}(h)+\beta_{n,(1)}(\ell)+o_{\mathbb{P}}(1),
$$

\bigskip \noindent with

$$
\beta_{n,(1)}(\ell)=\int_{0}^{1} \mathbb{G}_{n,(1)}(f_s) \ell(s) ds,
$$

\bigskip \noindent and for $i=1,...,k$, for non-random sizes $n_i$ becoming infinitely large,
$$
\sqrt{n}(I^{(i)}_{n_i}-I^{(i)})=\mathbb{G}_{n_i,(1)}(h_i)+\beta_{n_i,(1)}(\ell_i)+o_{\mathbb{P}}(1),
$$

\bigskip \noindent with

$$
\beta_{n_i,(1)}(\ell_i)=\mathbb{G}_{n_i,(1)}(f_s) \ell_i(s) ds.
$$

\bigskip \noindent Here we have simplified the notation and used $\mathbb{G}_{n_i,(1)}$ instead of $\mathbb{G}^{(i)}_{n_i,(1)}$ which is the functional empirical process based $m$ observations from the random variable $X^{(i)}$. We think that there will be no confusion because of the subscript $n_i$ that will remind us that we are on the $i^{th}$ subgroup.\\

\noindent To begin the proof, we remark that $n^{\ast }(\omega_{1})=(n_{1}^{\ast }(\omega _{1}),...,n_{K}^{\ast }(\omega _{1}))\rightarrow
_{\mathbb{P}_{1}}\{+\infty \}^{K}$ as $n=n_{1}^{\ast }(\omega_{1})+...+n_{K}^{\ast }(\omega _{1})\rightarrow \infty$.\\

\noindent We then get 
\begin{equation}
\sqrt{n}(I_{n}-I)=\mathbb{G}_{n_{i},(1)}(h)+\beta_{n,(1)}(\ell )+o_{\mathbb{P}%
}(1):=\gamma _{n}+o_{\mathbb{P}}(1)  \label{l1}
\end{equation}

\noindent and for any $1\leq i\leq K$, 
\begin{equation}
\sqrt{n_{i}^{\ast }}(I^{(i)}_{n_{i}^{\ast}}-I^{(i)})=\mathbb{G}_{n_{i}^{\ast},(1)}(h_{i})+\beta_{n_{i}^{\ast},(1)}(\ell _{i})+o_{\mathbb{P}}(1):=\gamma _{i,n_{i}^{\ast }}+o_{\mathbb{P}}(1)  \label{l2}
\end{equation}

\bigskip \noindent Now we use the intermediate centering coefficient  
\begin{equation*}
gd_{0,n}=I-\sum_{i=1}^{K}\frac{n_{i}^{\ast }}{n}I^{(i)}
\end{equation*}

\bigskip \noindent and, after some direct manipulations based on (\ref{l1}) and (\ref{l2}), to find 
\begin{equation*}
\left\vert \sqrt{n}(gd_{n}-gd_{0,n})-\left\{ \gamma
_{n}-\sum_{j=1}^{K}\left( \frac{n_{i}^{\ast }}{n}\right) ^{1/2}\gamma
_{i,n_{i}}\right\} \right\vert (\omega _{1},\omega _{2})=o_{\mathbb{P}_{1}\otimes \mathbb{P}_{2}}(1),
\end{equation*}

\bigskip \noindent as $n\rightarrow \infty $. Then, we have that $S_{n}^{\ast}$ is equal to
\begin{eqnarray*}
&&\gamma _{n}-\sum_{j=1}^{K}\left( \frac{n_{i}^{\ast }}{n}\right) ^{1/2}\gamma _{i,n_{i}^{\ast }}\\
&=&\mathbb{G}_{n,(1)}(h)-\sum_{j=1}^{K}\left( \frac{n_{i}^{\ast }}{n}\right)^{1/2}\mathbb{G}_{n_{i}^{\ast},(1)}(h_{i})+\beta_{n,(1)}(\ell)-
\sum_{j=1}^{K}\left(\frac{n_{i}^{\ast }}{n}\right) ^{1/2}\beta_{n_{i}^{\ast},(1)}(\ell_{i}).
\end{eqnarray*}

\bigskip \noindent We use Formula \eqref{Exp} and remark that
\begin{equation*}
\mathbb{G}_{n,(1)}(h)=\frac{1}{\sqrt{n}}\sum_{j=1}^{n}\left( h(X_{j})-\mathbb{E}h(X)\right) =\sqrt{n}\left( \frac{1}{n}\sum_{j=1}^{n}h(X_{j})-\mathbb{E}h(X)\right) 
\end{equation*}
\begin{equation*}
=:\sqrt{n}\left( \frac{1}{n}\sum_{j=1}^{n}h(X_{j})-\sum_{i=1}^{K}\frac{n_{i}^{\ast }}{n}\mathbb{E}h(X^{i})\right) +D^{\ast }(n,1),
\end{equation*}

\bigskip \noindent with 

\begin{equation*}
D^{\ast }(n,1)=\sum_{i=1}^{K}\frac{n_{i}^{\ast }-np_{i}}{\sqrt{np_{i}}}%
\mathbb{E}h(X^{(i)})\sqrt{p_{i}}.
\end{equation*}

\bigskip \noindent When conditioning on $n^{\ast}=n$, we denote 
 
\begin{equation*}
D(n,1)=\sum_{i=1}^{K}\frac{n_{i}-np_{i}}{\sqrt{np_{i}}}\sqrt{p_{i}}\mathbb{E}h(X^{(i)}),
\end{equation*}

\bigskip \noindent This leads to 

\begin{equation*}
S_{n}^{\ast }=\sqrt{n}\left( \frac{1}{n}\sum_{j=1}^{n}h(X_{j})-\sum_{i=1}^{K}%
\frac{n_{i}^{\ast }}{n}\mathbb{E}h(X^{(i)})\right) -\sum_{j=1}^{K}\left( \frac{%
n_{i}^{\ast }}{n}\right) ^{1/2}\mathbb{G}_{n_{i}^{\ast},(1)}(h_{i})
\end{equation*}%
\begin{equation*}
+\beta_{n,(1)}(\ell)-\sum_{j=1}^{K}\left( \frac{n_{i}^{\ast }}{n}\right)
^{1/2}\beta_{n_{i}^{\ast},(1)}(\ell_{i})+D^{\ast }(n,1).
\end{equation*}

\bigskip \noindent Now, by denoting 
\begin{equation*}
C^{\ast }(n,1)=\sqrt{n}\left( \frac{1}{n}\sum_{j=1}^{n}h(X_{j})-%
\sum_{i=1}^{K}\frac{n_{i}^{\ast }}{n}\mathbb{E}h(X^{(i)})\right)
-\sum_{i=1}^{K}\left( \frac{n_{i}^{\ast }}{n}\right) ^{1/2}\mathbb{G}_{n_{i}^{\ast},(1)}\left(h_{i}\right) ,
\end{equation*}

\noindent we have \label{c1}
\begin{eqnarray*}
&&C^{\ast }(n,1) \ \ \ \ \ \ (\textbf{c1})\\
&=&\sum_{i=1}^{K}\left( \frac{n_{i}^{\ast }}{n}\right)
^{1/2}\left\{ \frac{1}{\sqrt{n_{i}^{\ast }}}\sum_{j=1}^{n_{i}^{\ast
}}\left\{ \left( h-h_{i}\right) \left( X_{ij}\right) -\mathbb{E}\left(
h-h_{i}\right) (X^{(i)})\right\} \right\} .  \notag
\end{eqnarray*}

\noindent We get
\begin{equation}
S_{n}^{\ast }=C^{\ast }(n,1)+D^{\ast }(n,1)+\beta_{n,(1)}(\ell)-\sum_{j=1}^{K}\left( \frac{n_{i}^{\ast }}{n}\right) ^{1/2}\beta_{n_{i}^{\ast},(1)}(\ell_{i}).  \label{for1a}
\end{equation}

\noindent Further, we have 
\begin{eqnarray}
&&\sum_{j=1}^{K}\left( \frac{n_{i}^{\ast }}{n}\right) \beta_{n_{i}^{\ast},(1)}(\ell_{i}) \label{beta1}\\
&=&\frac{1}{\sqrt{n}}\sum_{i=1}^{K}\sum_{j=1}^{n_{i}^{\ast
}}[\mathbb{G}_{i,n_{i}^{\ast }}(X_{ij})-F_{i,(1)}(X_{ij}))]q_{i}(X_{ij}). \notag
\end{eqnarray}

\noindent But
\begin{equation*}
F_{(1)}(X_{ij})=\sum_{h=1}^{K}p_{h}F_{h,(1)}(X_{ij}),
\end{equation*}

\noindent and for $x\in \mathbb{R}$
\begin{equation*}
\mathbb{G}_{n,r,(1)}(x)=\frac{1}{n}\sum_{i=1}^{n} 1_{(X_{j}\leq x)}=\frac{1}{n}%
\sum_{i=1}^{K}\sum_{j=1}^{n_{i}^{\ast }}1_{(X_{ij}\leq x)}
\end{equation*}%
\begin{equation*}
=\sum_{i=1}^{K}\left( \frac{n_{i}^{\ast }}{n}\right) \frac{1}{n_{i}^{\ast }}%
\sum_{j=1}^{n_{i}^{\ast }} 1_{(X_{ij}\leq x)}=\sum_{i=1}^{K}\frac{%
n_{i}^{\ast }}{n}\mathbb{G}_{n_{i}^{\ast }}(x).
\end{equation*}

\noindent Thus 
\begin{equation*}
\beta_{n,(1)}(\ell)=\frac{1}{\sqrt{n}}\sum_{i=1}^{K}\sum_{j=1}^{n_{i}^{\ast }}%
\left[ \sum_{h=1}^{K}\left( \frac{n_{h}^{\ast }}{n}\right) G_{n_{h}^{\ast},r,(1)}(X_{ij})-p_{h}F_{h,(1)}(Y_{ij})\right] q (X_{ij}).
\end{equation*}

\bigskip \noindent From this, we put and subtract $\sum_{h=1}^{k}(\frac{n_{h}^{\ast }}{n}%
)F_{h,(1)}(X_{ij})$ to have 
\begin{equation*}
\beta_{n,(1)}(\ell)=\frac{1}{\sqrt{n}}\sum_{i=1}^{K}\sum_{j=1}^{n_{i}^{\ast }}%
\left[ \sum_{h=1}^{K}\left( \frac{n_{h}^{\ast }}{n}\right) G_{h,n_{h}^{\ast
}}(X_{ij})-\sum_{h=1}^{K}\left( \frac{n_{h}^{\ast }}{n}\right) F_{h,(1)}(X_{ij})%
\right] q (X_{ij})
\end{equation*}%
\begin{equation*}
+\frac{1}{\sqrt{n}}\sum_{i=1}^{K}\sum_{j=1}^{n_{i}^{\ast }}\left[
\sum_{h=1}^{K}\left( \frac{n_{h}^{\ast }}{n}-p_{h}\right) F_{h,(1)}(X_{ij})%
\right] q (X_{ij})
\end{equation*}

\begin{equation}
=\frac{1}{\sqrt{n}}\sum_{i=1}^{K}\sum_{j=1}^{n_{i}}\sum_{h=1}^{K}\left( 
\frac{n_{h}^{\ast }}{n}\right) \left\{
G_{h,n_{h}}(X_{ij})-F_{h,(1)}(X_{ij})\right\} q(X_{ij})  \label{beta2}
\end{equation}%
\begin{equation*}
+\frac{1}{\sqrt{n}}\sum_{i=1}^{K}\sum_{j=1}^{n_{i}}\left[ \sum_{h=1}^{K}%
\left( \frac{n_{h}^{\ast }}{n}-p_{h}\right) F_{h,(1)}(X_{ij})\right] q(X_{ij}).
\end{equation*}

\bigskip \noindent Now we put together (\ref{beta1}) and (\ref{beta2}), while separating the
two cases $h=i$ and $h\neq i$ in (\ref{beta2}) to get%
\begin{equation*}
\beta_{n,(1)}(\ell)-\sum_{j=1}^{K}\left( \frac{n_{i}^{\ast }}{n}\right)
^{1/2}\beta_{n_{i}^{\ast},(1)}(\ell_{i})
\end{equation*}%
\begin{equation*}
=\sum_{i=1}^{K}\left( \frac{n_{i}^{\ast }}{n}\right) ^{1/2}\left\{ \frac{1}{%
\sqrt{n_{i}^{\ast }}}\sum_{j=1}^{n_{i}^{\ast }}\left\{ G_{i,n_{i}^{\ast
}}(X_{ij})-F_{i,(1)}(X_{ij})\right\} \left( \frac{n_{i}^{\ast }}{n}q -q_{i}\right) (X_{ij})\right\} 
\end{equation*}%
\begin{equation*}
+\sum_{i=1}^{K}\left( \frac{n_{i}^{\ast }}{n}\right) ^{1/2}\sum_{h\neq i}^{K}%
\frac{n_{h}^{\ast }}{n}\left\{ \frac{1}{\sqrt{n_{i}^{\ast }}}%
\sum_{j=1}^{n_{i}^{\ast }}\left\{ G_{h,n_{h}^{\ast
}}(X_{ij})-F_{h,(1)}(X_{ij})\right\} q(X_{ij})\right\} 
\end{equation*}%
\begin{equation*}
+\frac{1}{\sqrt{n}}\sum_{i=1}^{K}\sum_{j=1}^{n_{i}^{\ast }}\left[
\sum_{h=1}^{K}\left( \frac{n_{h}^{\ast }}{n}-p_{h}\right) F_{h,(1)}(X_{ij})%
\right] q (X_{ij})
\end{equation*}

\begin{equation}
=:C^{\ast }(n,2)+C^{\ast }(n,3)+D^{\ast }(n,2),  \label{for1b}
\end{equation}

\bigskip \noindent with \label{c2}

\begin{eqnarray*}
&&C^{\ast}(n,2) \ \ \ \ \ \ (\textbf{c2}) \\
&=&\sum_{i=1}^{K}\left( \frac{n_{i}^{\ast }}{n}\right)
^{1/2}\left\{ \frac{1}{\sqrt{n_{i}^{\ast }}}\sum_{j=1}^{n_{i}^{\ast
}}\left\{ G_{i,n_{i}^{\ast }}(X_{ij})-F_{i,(1)}(X_{ij})\right\} \left( \frac{%
n_{i}^{\ast }}{n}q -q _{i}\right) (X_{ij})\right\},  \notag
\end{eqnarray*}

\bigskip \noindent and \label{c3} 
\begin{eqnarray*}
&&C^{\ast }(n,3) \ \ \ \ \ \ (\textbf{c3}) \\
&=&\sum_{i=1}^{K}\left( \frac{n_{i}^{\ast }}{n}\right)
^{1/2}\sum_{h\neq i}^{K}\frac{n_{h}^{\ast }}{n}\left\{ \frac{1}{\sqrt{%
n_{i}^{\ast }}}\sum_{j=1}^{n_{i}^{\ast }}\left\{ G_{h,n_{h}^{\ast
}}(X_{ij})-F_{h,(1)}(X_{ij})\right\} q(X_{ij})\right\}. \notag  
\end{eqnarray*}

\noindent We arrive, by comparing (\ref{for1a}) and (\ref{for1b}), at%
\begin{equation}
S_{n}^{\ast }=C^{\ast }(n,1)+C^{\ast }(n,2)+C^{\ast }(n,3)+D^{\ast
}(n,1)+D^{\ast \ast }(n,2).  \label{sn}
\end{equation}

\bigskip \noindent Let us have a look at 
\begin{equation*}
D^{\ast \ast }(n,2)=\sqrt{n}\sum_{h=1}^{K}\left( \frac{n_{h}^{\ast }}{n}%
-p_{h}\right) \left\{ \sum_{i=1}^{K}\left( \frac{n_{i}^{\ast }}{n}\right) 
\frac{1}{n_{i}^{\ast }}\sum_{j=1}^{n_{i}^{\ast }}F_{h,(1)}(X_{ij})q(X_{ij})\right\} .
\end{equation*}

\bigskip \noindent By the weak law of large numbers
\begin{equation*}
\left\{ \sum_{i=1}^{K}\left( \frac{n_{i}^{\ast }}{n}\right) \frac{1}{%
n_{i}^{\ast }}\sum_{j=1}^{n_{i}^{\ast }}F_{h,(1)}(X_{ij})q (X_{ij})\right\}
\rightarrow _{\mathbb{P}}\sum_{i=1}^{K}p_{i}\mathbb{E}F_{h,(1)}(X^{i})q
(X^{i})=H_{h}.
\end{equation*}

\bigskip \noindent That is
\begin{equation*}
D^{\ast \ast }(n,2)=\sum_{h=1}^{K}\left( \frac{n_{h}^{\ast }-np_{h}}{\sqrt{%
np_{h}}}\right) H_{h}\sqrt{p_{h}}+o_{\mathbb{P}}(1).
\end{equation*}

\begin{equation*}
=:D^{\ast }(n,2)+o_{\mathbb{P}}(1).
\end{equation*}%

\bigskip \noindent Finally, we have for all $n\geq 1$,

\begin{equation}
gd_{n}^{\ast }=S_{n}^{\ast }+\sqrt{n}(gd_{0,n}-gd).  \label{sntosnt}
\end{equation}

\noindent Hence

\begin{equation*}
gd_{n}^{\ast }=C^{\ast }(n,1)+C^{\ast }(n,2)+C^{\ast }(n,3)
\end{equation*}
\begin{equation*}
+D^{\ast }(n,1)+D^{\ast }(n,2)-\sum_{i=1}^{K}\left( \frac{n_{i}^{\ast
}-np_{i}}{\sqrt{np_{i}}}\right) I^{(i)}\sqrt{p_{i}}+o_{\mathbb{P}}(1),
\end{equation*}%
\begin{equation}
=:C^{\ast }(n)+D^{\ast }(n)+o_{\mathbb{P}}(1),  \label{reste}
\end{equation}

\bigskip \noindent with
\begin{equation}
C^{\ast }(n)=C^{\ast }(n,1)+C^{\ast }(n,2)+C^{\ast }(n,3)  \label{c}
\end{equation}

\bigskip \noindent  and
\begin{equation*}
D^{\ast }(n)=D^{\ast }(n,1)+D^{\ast }(n,2)-\sum_{i=1}^{K}\left( \frac{%
n_{i}^{\ast }-np_{i}}{\sqrt{np_{i}}}\right) I_{(i)}\sqrt{p_{i}}
\end{equation*}%
\begin{equation*}
=\sum_{i=1}^{K}\left( \frac{n_{i}^{\ast }-np_{i}}{\sqrt{np_{i}}}\right)
(H_{i}+\mathbb{E}h(X^{(i)})-I^{(i)})\sqrt{p_{i}}
\end{equation*}
\begin{equation*}
=:\sum_{i=1}^{K}\left( \frac{n_{i}^{\ast }-np_{i}}{\sqrt{np_{i}}}\right)
F_{i}\sqrt{p_{i}}.
\end{equation*}%

\bigskip \noindent We have now to prove that $gd_{n}^{\ast }=\sqrt{n}(gd_{n}-gd)$\
weakly converges to a $N(0,\vartheta _{1}^{2}+\vartheta _{2}^{2})$ random
variable. For this it suffices, based on (\ref{reste}), to prove that $%
S_{n}^{\ast \ast }=C^{\ast }(n)+D^{\ast }(n)$ converges to $N(0,\vartheta
_{1}^{2}+\vartheta _{2}^{2})$. Now put 
\begin{equation*}
\mathbb{N}(K)=\{\overline{n}=(n_{1},...n_{K}),n_{i}\geq
0,n_{1}+...,n_{K}=n\}.
\end{equation*}

\bigskip \noindent Since $n^{\ast }=(n_{1}^{\ast },...n_{K}^{\ast })\rightarrow
_{P_{1}}\{\infty \}^{K},$ we find for a fixed $\varepsilon >0$, $K$ positive
numbers $N_{i}$ $(1\leq i\leq K)$ such that for $n_{i}\geq N_{i}$ $(1\leq
i\leq K),$ which implies that $n\geq N=N_{1}+...+N_{K},$%
\begin{equation*}
\mathbb{P}(\exists (1\leq i\leq K),n_{i}^{\ast }<N_{i})<\varepsilon .
\end{equation*}

\bigskip \noindent Let

\begin{equation*}
\mathbb{N}(K,1)=\mathbb{N}(K)\cap \{\overline{n}=(n_{1},...n_{K}),\exists
(1\leq i\leq K),n_{i}<N_{i}\}
\end{equation*}%

\bigskip \noindent and $N(K,2)=N(K)\diagdown N(K,1).$ We remark that conditionally on $(n^{\ast
}=\overline{n})$, $C^{\ast }(n)$ becomes $C(n),$ does not depend on $\omega
_{1}$ and only include the independent random variables $\{X_{i,j},1\leq
j\leq n_{i},1\leq i\leq K\}$. From Lemma \ref{lem2} below, we have 
\begin{equation*}
C(n)\rightarrow \mathcal{N}(0,\vartheta _{1}^{2}).
\end{equation*}

\bigskip \noindent Also conditionally on $(n^{\ast }=\overline{n})$, $D^{\ast }(n)$ becomes $%
D(n)$ and we denote it $D(n)$. Now for $h^{2}=-1,$%
\begin{equation*}
\psi _{S_{n}^{\ast \ast }}(t)=\mathbb{E}(\exp (htS_{n}^{\ast \ast }))
\end{equation*}%
\begin{equation*}
=\sum_{\overline{n}\in \mathbb{N}(K)}\mathbb{P}(n^{\ast }=\overline{n})%
\mathbb{E}(\exp (htC^{\ast }(n)+htD^{\ast }(n))\diagup (n^{\ast }=\overline{n%
}))
\end{equation*}%
\begin{equation*}
=\sum_{\overline{n}\in \mathbb{N}(K)}\mathbb{P}(n^{\ast }=\overline{n})%
\mathbb{E}(\exp (htD(n))\text{ }\mathbb{E}(\exp (htC^{\ast }(n))\diagup
(n^{\ast }=\overline{n})).
\end{equation*}

\bigskip \noindent Recall that, by the classical limiting law of the multinomial $K$-vector, 
\begin{equation*}
D^{\ast }(n)\rightarrow D=\sum_{i=1}^{K}Z_{i}F_{i}\sqrt{p_{i}},
\end{equation*}

\bigskip \noindent where $(Z_{1},...,Z_{K})^{t}$ is a Gaussian vector with $Var(Z_{i})=1-p_{i}$
and $Cov(Z_{i},Z_{j})=-\sqrt{p_{i}p_{j}},$ for $i\neq j.$ Then 
\begin{equation*}
D^{\ast }(n)\rightarrow \mathcal{N}(0,\vartheta _{2}^{2}),
\end{equation*}

\bigskip \noindent with
 
\begin{equation*}
\vartheta _{2}^{2}=\sum_{h=1}^{K}F_{h}^{2}p_{h}(1-p_{h})-\sum_{1\leq h\neq
k\leq K}F_{h}F_{k}p_{h}p_{k}
\end{equation*}%
\begin{equation*}
=\sum_{h=1}^{K}F_{h}{}^{2}p_{h}-\left( \sum_{h=1}^{K}F_{h}p_{h}\right) ^{2}.
\end{equation*}

\bigskip \noindent We remark that this is the variance of the function $F_{h}$ of $h\in \lbrack
1,K]$ with respect to the probability measure $\sum_{1\leq h\leq
K}p_{h}\delta _{h}$.\\ 

\noindent Put now 
\begin{equation*}
\mathbb{N}(K,1)=\mathbb{N}(K)\cap \{\overline{n}=(n_{1},...n_{K}),\exists
(1\leq i\leq K),n_{i}<N_{i}\}
\end{equation*}

\bigskip \noindent and $N(K,2)=N(K)\diagdown N(K,1)$. Then 
\begin{equation*}
\sum_{\overline{n}\in \mathbb{N}(K)}\exp (htD(n))\mathbb{P}(n^{\ast }=%
\overline{n})\mathbb{E}(\exp (htC(n))))=B(n,1)+B(n,2)
\end{equation*}

\bigskip \noindent with 
\begin{equation*}
\left\vert B(n,1)\right\vert =\left\vert \sum_{\overline{n}\in \mathbb{N}%
(K,1)}\exp (htD(n))\mathbb{P}(n^{\ast }=\overline{n})\mathbb{E}(\exp
(htC(n)))\right\vert 
\end{equation*}
\begin{equation}
\leq \mathbb{P}(\exists (1\leq i\leq K),n_{i}^{\ast }<N_{i})\rightarrow 0,
\label{approx2}
\end{equation}

\bigskip \noindent and 
\begin{equation}
\left\vert B(n,2)-\sum_{\overline{n}\in \mathbb{N}(K,2)}\exp (-(\vartheta
_{1}t)^{2}/2)\exp (htD(n))\mathbb{P}(n^{\ast }=\overline{n})\right\vert 
\label{approx3}
\end{equation}
\begin{equation*}
\leq \varepsilon \sum_{\overline{n}\in \mathbb{N}(K,2)}\mathbb{P}(n^{\ast }=%
\overline{n})\leq \varepsilon .
\end{equation*}

\bigskip \noindent Finally, for 
\begin{equation}
B^{\ast }(n,2)=\sum_{\overline{n}\in \mathbb{N}(K,2)}\exp (-(\vartheta
_{1}t)^{2}/2)\exp (htD(n))\mathbb{P}(n^{\ast }=\overline{n}),
\label{approx4}
\end{equation}

\bigskip \noindent we are able to use (\ref{approx4}) and to get 
\begin{equation*}
\lim \sup_{n\rightarrow \infty }\left\vert B^{\ast }(n,2)-\sum_{\overline{n}%
\in \mathbb{N}(K)}\exp (htD(n))\mathbb{P}(n^{\ast }=\overline{n})\mathbb{E}%
(\exp (-(\vartheta _{1}t)^{2}/2))\right\vert =0.
\end{equation*}

\bigskip \noindent But 
\begin{equation}
\mathbb{E}\exp (thD^{\ast }(n))=\sum_{\overline{n}\in \mathbb{N}(K)}\exp
(htD^{\ast }(n)/(n^{\ast }=\overline{n}))\mathbb{P}(n^{\ast }=\overline{n})
\end{equation}
\begin{equation*}
=\sum_{\overline{n}\in \mathbb{N}(K)}\exp (htD(n))\mathbb{P}(n^{\ast }=%
\overline{n})\rightarrow \exp (-(\vartheta _{2}t)^{2}/2))
\end{equation*}

\bigskip \noindent By putting together the previous formulas, and by letting $\varepsilon
\downarrow 0,$ we arrive at 
\begin{equation*}
\psi _{d_{n}^{\ast \ast }}(t)\rightarrow \exp (-(\vartheta
_{1}^{2}+\vartheta _{2}^{2})t^{2}/2).
\end{equation*}

\bigskip \noindent This proves the asymptotic normality of $dg_{n}^{\ast }$ of the
theorem corresponding to $S_{n}^{\ast \ast }$. That of $dg_{n,0}^{\ast }$
corresponds to $S_{n}^{\ast }$. This latter is achieved by omitting the term 
$\sqrt{n}\sum_{i=1}^{K}(\frac{n_{i}^{\ast }}{n}-p_{i})I^{(i)}$ in (\ref%
{sntosnt}). This leads to $M_{h}$ obtained from $F_{h}$ by dropping $I^{(i)}$. This completes the proofs.\newline

\noindent We now prove this lemma used in the proof. 

\begin{lemma}
\label{lem2} Let $C(n)=C(n,1)+C(n,2)+C(n,3)$, where the $C(n,i)$ are
respectively defined in Formula (c1) (page \pageref{c1}), Formula (c2) (page \pageref{c2}) and Formula (c3) (page \pageref{c3}) for $i=1,2,3$. Then, as $n\rightarrow +\infty ,$%
\begin{equation*}
C(n)\leadsto \mathcal{N}(0,\vartheta _{1}^{2}).
\end{equation*}
\end{lemma}

\bigskip \noindent Recall that \label{ca}
\begin{equation}
C(n)=C(n,1)+C(n,2)+C(n,3).  \ \ (ca)
\end{equation}

\bigskip \noindent \textbf{At this step, new tools are introduced and the attention of the reader is drawn}. Using the continuity of the the \textit{cdf} $F_{i,(1)}$'s, we are going to use the functional empircal process based on the independent and 
$(0,1)$-uniform random variables $\{F_{i,(1)}(X_{i,j}),1\leq i\leq n_{i}\}$ for each $1\leq i\leq K\}$. The remainder of the proof uses this frame. So let $G_{n_{i}}(i,f)$, $\mathbb{U}_n(i,\circ)$ and  $\mathbb{V}_n(i,\circ)$, be the functional empirical process,  the empirical \textit{cdf} and quantile functions based on $\{F_{i,(1)}(X_{i,j}),1\leq i\leq n_{i}\}$ for each $1\leq i\leq K\}$. Similary we define the functional empirical process,  the empirical \textit{cdf} and quantile functions based on the whole sample 
$$
\{F_{(1)}(X_{i,j}), \ 1\leq i\leq K, 1\leq i\leq n_{i}\}
$$

\noindent by dropping the label $i$ in the definition relative to group $i \in \{1,...,K\}$.\\

\noindent We will consider the three terms in Formula (ca) (page \pageref{ca}), that is the $C(n,i)$, $1\leq i\leq 3$, defined in Formula (c1) (page \pageref{c1}), Formula (c2) (page \pageref{c2}) and in Formula (c3) (page \pageref{c3}), and prove that each of them converges to a random variable $C(i)$ depending on the limiting Gaussian processes $G(i,\cdot )$ of $G_{n_{i}}(i,\cdot )$. This is enough to prove the asymptotic normality. The
variance $\vartheta _{1}^{2}$ will be nothing else but that of $%
C(1)+C(2)+C(3)$. Firstly, we treat $C(n,1).$ Remark that conditionally on $%
(n^{\ast }=\overline{n}),$ the random sequences $\{X_{i,j},1\leq i\leq
n_{i},1\leq i\leq K\}$ are independent and only depend on the $\omega
_{2}\in \Omega _{2}.$ We have%
\begin{equation*}
\sum_{i=1}^{K}\left( \frac{n_{i}}{n}\right) ^{1/2}\mathbb{G}_{n_{i}^{\ast},(1)}(h_{i})=%
\frac{1}{\sqrt{n}}\left[ \sum_{i=1}^{K}\sum_{j=1}^{n_{i}}h_{i}(X_{ij})-%
\sum_{i=1}^{K}n_{i}\mathbb{E}(h_{i}(X^{(i)}))\right] 
\end{equation*}%
\begin{equation*}
=\sqrt{n}\left[ \frac{1}{n}\sum_{i=1}^{K}\sum_{j=1}^{n_{i}}h_{(i)}(X_{ij})-%
\sum_{i=1}^{K}\left( \frac{n_{i}}{n}\right) \mathbb{E}\left(
h_{i}(X^{(i)})\right) \right] ,
\end{equation*}

\bigskip \noindent and
\begin{equation*}
\alpha _{n}(h,1)=\sqrt{n}\left( \frac{1}{n}\sum_{j=1}^{n}h(X_{j})-%
\sum_{i=1}^{K}\left( \frac{n_{i}}{n}\right) \mathbb{E}\left( h(X^{(i)})\right)
\right) 
\end{equation*}
\begin{equation*}
=\sqrt{n}\left( \frac{1}{n}\sum_{i=1}^{K}\sum_{j=1}^{n_{i}}h(X_{ij})-%
\sum_{i=1}^{K}\left( \frac{n_{i}}{n}\right) \mathbb{E}\left( h(X^{(i)})\right)
\right) .
\end{equation*}

\bigskip \noindent Then, by Formula (c1) (page \pageref{c1}) and replacing $n_{i}^{\ast }$ by $n_{i}$, $i=1,...,K$,
we get

\begin{equation*}
C(n,1)=\alpha _{n}(h,1)-\sum_{i=1}^{K}\left( \frac{n_{i}}{n}\right) \mathbb{G}_{n_{i},(1)}(h_{i})
\end{equation*}%

\begin{equation}
=\sum_{i=1}^{K}\left( \frac{n_{i}}{n}\right) ^{1/2}\left\{ \frac{1}{\sqrt{%
n_{i}}}\sum_{j=1}^{n_{i}}\left\{ \left( h-h_{i}\right) (X_{ij})-\mathbb{E}%
\left( h-h_{i}\right) (X^{(i)}))\right\} \right\} .  \label{c1a}
\end{equation}%

\noindent This implies that 

\begin{equation*}
C(n,1)=\sum_{i=1}^{K}\left( \frac{n_{i}}{n}\right) ^{1/2}\mathbb{G}%
_{n_{i}}\left( i,\left( h-h_{i}\right) F_{i,(1)}^{-1}\right) .
\end{equation*}%

\bigskip \noindent We finally have that
\begin{equation*}
C(n,1)\rightarrow C(1)=\sum_{i=1}^{K}p_{i}^{1/2}\mathbb{G}%
(i,(h-h_{i})F_{i,(1)}^{-1}).
\end{equation*}

\bigskip \noindent Since the $\mathbb{G}\left( i,\left( h-h_{i}\right) F_{i,(1)}{i}^{-1}\right) $ are
independent, centered and Gaussian, we get that 
\begin{equation*}
A_{1}=\mathbb{E}C^{2}(1)=\sum_{i=1}^{K}p_{i}\mathbb{EG}%
^{2}(i,(h-h_{i})F_{i,(1)}^{-1})
\end{equation*}%
\begin{equation*}
=\sum_{i=1}^{K}p_{i}\left\{ \mathbb{E}(h-h_{i})^{2}(X^{(i)})-(\mathbb{E}%
(h-h_{i})(X^{i}))^{2}\right\} .
\end{equation*}

\bigskip \noindent Then we arrive 

\begin{equation*}
A_{1}=\sum_{i=1}^{K}p_{i}\left\{ \int_{0}^{1}(\overline{h}-\overline{h}_{i})^{2}(F_{i,(1)}^{-1}(t))dt-\left( \int_{0}^{1}(\overline{h}-h_{i})(F_{i,(1)}^{-1}(t))dt\right) ^{2}\right\} .
\end{equation*}

\bigskip \noindent Secondly, one has%
\begin{equation*}
C(n,2)=\sum_{i=1}^{K}\left( \frac{n_{i}}{n}\right) ^{1/2}\left\{ \frac{1}{%
\sqrt{n_{i}}}\sum_{j=1}^{n_{i}}\left\{\mathbb{G}_{n_{i},(1)}(X_{ij})-F_{i,(1)}(X_{ij})\right\} \left( \frac{n_{i}}{n}q -q_{i}\right) (X_{ij})\right\} .
\end{equation*}

\bigskip \noindent We have
\begin{equation*}
\frac{1}{\sqrt{n_{i}}}\sum_{j=1}^{n_{i}}\left\{
\mathbb{G}_{n_{i}}(X_{ij})-F_{i,(1)}(X_{ij})\right\} \left( \frac{n_{i}}{n}q -q
_{i}\right) (X_{ij})
\end{equation*}
\begin{equation*}
=\int_{0}^{1}-\varepsilon _{n_{i}}(i,s)(p_{i}q -q
_{i})(F_{i,(1)}^{-1}(s))ds+o_{\mathbb{P}}(1)
\end{equation*}%
\begin{equation*}
=\int_{0}^{1}\mathbb{G}_{n_{i}}(i,s)(p_{i}q -q _{i})(F_{i,(1)}^{-1}(s))ds+o_{%
\mathbb{P}}(1)
\end{equation*}
\begin{equation*}
\rightarrow \int_{0}^{1}\mathbb{G}(i,s)(p_{i}q -q _{i})(F_{i,(1)}^{-1}(s))ds,
\end{equation*}

\bigskip \noindent and thus
\begin{equation}
C(n,2)\rightarrow C(2)=\sum_{i=1}^{K}p_{i}^{1/2}\int_{0}^{1}\mathbb{G}%
(i,s)(p_{i}q -q _{i})(F_{i,(1)}^{-1}(s))ds.  \label{l02}
\end{equation}

\bigskip \noindent Finally, we have 
\begin{equation*}
C(n,3)=\sum_{i=1}^{K}\left( \frac{n_{i}}{n}\right) ^{1/2}\sum_{h\neq i}^{K}%
\frac{n_{h}}{n}\left\{ \frac{1}{\sqrt{n_{i}}}\sum_{j=1}^{n_{i}}\left\{
G_{h,n_{h}}(X_{ij})-F_{h,(1)}(X_{ij})\right\} q(X_{ij})\right\} .
\end{equation*}

\bigskip \noindent But, for each fixed $i\in \{1,..,K\},$%
\begin{equation*}
\frac{1}{\sqrt{n_{i}}}\sum_{j=1}^{n_{i}}\left\{
G_{h,n_{h}}(X_{ij})-F_{h,(1)}(X_{ij})\right\} q(X_{ij})
\end{equation*}
\begin{equation*}
=\int_{0}^{1}\sqrt{n_{i}}\left\{
G_{h,n_{h}}(F_{i,(1)}^{-1}(V_{n_{i}}(i,s)))-F_{h,(1)}(F_{i,(1)}^{-1}(V_{n_{i}}(i,s)))%
\right\} \times q (F_{i,(1)}^{-1}(V_{n_{i}}(i,s)))ds.
\end{equation*}

\noindent By the assumptions, the functions $q$ and $F_{(1)}$ are continuous on such compact sets. Thus%

\begin{eqnarray*}
&&\frac{1}{\sqrt{n_{i}}}\sum_{j=1}^{n_{i}}[G_{h,n_{h}}(X_{ij})-F_{i,(1)}(X_{ij})]
q(X_{ij})\\
&=&\sqrt{\frac{n_{i}}{n_{h}}}\int_{0}^{1}\mathbb{G}%
_{n_{h}}(h,F_{i,(1)}(F_{i,(1)}^{-1}(V_{n_{i}}(i,s)))\times q(F_{i,(1)}^{-1}(V_{n_{i}}(i,s)))ds\\
&=&\sqrt{\frac{n_{i}}{n_{h}}}\int_{0}^{1}\mathbb{G}%
_{n_{h}}(h,F_{h,(1)}(F_{i,(1)}^{-1}(V_{n_{i}}(i,s)))\times q(F_{i,(1)}^{-1}(s))ds+o_{%
\mathbb{P}}(1)\\
&=&\sqrt{\frac{n_{i}}{n_{h}}}\int_{0}^{1}\mathbb{G}%
_{n_{h}}(h,F_{h,(1)}(F_{i,(1)}^{-1}(s))\times q(F_{i,(1)}^{-1}(s))ds+R_{n}+o_{\mathbb{%
P}}(1),
\end{eqnarray*}

\bigskip \noindent with
\begin{eqnarray*}
R_{n}\\
&=&\int_{0}^{1}\left\{ \mathbb{G}_{n_{h}}(h,F_{h,(1)}(F_{i,(1)}^{-1}(V_{n_{i}}(i,s)))-%
\mathbb{G}_{n_{h}}(h,F_{h,(1)}(F_{i,(1)}^{-1}(s))\right\} \times q
(F_{i,(1)}^{-1}(s))ds.
\end{eqnarray*}

\bigskip \noindent Based on the the assumption that, for any $(i,h) \in \{1,...,K\}^2$,

\begin{equation*}
\sup_{s\leq 1}\left\vert
F_{h,(1)}(F_{i,(1)}^{-1}(V_{n_{i}}(i,s)))-F_{h,(1)}(F_{i,(1)}^{-1}(s))\right\vert
=a_{n}\rightarrow 0.
\end{equation*}

\bigskip \noindent We obtain here a continuous modulus of the uniform empirical process (see \cite{shwell}, page 531) and then%
\begin{equation*}
\sup_{0\leq s\leq 1}\left\vert \left\{ \mathbb{G}%
_{n_{h}}(h,F_{h,(1)}(F_{i,(1)}^{-1}(V_{n_{i}}(i,s)))-\mathbb{G}%
_{n_{h}}(h,F_{i,(1)}(F_{i,(1)}^{-1}(s))\right\} \right\vert 
\end{equation*}%
\begin{equation*}
=O(\sqrt{-a_{n}\log a_{n}}).
\end{equation*}

\bigskip \noindent We finally get
\begin{equation*}
R_{n}=O\left( \sqrt{-a_{n}\log a_{n}}\right) \int_{0}^{1}q
(F_{i,(1)}^{-1}(s))ds\rightarrow 0
\end{equation*}

\bigskip \noindent and we arrive at
\begin{eqnarray}
&&C(n,3)\rightarrow C(3) \label{ll03} \\
&=&\sum_{i=1}^{K}p_{i}\sum_{h\neq i}^{K}\sqrt{p_{h}}%
\int_{0}^{1}\mathbb{G(}h,F_{h,(1)}(F_{i,(1)}^{-1}(s))\times q (F_{i,(1)}^{-1}(s))ds. \notag
\end{eqnarray}

\noindent Now, we are going to compute the variance $\vartheta _{1}^{2}$ based on the
independent functional Brownian bridges $\mathbb{G}(i,\cdot )$ which are limits of the
functional empirical process $\mathbb{G}_{n}(i,\cdot )$ \ respectively associated
with $\{F_{i,(1)}(X_{i,j}),1\leq i\leq n_{i}\}$, $i=1,..,K.$ Straightforward
calculations give what comes. First%
\begin{equation*}
A_{1}=\mathbb{E}C^{2}(1)=\sum_{i=1}^{K}p_{i}\mathbb{EG}%
^{2}(i,(h-h_{i})F_{i,(1)}^{-1}).
\end{equation*}

\noindent In order to lessen the expressions, we write for $i \in \{1,\cdots,K\}$,

$$
h^{\ast}_{i}(\circ)=(h-h_{i})(F_{i,(1)}^{-1}(\circ)), \ and \ c_{i}(\circ)=(p_{i}q -q _{i})\left(F_{i,(1)}^{-1}(\circ)\right). \ \label{changeNot}
$$

\noindent \noindent Next for
\begin{equation*}
C(2)=\sum_{i=1}^{K}p_{i}^{1/2}\int_{0}^{1}\mathbb{G}(i,s)(p_{i}q -q
_{i})(F_{i,(1)}^{-1}(s))ds
\end{equation*}

\bigskip \noindent we have
\begin{equation*}
A_{2}=\mathbb{E}(C^{2}(2))=\sum_{i=1}^{K}p_{i}\int_{0}^{1}\int_{0}^{1}(s%
\wedge t-st)c_{i}(t)c_{i}(s)dsdt
\end{equation*}
\begin{equation*}
=\sum_{i}^{K}p_{i}\int_{0}^{1}\int_{0}^{1} (s\wedge t-st)(p_{i}q-q_{i})(F_{i,(1)}^{-1}(s))(p_{i}q-
q_{i})(F_{i,(1)}^{-1}(t))dsdt,
\end{equation*}

\noindent  * Now for
\begin{equation*}
C(3)=\sum_{i=1}^{K}p_{i}{}\sum_{h\neq i}^{K}\sqrt{p_{h}}\int_{0}^{1}\mathbb{%
G(}h,F_{h,(1)}(F_{i,(1)}^{-1}(s)))\times q (F_{i,(1)}^{-1}(s))ds,
\end{equation*}

\bigskip \noindent we have%
\begin{equation*}
A_{3}=\mathbb{E}(C^{2}(3))
\end{equation*}
\begin{equation*}
=\mathbb{E}\left\{ \sum_{i=1}^{K}p_{i}^{2}\left( \sum_{h\neq
i}^{K}K_{i,h}\right) ^{2}+\sum_{i=1}^{K}\sum_{j\neq i}^{K}p_{i}p_{j}\left(
\sum_{h\neq i}^{K}K_{i,h}\right) \left( \sum_{h^{\prime }\neq
j}^{K}K_{j,h^{\prime }}\right) \right\} .
\end{equation*}

\noindent Put 
\begin{equation*}
K_{i,h}=\sqrt{p_{h}}\int_{0}^{1}\mathbb{G(}h,F_{h,(1)}(F_{i,(1)}^{-1}(s)))\times q
(F_{i,(1)}^{-1}(s))ds,
\end{equation*}

\bigskip \noindent Let us split $A_{3}$ into  
\begin{equation*}
A_{31}=\mathbb{E}\left( \sum_{i=1}^{K}p_{i}^{2}\left( \sum_{h\neq
i}^{K}K_{i,h}\right) ^{2}\right) 
\end{equation*}

\bigskip \noindent and  

\begin{equation*}
A_{32}=\mathbb{E}\left( \sum_{i=1}^{K}\sum_{j\neq i}^{K}p_{i}p_{j}\left(
\sum_{h\neq i}^{K}K_{i,h}\right) \left( \sum_{h^{\prime }\neq
j}^{K}K_{j,h^{\prime }}\right) \right) .
\end{equation*}

\bigskip \noindent Now by using the independence of the centered stochastic process $%
G(h,\cdot )$ for differents values of $h\in \{1,...,K\}$, one gets  
\begin{equation*}
A_{31}=\mathbb{E}\left( \sum_{i=1}^{K}p_{i}^{2}\left( \sum_{h\neq
i}^{K}K_{i,h}\right) ^{2}\right) 
\end{equation*}

\bigskip \noindent and then  
\begin{equation*}
A_{31}=\sum_{i=1}^{K}p_{i}^{2}\sum_{h\neq i}^{K}p_{h}\newline
\int_{0}^{1}\int_{0}^{1}\left[ {F_{h,(1)}(F_{i,(1)}^{-1}(s))\wedge
F_{h,(1)}(F_{i,(1)}^{-1}(t))}\right. 
\end{equation*}%
\begin{equation*}
\left. {-F_{h,(1)}(F_{i,(1)}^{-1}(s))F_{h,(1)}(F_{i,(1)}^{-1}(t))}\right] q%
(F_{i,(1)}^{-1}(s))q(F_{i,(1)}^{-1}(t))dsdt.
\end{equation*}

\bigskip \noindent Next, one has  
\begin{equation*}
A_{32}=\mathbb{E}\sum_{i=1}^{K}p_{i}\sum_{j\neq i}^{K}p_{j}\sum_{h\neq
i}^{K}p_{h}^{1/2}\sum_{h^{\prime }\neq j}^{K}p_{h^{\prime
}}^{1/2}\int_{0}^{1}\int_{0}^{1}
\end{equation*}%
\begin{equation*}
\mathbb{G(}h,F_{h,(1)}(F_{i,(1)}^{-1}(s))\mathbb{G}(h^{\prime },G_{h^{\prime
}}(F_{j,(1)}^{-1}(t)))q (F_{i,(1)}^{-1}(s))q (F_{j,(1)}^{-1}(t))dtds
\end{equation*}

\begin{equation*}
=\sum_{i=1}^{K}p_{i}\sum_{j\neq i}^{K}p_{j}\sum_{h\notin
\{i,j\}}^{K}p_{h}\int_{0}^{F_{i,(1)}(Z)}\int_{0}^{F_{j,(1)}(Z)}\newline
\left[ {F_{j,(1)}(F_{i,(1)}^{-1}(s))\wedge F_{h,(1)}(F_{j,(1)}^{-1}(t))}\right. 
\end{equation*}%
\begin{equation*}
\left. {-F_{j,(1)}(F_{i,(1)}^{-1}(s))F_{h,(1)}(F_{j,(1)}^{-1}(t))}\right] q(F_{i,(1)}^{-1}(s))q(F_{i,(1)}^{-1}(t))ds \ dt.
\end{equation*}

\noindent Now we have 
\begin{equation*}
C(1)C(2)=\left( \sum_{i=1}^{K}p_{i}^{1/2}\mathbb{G}(i,h^{\ast}_{i})\right)
\left( \sum_{i=1}^{K}p_{i}^{1/2}\int_{0}^{1}\mathbb{G}(i,s)c_{i}(s)\text{ }%
ds\right) 
\end{equation*}
\begin{equation*}
=\sum_{i=1}^{K}p_{i}^{1/2}\sum_{j=1}^{K}p_{j}^{1/2}\int_{0}^{1}\mathbb{G}%
(i,s)c(s)\mathbb{G}(j,h^{\ast}_{j})\text{ }c_{i}(s)\text{ }ds.
\end{equation*}

\noindent And we get 
\begin{equation*}
B_{1}=\mathbb{E}C(1)C(2)=\sum_{i=1}^{K}p_{i}\int_{0}^{1}\mathbb{E}(\mathbb{G}%
(i,s)\mathbb{G}(i,\ell _{i})\text{ }c_{i}(s)ds
\end{equation*}
\begin{equation*}
=\sum_{i=1}^{K}p_{i}\int_{0}^{1}\left\{ \int_{-\infty
}^{F_{i,(1)}^{-1}(s)}(h-h_{i})(y)dF_{i,(1)}(y)-s\mathbb{E}(h-h_{i})(X^{(i)})\right\} c_{i}(s)ds
\end{equation*}
\begin{equation*}
=\sum_{i=1}^{K}p_{i}\int_{0}^{1} \left\{ {\int_{0}^{s}(%
h-h_{i})(F_{i,(1)}^{-1}(t))dt}\right. 
\end{equation*}
\begin{equation*}
\left. {-s\int_{0}^{1}(h-h_{i})(F_{i,(1)}^{-1}(t))dt}%
\right\} (p_{i}q-q_{i})(F_{i,(1)}^{-1}(s))ds.
\end{equation*}

\bigskip \noindent We have next 
\begin{equation*}
C(2)C(3)=\left( \sum_{i=1}^{K}p_{i}^{1/2}\int_{0}^{1}\mathbb{G}%
(i,s)c_{i}(s)ds\right) 
\end{equation*}
\begin{equation*}
\times \left( \sum_{i=1}^{K}p_{i}{}\sum_{h\neq i}^{K}p_{h}^{1/2}\int_{0}^{1}%
\mathbb{G(}h,F_{h,(1)}(F_{i,(1)}^{-1}(s))\times q (F_{i,(1)}^{-1}(s))ds\right) 
\end{equation*}%
\begin{equation*}
=\sum_{i=1}^{K}p_{i}^{1/2}\sum_{j=1}^{K}p_{j}\sum_{h\neq
j}^{K}p_{h}^{1/2}\int_{0}^{1}\int_{0}^{1}\mathbb{G}\left( i,s\right) \mathbb{%
G}(h,F_{h,(1)}(F_{j,(1)}^{-1}(t))c_{i}(s)q (F_{j,(1)}^{-1}(t))\mathbb{)}dsdt.
\end{equation*}

\bigskip \noindent It is derived from what above that 
\begin{equation*}
B_{2}=\mathbb{E}C(2)C(3)=\sum_{j=1}^{K}p_{j}\sum_{i\neq
j}^{K}p_{i}\int_{0}^{1} \int_{0}^{1}
\end{equation*}%
\begin{equation*}
\lbrack s\wedge F_{i,(1)}(F_{j,(1)}^{-1}(t))-sF_{i,(1)}(F_{j,(1)}^{-1}(t))]\times (p_{i}%
q-q_{i})(F_{i,(1)}^{-1}(s))q%
(F_{j,(1)}^{-1}(t))dsdt.
\end{equation*}

\noindent Now finally for  
\begin{equation*}
C(1)C(3)=\left( \sum_{i=1}^{K}p_{i}^{1/2}\mathbb{G}(i,\ell _{i})\right) 
\end{equation*}
\begin{equation*}
\times \left( \sum_{i=1}^{K}p_{i}{}\sum_{h\neq i}^{K}p_{h}^{1/2}\int_{0}^{1}%
\mathbb{G(}h,F_{h,(1)}(F_{i,(1)}^{-1}(s))\times q (F_{i,(1)}^{-1}(s))ds\right) 
\end{equation*}%
\begin{equation*}
=\sum_{i=1}^{K}p_{i}^{1/2}\sum_{j=1}^{K}p_{j}\sum_{h\neq
j}^{K}p_{h}^{1/2}\int_{0}^{1}\mathbb{G(}h,F_{h,(1)}(F_{j,(1)}^{-1}(s))\mathbb{G}%
(i,h^{\ast}_{i})\times q (F_{j,(1)}^{-1}(s))ds,
\end{equation*}

\bigskip \noindent where the $h _{\ast}$'s  are defined in (\ref{changeNot}), we have%
\begin{eqnarray*}
B_{3}&=&\mathbb{E}C(1)C(3)\\
&=&\sum_{j=1}^{K}p_{j}\sum_{i\neq j}^{K}p_{i}\int_{0}^{1}\mathbb{E}\left\{ 
\mathbb{G}(i,h^{\ast}_i)\mathbb{G(}i,F_{i,(1)}(F_{j,(1)}^{-1}(s))\right\} \times q
(F_{i,(1)}^{-1}(s))ds\\
&=&\sum_{j=1}^{K}p_{j}\sum_{i\neq j}^{K}p_{i}\int_{0}^{1}\left\{ {\int_{0}^{1}(h-h
_{i})(F_{i,(1)}^{-1}(t))dt}\right.\\
&&\left. {-F_{i,(1)}(F_{j,(1)}^{-1}(s))\int_{0}^{1}(h-h
_{i})(F_{i,(1)}^{-1}(t))dt}\right\} q(F_{j,(1)}^{-1}(s))ds.
\end{eqnarray*}

\bigskip \noindent We have now finished the variance computation, that is 
\begin{equation*}
\vartheta _{1}^{2}=A_{1}+A_{2}+A_{3}+2(B_{1}+B_{2}+B_{3})
\end{equation*}

\chapter[Absolute and Relative Variations]{Asymptotic Laws of indices, of their absolute and relative variation of indices} \label{hfep_gateway_variation} 

\noindent In all this chapter, we use limiting results on variance-covariances of finite linear combinations of the margins of a same sequences of stochastic processes whose finite-distributions converge to those of a Gaussian processes.

\section{Asymptotic Laws of indices} \label{hfep_gateway_variation_sec1}

\noindent Suppose we deal with an index $I$. Suppose that general representation (GRI) in Section \ref{hfep_gateway_sec4} in Chapter \ref{hfep_gateway_intro} holds for the sampled indice $I_n$, that is $h(X)$ is square integrable and that conditions $(Re1)$ and $(Re2)$ of Theorem \ref{theoGenRes} (Section \ref{hfep_gateway_sec4} in Chapter \ref{hfep_gateway_intro}) also are satisfied for $\ell$. We refer to these conditions as (\textbf{HFEP1}).\\

\begin{theorem} (General law of Indice) \label{GLI}
Suppose that Assumptions (\textbf{HFEP1}) hold. Then we have as $n\rightarrow +\infty$,

$$
I^{*}_{n}=\sqrt{n}(I_{n}-I) \rightsquigarrow \mathcal{N}(0, \Gamma),
$$

\bigskip \noindent where $\Gamma =\gamma_1 + \gamma_2 +2 \gamma_3$, with

$$
\Gamma_{(1)}(h,h)=\int (h(x)-\mathbb{E}(h(X)))^2 dF_{(1)}(x)
$$

\noindent and

$$
\gamma_1=\Gamma_{(1)}(h,h), \ \gamma_2=\int_{0}^{1} \int_{0}^{1} \Gamma_{(1)}(f_s,f_t)dsdt \ and \ \gamma_3=\int_{0}^{1} \Gamma_{(1)}(h,f_s)ds. 
$$
\end{theorem}

\bigskip \noindent \textbf{Remark}. Later, we will deal with different indices. In that situation the variance $\Gamma$ for the specific index $I$ will be denoted
$$
\Gamma^{(I)}=\Gamma^{(I)}(h,\ell). \ \ \  \text{(Var-I)}
$$

\noindent \textbf{Proof}. The proof easily comes from the preliminaries in Chapter \ref{hfep_gateway_sec1}, especially in Section \ref{hfep_gateway_sec3}. We simply say that under the assumption and the (GRI) representation that $I^{*}_{n}=\sqrt{n}(I_{n}-I)$ weakly converges to a Gaussian variable and, by using Formula \ref{Gamma1} and strait computations, we have that the asymptotic variance is

$$
\Gamma =\gamma_1 + \gamma_2 +2 \gamma_3,
$$

\noindent where

$$
\gamma_1=\Gamma_{(1)}(h,h), \ \gamma_2=\int_{0}^{1} \int_{0}^{1} \Gamma_{(1)}(f_s,f_t)dsdt \ and \gamma_3=\int_{0}^{1} \Gamma_{(1)}(h,f_s)ds. \blacksquare.
$$

\section{Asymptotic Laws of variations of an index}

\noindent Let us place ourselves in the bidimensional space created Section \ref{hfep_gateway_sec3}, Chapter \ref{hfep_gateway_intro}. Let us suppose that the index $I$ is measured for from
a sample of observations of the couple $Y=(X^{(1)}, X^{(2)})$. We get the statistics $I_{n}^{(i)}$ for times $t=1$ and $t=2$. We are interested in finding the asymptotic laws of the
the variation $\Delta I_{n}=I_{n}^{(2)}-I_{n}^{(1)}$ of $I_{n}$ from times $t=1$ and $t=2$.\\

\noindent Let us begin to suppose that the square integrability conditions required for the convergence of the empirical processes based on $X^{(1)}$ and $X^{(2)}$, and that conditions $(Re1)$ and $(Re2)$ of Theorem \ref{theoGenRes} (Section \ref{hfep_gateway_sec4} in Chapter \ref{hfep_gateway_intro}) based on $X^{(1)}$ and $X^{(2)}$ and the appropriate function $\ell$ hold. We refer to these conditions by 
(\textbf{HFEP2}). So we may write the indice's general representation (GRI) in Section \ref{hfep_gateway_sec5} 
in Chapter \ref{hfep_gateway_intro} for both times to get (GR1) \label{GRI} : \\

$$
\sqrt{n}(I_{n}^{(i)} - \mathbb{E}h_i(X))=\mathbb{G}_{n,(1),(i)}(h_i)+\beta_{n,(1),(i)}(\ell_i)+o_{\mathbb{P}}(1),  \ \ i=1,2
$$

\bigskip \noindent where $\mathbb{G}_{n,(1),(i)}$ and $\beta_{n,(1),(i)}$ are respectively the one dimensional $fep$ and residual empirical process based on the $n$-sized sample from $X^{(i)}$. To simplify, we drop the subscript in $\beta_{n,(1),(i)}$ to only write $\beta_{n,(1)}$, and where $\ell_i(s)=q_i(F_{(2),i}^{-1}(s)), \ s \in (0,1)$. Denote

$$
h^{(1)}(x,y)=h_1(x) \ and \ h^{(2)}(x,y)=h_2(y), \ (x,y) \in \mathbb{R}^2; 
$$

$$
f^{(i)}_{s}(x,y)=1_{(x\leq F_{(2),1}^{-1}(s))}, \ s \in (0,1).
$$

\bigskip \noindent and

$$
f^{(2)}_{s}(x,y)=1_{(y\leq F_{(2),2}^{-1}(s))}, \ s \in (0,1).
$$

\bigskip \noindent We will use the following transform for any function $g$ of \ $(x,y) \in \mathbb{R}^2$ :

\begin{equation}
\widetilde{g}(s,t)=g\left(F^{-1}_{(2),1}(s),F^{-1}_{(2),2}(t)\right), \ (s,t) \in [0,1]^2.  \label{tildaOp}
\end{equation}

\bigskip \noindent But we may express (GRIS) using the bi-dimensional $fep$ based on the $n$-sized sample from $Y=(X^{(1)}, X^{(2)})$ through (GRI2) \label{GRI2}

$$
\sqrt{n}(I_{n}^{(i)} - \mathbb{E}h_i(X))=\mathbb{G}_{n,(2),(i)}(h^{(i)})+\int_{0}^{1} \mathbb{G}_{n,(1)}(f^{(i)}_s)\,\ell_i(s)\,ds + o_p(1), \ (GRIS)
$$

\noindent $i=1,2$, which, by the notations in Section \ref{hfep_gateway_sec3}, is (GRI2): \label{GRI2}

$$
I_{n}^{\ast}(i)=\sqrt{n}(I_{n}^{(i)} - \mathbb{E}h_i(X))=\mathbb{G}_{n,u,(2),(i)}(\widetilde{h}^{(i)})+\int_{0}^{1} \mathbb{G}_{n,u,(1)}(\widetilde{f}^{(i)}_s)\,\ell(s)\,ds + o_p(1), \  (GRIS)
$$

\noindent  $i=1,2$. Let us remark that

$$
\widetilde{f}^{(i)}_s=1_{[0,s]}, \ s \in (0,1), \ i=1,2.
$$

\bigskip \noindent The asymptotic covariance $\Gamma_{12}$ between $(I_{n}^{\ast}(1)$ and $I_{n}^{\ast}(2))$ is obtained from the combination between Formula (GRIS) just above and Formula (GammaStar) (page \pageref{GammaStar}) in Section \ref{hfep_gateway_sec3} in \ref{hfep_gateway_sec1} following these notations.\\

$$
\widetilde{\gamma}^{(12)}(s,t)=\Gamma^{\ast}(\widetilde{f}^{(1)}_s,\widetilde{f}^{(2)}_s)=\int_{0}^{s} \int_{0}^{t} dC(u,v) dudv -st,
$$

$$
\widetilde{\gamma}^{(12)}(s,t)=C(s,t)-st,
$$

\bigskip \noindent We also need

$$
\widetilde{\gamma}^{(1)}(s)=\widetilde{\Gamma}_{(2)}(\widetilde{h}^{(1)},\widetilde{f}^{(2)}_s)=\int_{0}^{s} \widetilde{h}^{(1)}(u,v) dC(u,v) dudv -s\int_{0} h_1(F_{(2),1}^{-1}(u)) \ du,
$$

$$
\widetilde{\gamma}^{(2)}=\widetilde{\Gamma}_{(2)}(\widetilde{f}^{(1)}_s,\widetilde{h}^{(2)})=\int_{0}^{s} \widetilde{h}^{(2)}(u,v) dC(u,v) dudv -s\int_{0} h_2(F_{(2),2}^{-1}(u)) \ du.
$$

\bigskip \noindent Then the asymptotic co-variance $\Gamma=(\Gamma_{ij}, \ 1\leq i \leq 2, \ 1\leq i \leq 2)$ of $(I_{n}^{\ast}(1), \ I_{n}^{\ast}(2))$ is given by :

$$
\gamma_{11}=\widetilde{\Gamma}_{(2)}(\widetilde{h}^{(1)},\widetilde{h}^{(2)})
$$

$$
\gamma_{22}=\int_{0}^{1} \int_{0}^{1} \widetilde{\gamma}^{(12)}(s,t)\ell_1(s) \ell_2(t) ds dt=\int_{0}^{1} \int_{0}^{1} (C(s,t)-st) \ell_1(s) \ell_2(t) ds dt
$$

\bigskip \noindent and

$$
\gamma_{12}=\int_{0}^{1} \widetilde{\gamma}^{(1)} \ell_2(s)ds \ and  \gamma_{21}=\int_{0}^{1} \widetilde{\gamma}^{(2)} \ell_1(s)ds.
$$

\noindent By using the product of factors in (GRIS) for $i=1,2$ and by using the function $\Gamma^\ast$, we arrive at

$$
\Gamma^{(12)}=\sum_{1\leq i,j\leq 2} \gamma_{ij}.
$$

\bigskip \noindent As to the asymptotic variances of $I_n^{(i)}$, $i=1,2$, we find it as in Theorem \ref{GLI}, by

$$
\Gamma^{(i)} =\gamma^{(i)}_1 + \gamma^{(i)}_2 +2 \gamma^{(i)}_3,
$$

\bigskip \noindent  with

$$
\Gamma^{(i)}_{(1)}(h^{(i)},h^{(i)})=\int (h^{(i)}(x)-\mathbb{E}(h^{(i)}(X)))^2 dF_{(2),i}(x),
$$

$$
\gamma^{(i)}_1=\Gamma^{(i)}_{(1)}(h^{(i)},h^{(i)}), \ \gamma^{(i)}_2=\int_{0}^{1} \int_{0}^{1} \Gamma^{(i)}_{(1)}(f^{(i)}_s,f^{(i)}_t)dsdt,
$$

\noindent and

$$
\gamma^{(i)}_3=\int_{0}^{1} \Gamma^{(i)}_{(1)}(h^{(i)},f^{(i)}_s)ds. 
$$

\noindent for $i=1,2$.\\

\bigskip \noindent With these notations, we are able to give the general result :

\begin{theorem} (General law of Variation of Indices) \label{GLVI}
Suppose that the Assumptions (HFEP2) hold and denote $\Delta I=I_2-I_1$. Then we have as $n\rightarrow +\infty$ 

$$
\Delta I^{*}_{n}=\sqrt{n}(\Delta I_{n}- \Delta I) \rightsquigarrow \mathcal{N}(0, \Delta\Gamma),
$$

\bigskip \noindent where $\Delta \Gamma =\Gamma^{(1)} + \Gamma^{(2)}+2\Gamma^{(12)}$.
\end{theorem}

\bigskip \noindent \textbf{Proof}. The proof follows the same lines as in the proof of Theorem \ref{GLI}, by remarking that $\Delta I^{*}_{n}=\sqrt{n}(\Delta I_{n}- \Delta I)$ is still a finite linear combinations of the margins of a same sequences of stochastic processes whose finite-distributions converge to those. of a Gaussian processes. The remainder is a matter of computations which are featured above.

\section{Asymptotic Laws of relative variations of an indice}

\noindent Following the results of the previous section, we use the Delta method and the same principles of finite linear combinations of the margins of a same sequences of stochastic processes whose finite-distributions converge to those. of a Gaussian processes to get the law of the relative variation of $I$ 
$$
\Delta RI_{n}=\frac{I_{n}^{(2)}-I_{n}^{(1)}}{I_{n}^{(1)}}.
$$ 

\noindent We have

\begin{theorem} (General law of Relative Variation of Indices) \label{GLRVI} Suppose that the Assumptions (HFEP2) hold and $\Delta I=(I_2-I_1)/I_1$ and

$$
\gamma_{4}=1/I_1 \  and \ \gamma_{5}=\Delta I/I_{1}^{2}.
$$

\noindent Then we have, as $n\rightarrow +\infty$, 

$$
\Delta RI^{*}_{n}=\sqrt{n}(\Delta RI_{n}- \Delta RI) \rightsquigarrow \mathcal{N}(0, \Delta R\Gamma^2),
$$

\bigskip \noindent where $\Delta R\Gamma =\gamma_{5}(\gamma_{5}\Gamma^{(1)}-2\gamma_{4})+\gamma_{4}^2 \Delta R\Gamma$.
\end{theorem}

\bigskip \noindent By using the delta method (see for Chapter 4 in \cite{ips-wcia-ang}, for example), we have that

$$
\Delta RI^{*}_{n}=\sqrt{n}(\Delta RI_{n}- \Delta RI)=\frac{1}{I_1} \Delta I^{*}_{n} - \frac{\Delta I}{I_1^2} \sqrt{n}(I^{(1)}_{n}-I_1)+o_{P}(1). 
$$

\noindent We already denote $\gamma_{4}=1/I_1$ and $\gamma_{5}=\Delta I/I_{1}^{2}$. The computations of the variance-covariances imply that the asymptotic variance of $\Delta RI^{*}_{n}$ is

$$
\gamma_{5}^2 \Gamma^{(1)} + \gamma_{4}^2 \Delta R\Gamma -2\gamma_{4} \gamma_{5} (\Gamma_{(12)}- (\Gamma^{(1)})^2)
$$

\noindent which is

$$
\gamma_{5}(\gamma_{5}\Gamma^{(1)}-2\gamma_{4})+\gamma_{4}^2 \Delta R\Gamma.
$$

\noindent $\blacksquare$.\\

\bigskip \noindent Let us finish by emphasizing the importance of knowing the law of $\Delta RI_{n}$. It is useful to check whether a  Millennium Development Goals (MDG) is achieved. For example, the poverty reduction MDG is expressed as to have a poverty measure $I$ to be reduced by a fixed rate $r$ from a time $t=1$ to a $t=2$. For poverty, $r$ was set to $50\%$ at 2015. One has to check that 

$$
\Delta RI_{n} \leq -r.
$$

\bigskip \noindent A way to answer to this requirement is to find cover of $\Delta RI_{n}$, say at $95\%$ of the form

$$
\mathbb{P}(\Delta RI_{n} \leq A)\geq 95\%
$$ 

\noindent and

$$
A \leq r.
$$

\bigskip \noindent Of course, the exact law of $\Delta RI_{n}$ allows a precise answer to the problem. Since we do not know it, we may try a use an approximated solution from the asymptotic law of  
$\Delta RI_{n}$.\\

\chapter{Mutual Asymptotic Influence between indices} \label{hfep_gateway_influence}

\noindent Here, we face the question of mutual influence between two indices. Usually, this question may be of interest if we want to know if a growth, in Economics, is fair or not. Fairness means here that all the population concerned by the growth, of the worst off of them, make benefice of that grow, what we call pro-poor growth. But in general, given two indices based on the same set of variables, we may also see if they evolve together in the same direction or not, and how much they evolve relatively each other.\\

\noindent We are going see in the lines below the influence of two different indices based on the same random variable between them at a fixed time and that of their absolute and/or relative variations. To begin, suppose that we have two indices $I$ and $J$.\\

\noindent In a one-dimensional frame, we consider their measures $I_n$ and $J_n$ from the $n$-size sample $X_1$, ..., $X_n$, $n\geq 1$, with underlying $cdf$ $F_{(1)}$. We suppose that Assumptions (\textbf{HFEP1}) holds for both indices so that we have for them, the indice's general representation (GRI) in Section \ref{hfep_gateway_sec4} in Chapter \ref{hfep_gateway_intro} in the from : \\

$$
\sqrt{n}(I_{n} - I)=\mathbb{G}_{n,(2)}(h)+\int_{0}^{1} \mathbb{G}_{n,(1)}(f_s)\,\ell(s)\,ds + o_p(1), \ (GRI-I)
$$

\bigskip \noindent and

$$
\sqrt{n}(J_{n} - J)=\mathbb{G}_{n,(2)}(g)+\int_{0}^{1} \mathbb{G}_{n,(1)}(f_s)\,\nu(s)\,ds + o_p(1), \ (GRI-J)
$$

\bigskip \noindent where for there exist two measurable function $p(x)$ and $p(x)$ of $x\in \mathbb{R}$ such that $\ell(s)=q(F_{(1)}^{-1}(s))$ and $\nu(s)=p(F_{(1)}^{-1}(s))$, for $s\in (0,1)$.\\

\bigskip \noindent In a two-dimensional frame, we still use the created Section \ref{hfep_gateway_sec3} in Chapter \ref{hfep_gateway_intro}. Assuming Assumptions (\textbf{HEFP2}) hold for both $I$ and $J$ By using the notations in Chapter \ref{hfep_gateway_variation} and in Formulas (GRI-I) and (GRI-J) above, we have for time $i=1$ and time $i=2$,

$$
\sqrt{n}(I_{n}^{(i)} - I^{i})=\mathbb{G}_{n,(2)}(h^{i})+\int_{0}^{1} \mathbb{G}_{n,(1)}(\tilde{f}^{i}_s)\,\\,ds+o_{\mathbb{P}}(1),  \ \ i=1,2 (GRIS-I)
$$

$$
\sqrt{n}(J_{n}^{(i)} - J^{i})=\mathbb{G}_{n,(1)}(h^{(i)})+\int_{0}^{1} \mathbb{G}_{n,(2)}(\tilde{f}^{(i)}_s)\,\nu_i(s)\,ds + o_p(1), \ (GRIS-J)
$$

\bigskip \noindent In the sequel, the full details of the computations will not be given. Once the representations are given, we suppose the reader will be able to make some direct and easy computations
to derive the results. The most essential arguments and notations are Chapter \ref{hfep_gateway_intro}.  

\newpage
\section{Mutual influence of two simple indices}

\begin{theorem} \label{MISI} Suppose Assumptions (HFEP1) are satisfied for two indices $I$ and $J$, then we have as $n \rightarrow +\infty$,

$$
(I_n^{*}, J_n^{*})  \rightsquigarrow \mathcal{N}\left(0, \left(  \begin{tabular}{lr}
$\Gamma^{(I)}$ & $\Gamma^{(I,J)}$\\
$\Gamma^{(I,J)}$ & $\Gamma^{(J)}$ 		
\end{tabular}
 \right) \right)
$$

\bigskip \noindent where $\Gamma^{(I)}$ dans $\Gamma^{(J)}$ are described in Formula in $(Var-I)$ in Chapter \ref{hfep_gateway_variation}, and 

\begin{eqnarray*}
\Gamma^{(I,J)}&=&\Gamma_{(1)}(h,g)+\int_{0}^{1}\int_{0}^{1} \Gamma_{(1)}(f_s,f_t) \ell(s) \nu(t) ds \ dt\\
&+& \int_{0}^{1} \Gamma_{(1)}(h,f_t) \nu(t) \ ds + \int_{0}^{1} \Gamma_{(1)}(f_s,g) \ell(s) ds\\
\end{eqnarray*}
\end{theorem}

\newpage
\section{Mutual influence of variations of indices}

\begin{theorem} \label{MIVI} Suppose Assumptions (HFEP2) are satisfied for two indices $I$ and $J$, then we have as $n \rightarrow +\infty$,

$$
(\Delta I_n^{*}, \Delta J_n^{*})  \rightsquigarrow \mathcal{N}\left(0, \left(  \begin{tabular}{lr}
$\Delta \Gamma^{(I)}$ & $\Delta \Gamma^{(I,J)}$\\
$\Delta \Gamma^{(I,J)}$ & $\Delta \Gamma^{(J)}$ 		
\end{tabular}
 \right) \right)
$$

\bigskip \noindent where $\Delta \Gamma^{(I)}$ and $\Delta \Gamma^{(J)}$ are described in Theorem \ref{GLVI} in Chapter \ref{hfep_gateway_variation}, and 

\begin{eqnarray*}
\Delta \Gamma^{(I,J)}=\Delta \Gamma^{(I,J)}_{11}+\Delta \Gamma^{(I,J)}_{22}-\Delta \Gamma^{(I,J)}_{12} - \Delta \Gamma^{(I,J)}_{21},
\end{eqnarray*}

\noindent where for $i=1,2$,

\begin{eqnarray*}
\Gamma^{(I,J)}_{ii}&=&\widetilde{\Gamma}_{(2)}(\widetilde{h}^{(i)},\widetilde{g}^{(i)})+\int_{0}^{1}\int_{0}^{1} (C(s,t)-st) \ell_i(s) \nu_i(t) ds \ dt\\
&+& \int_{0}^{1}  \nu_i(s) \left(\int_{0}^{s} \widetilde{h}^{(i)}(t)-\mathbb{E}\widetilde{h}^{(i)}(X^{(i)}) \ dt \right) \ ds\\
&+& \int_{0}^{1}  \ell_i(s) \left(\int_{0}^{s} \widetilde{g}^{(i)}(t)-\mathbb{E}\widetilde{g}^{(i)}(X^{(i)}) \ dt \right) \ ds
\end{eqnarray*}

\begin{eqnarray*}
\Gamma^{(I,J)}_{12}&=&\widetilde{\Gamma}_{(2)}(\widetilde{h}^{(1)},\widetilde{g}^{(2)})+\int_{0}^{1}\int_{0}^{1}  \widetilde{\Gamma}_{(2)}(\widetilde{f}^{(1)}_s,\widetilde{g}^{(2)}) \ell_1(s) \nu_i(t) ds \ dt\\
&+& \int_{0}^{1} \widetilde{\Gamma}_{(2)}(\widetilde{h}^{(1)},\widetilde{f}^{(2)}_s) \nu_2(s) ds+ 
\int_{0}^{1}  \widetilde{\Gamma}_{(2)}(\widetilde{g}^{(2)},\widetilde{f}^{(1)}_s) \ell_1(s) \ ds
\end{eqnarray*}

\noindent and

\begin{eqnarray*}
\Gamma^{(I,J)}_{21}&=&\widetilde{\Gamma}_{(2)}(\widetilde{g}^{(2)},\widetilde{h}^{(1)})+\int_{0}^{1}\int_{0}^{1}  \widetilde{\Gamma}_{(2)}(\widetilde{f}^{(1)}_s,\widetilde{g}^{(2)}) \ell_2(s) \nu_1(t) ds \ dt\\
&+& \int_{0}^{1} \widetilde{\Gamma}_{(2)}(\widetilde{h}^{(2)},\widetilde{f}^{(1)}_s) \nu_1(s) ds+ 
\int_{0}^{1}  \widetilde{\Gamma}_{(2)}(\widetilde{g}^{(1)},\widetilde{f}^{(2)}_s) \ell_2(s) \ ds
\end{eqnarray*}

\end{theorem}

\newpage
\section{Mutual influence of Relative Variations of indices}

\noindent Let us denote as previously

$$
\widetilde{\gamma}_{4,I}=\frac{1}{I_1}, \ \widetilde{\gamma}_{5,I}=\frac{\Delta I}{I_1}, \ \widetilde{\gamma}_{4,J}=\frac{1}{J_1}, \ and \widetilde{\gamma}_{5,J}=\frac{\Delta J}{J_1}.
$$

\bigskip \noindent As in the proof of Theorem \ref{GLRVI}, we have

$$
\Delta RI^{*}_{n}=\widetilde{\gamma}_{4,I} \Delta I^{*}_{n} - \widetilde{\gamma}_{5,I} \sqrt{n}(I^{(1)}_{n}-I^{(2)})+o_{P}(1). 
$$

\bigskip \noindent and

$$
\Delta RJ^{*}_{n}=\widetilde{\gamma}_{4,J} \Delta J^{*}_{n} - \widetilde{\gamma}_{5,J} \sqrt{n}(J^{(1)}_{n}-J^{(1)})+o_{P}(1). 
$$

\bigskip \noindent Doing the right the computations leads to

\begin{theorem} \label{MIRVI} Suppose Assumptions (HFEP2) are satisfied for two indices $I$ and $J$, then we have as $n \rightarrow +\infty$,

$$
(\Delta RI_n^{*}, \Delta RJ_n^{*})  \rightsquigarrow \mathcal{N}\left(0, \left(  \begin{tabular}{lr}
$\Delta R\Gamma^{(I)}$ & $\Delta R\Gamma^{(I,J)}$\\
$\Delta R\Gamma^{(I,J)}$ & $\Delta R\Gamma^{(J)}$ 		
\end{tabular}
 \right) \right)
$$

\bigskip \noindent where $\Delta R\Gamma^{(I)}$ and $\Delta R\Gamma^{(J)}$ are described in Theorem \ref{GLRVI} in Chapter \ref{hfep_gateway_variation}, and

\begin{eqnarray*}
\Delta R\Gamma^{(I,J)}&=&\widetilde{\gamma}_{4,I}\widetilde{\gamma}_{4,J} \Delta \Gamma^{(I,J)}-\widetilde{\gamma}_{4,I}\widetilde{\gamma}_{5,J} (\Gamma^{(I,J)}_{21}-\Gamma^{(I,J)}_{11})\\
&-&\widetilde{\gamma}_{4,J}\widetilde{\gamma}_{5,I} (\Gamma^{(I,J)}_{12}-\Gamma^{(I,J)}_{11})+\widetilde{\gamma}_{5,I}\widetilde{\gamma}_{5,J} \Gamma^{(I,J)}_{11}.
\end{eqnarray*}
\end{theorem}

\noindent As announced, we include in this portal a second part with the aim to show how to apply the results of the gateway in some important example before we move to the handbook.

\part{Applications and Examples}
\chapter*{Introduction to Part II}  \label{introPart2}

\noindent In this part, we will give some examples of \textit{GRI}'s of noticeable statistics. Some will be reports of existing results and hence given without proofs. Others will be proved here. The results given here will be consigned and will be used by coming works. So we want to begin by the most basic statistics which are moments statistics.\\

\noindent The examples given here are :\\

\noindent (a) The moments estimators and the normalized moments estimators.\\

\noindent (b) The general poverty index in Welfare Analysis.\\

\noindent (c) The Takayama poverty index in Welfare Analysis.\\

\noindent We make profit of this introduction to present a technical result which has been proved to be useful in many situations and which may be useful to check condition Condition (CRe2) in page \pageref{cre2}. Here is the lemma.\\

\begin{lemma} \label{lemTech01} Let $(A_n)_{n\geq 1}$ and $(B_n(\eta)_{(n\geq 1, \eta\in T)}$, where $T\neq \emptyset$ be two families of non-negative real-valued random variables defined on the same probability space $(\Omega, \mathcal{A},\mathbb{P})$ such that :\\

$$
\forall \varepsilon>0, \exists \eta_{0} \in T, \ \exists n_0\geq 1, \ \forall n\geq n_0, \ \mathbb{P}(A_n > B_n(\eta_0))\leq \varepsilon
$$

\bigskip \noindent and, as $n\rightarrow +\infty$,

$$
\forall \eta \in T, \ B_n(\eta) \rightarrow_{\mathbb{P}} 0 \ or \ \mathbb{E}B_n(\eta) \rightarrow 0.
$$

\bigskip \noindent Then $A_n \rightarrow_{\mathbb{P}} 0$, $n\rightarrow +\infty$. 
\end{lemma}

\bigskip \noindent \textbf{proof}.  Assume that the hypotheses of the lemma hold. Fix $0\delta>0$ and $0<\varepsilon<\delta$. Then there exists $\eta_0$ such that $B_n(\eta) \rightarrow_{\mathbb{P}} 0$ as $n\rightarrow +\infty$ and $\mathbb{P}(A_n > B_n(\eta_0))\leq \varepsilon$ for $n$ large enough. Hence 

\begin{eqnarray*}
\mathbb{P}(A_n > \delta)&=&\mathbb{P}((A_n > \delta)\cup (B_n(\eta_0) \leq \varepsilon)) + \mathbb{P}((A_n >
\delta)\cup (B_n(\eta) \leq \varepsilon))\\
&\leq &\mathbb{P}((A_n > \delta)\cup (B_n(\eta_0) \leq \varepsilon)) + \mathbb{P}((A_n > \delta)\cup (B_n(\eta) \leq \varepsilon))\\
&\leq&\mathbb{P}(A_n > B_n(\eta_0)) + \mathbb{P}(B_n(\eta_0) \leq \varepsilon)\\
&\leq& \varepsilon  + \mathbb{P}(B_n(\eta_0) \leq \varepsilon).
\end{eqnarray*}

\bigskip \noindent Hence for all any $0\varepsilon \in ]0,\delta[$, we have

$$
\limsup_{n\rightarrow +\infty} \mathbb{P}(A_n > \delta) \leq \varepsilon.
$$

\bigskip \noindent The proof of the lemma is finished by letting $\varepsilon \searrow 0$. $\square$

\chapter{Moments Estimation of moments} \label{hfep_part1_moEstim}

\section{Asymptotic representations of the empirical moments}

\noindent We are going to provide asymptotic representations of the non-centered moments 
$$
m_{\ell }=\mathbb{E}(X^{\ell }),
$$ 

\noindent with $m_{1}\equiv m$ and the centered moments 
$$
\mu _{\ell }=\mathbb{E}(X-m_{1})^{\ell },
$$ 

\bigskip \noindent where $\ell \geq 1$ whenever they exist, in the Gaussian field described in the Gateway. Their plug-in estimators are respectively 
\begin{equation*}
m_{n,\ell }=\sum_{i=1}^{n}X_{i}^{\ell}, \ \ell \geq 1.
\end{equation*}

\bigskip \noindent and

\begin{equation*}
\mu _{n,\ell }=\frac{1}{n}\sum_{i=1}^{n}\left( X_{i}-\overline{X}\right) ^{\ell}, \ \ell \geq 1.
\end{equation*}

\bigskip \noindent Let us put $\mu _{2}=\sigma ^{2}$ and $m_{1}=m$ and $h_{\ell }(x)=x^{\ell
},x\in \mathbb{R}$ and the following functions :

\begin{equation}
A(\ell )=h_{\ell }+\sum_{p=0}^{\ell -1}C_{\ell }^{p}(-1)^{\ell -p}\left(
m_{1}^{\ell -p}h_{p}+(\ell -p)m_{1}^{\ell -p-1}m_{p}h_{1}\right), 
\label{hfep_part1_moEstimno1}
\end{equation}

\begin{equation}
B(p)=\sigma^{-(2p-1)}\left( A(2p-1)-\frac{1}{2}(2p-1)\sigma
^{-2}\mu_{2p-1}A(2) \right)  \label{hfep_part1_moEstimno2}
\end{equation}

\bigskip \noindent and 

\begin{equation}
C(p)=\sigma ^{-2p}\left( A(2p)-p\sigma ^{-2}\mu _{2p}A(2)\right) 
\label{hfep_part1_moEstimno3}
\end{equation}

\noindent we have the following results which were proved first in \cite{loJB01}. 

\begin{theorem} \label{hfep_part1_moEstimtheoremMoments} Let $\ell \geq 1$ and assume that $\int x^{2\ell
}dF_{(1)}(x)<\infty$, then  
\begin{eqnarray*}
\sqrt{n}\left( \mu _{n,\ell }-\mu _{\ell }\right) &=&\mathbb{G}_{n}\left(
A(\ell )\right) +o_{p}(1)\\
&=&\mathbb{G}\left(A(\ell )\right)\rightsquigarrow \mathcal{N}(0,\mathbb{V}ar(A(\ell )(X)). \notag
\end{eqnarray*}
\end{theorem}

\bigskip \noindent \textbf{Proof}.  we have

\begin{eqnarray*}
\mu_{n,\ell}&=&\sum_{p=0}^{\ell}C_{\ell}^{p}\left( -\overline{X}\right)
^{\ell-p}\left( \frac{1}{n}\sum_{i=1}^{n}X_{i}^{p}\right)\\
&=&\sum_{p=0}^{\ell} C_{\ell}^{p}\left( -1\right) ^{\ell-p}\left(m_{1}+\frac{%
\mathbb{G}_{n}(h_{1})}{\sqrt{n}}\right) ^{\ell-p}\left( m_{p}+\frac{\mathbb{G%
}_{n}(h_{p})}{\sqrt{n}}\right)\\
&=&\left( m_{\ell}+\frac{\mathbb{G}_{n}(h_{\ell})}{\sqrt{n}}\right) +
\sum_{p=0}^{\ell-1}C_{\ell}^{p}\left( -1\right)
^{\ell-p}\left(m_{1}^{\ell-p}+(\ell-p) m_{1}^{\ell-p-1}\frac{\mathbb{G}%
_{n}(h_{1})}{\sqrt{n}}+o_{p}(n^{-1/2})\right)\\
&\times& \left( m_{p}+\frac{\mathbb{G}_{n}(h_{p})}{\sqrt{n}}\right)\\
&=&m_{\ell}+h_{\ell} + \sum_{p=0}^{\ell-1}C_{\ell}^{p} (-1)^{\ell-p} \left(
m_{1}^{\ell-p} m_{p} + \frac{\mathbb{G}_{n}(A_{\ell})}{\sqrt{n}} \right) +
o_{p}(n^{-1/2}),
\end{eqnarray*}

\bigskip \noindent where $A(\ell )$ is defined in (\ref{hfep_part1_moEstimno1}) and where we used that the linearity of the empirical functional process. By observing that $\mu_{\ell }=\sum_{p=0}^{\ell }C_{\ell }^{p}\left( -m_{1}\right) ^{\ell
-p}\left( m_{p}\right) $, we finally obtain 
\begin{equation}
\sqrt{n}\left( \mu _{n,\ell }-\mu _{\ell }\right) =\mathbb{G}_{n}\left(
A(\ell )\right) +o_{p}(1).
\end{equation}

\bigskip \noindent Now, we may do some algebra to find estimators of normalized moments including
skewness and kurtosis.

\section{Estimation of normalized moments}

\noindent This section is an example of what can be done once we have established a \textit{GRI}. We are going to combine the obtained
representations to represent the normalized centered empirical moments (NCM), defined by, 

\begin{equation}
b_{p}=\frac{\mathbb{E}\left( (X-m)^{2p-1}\right) }{\sigma ^{(2p-1)}},
\end{equation}

\bigskip \noindent and

\begin{equation}
a_{p}=\frac{\mathbb{E}\left( (X-m)^{2p}\right) }{\sigma ^{2p}},
\end{equation}

\bigskip \noindent where $p\geq 2$ whenever they exist, and consider their plug-in estimators called \textit{normalized centered
empirical moments} (NCEM), 
\begin{equation}
b_{n,p}=\frac{\mu _{n,2p-1}}{\mu _{n,2}^{\left( 2p-1\right) /2}}\text{ and }%
a_{n,p}=\frac{\mu _{n,2p}}{\mu _{n,2}^{p}},\text{ }p\geq 2,
\end{equation}

\bigskip \noindent We have the following results below.

\begin{theorem} \label{hfep_part1_moEstimhefp_ncem} Let $p\geq 1$ and assusme that $\int x^{2k}dG(x)<\infty ,$ then 
\begin{equation}
\sqrt{n}((b_{n,p}-b_{p}),(a_{n,2}-a_{p}))=(\mathbb{G}_{n}(B(p)),\mathbb{G}%
_{n}(C(p)))+o_{\mathbb{P}}(1).
\end{equation}
\end{theorem}

\bigskip \noindent \textbf{Proof}. This proof is a continuation of that of \ref{hfep_part1_moEstimtheoremMoments}. Then
the law of $b_{n,p}$ is given by

\begin{equation*}
\sqrt{n}\left( b_{n,p}-b_{p}\right) =\frac{1}{\mu _{n,2}^{\left( 2p-1\right)
/2}}\sqrt{n}\left( \mu_{n,2p-1}-\mu_{2p-1}\right)
\end{equation*}

\begin{equation*}
-\frac{\mu_{2p-1}}{\mu _{n,2}^{\left( 2p-1\right) /2}\mu _{2}^{\left(
2p-1\right) /2}}\sqrt{n}\left( \mu _{n,2}^{\left( 2p-1\right) /2}-\mu
_{2}^{\left( 2p-1\right) /2}\right) .
\end{equation*}

\noindent By the delta-method, we have

\begin{equation*}
\mu _{n,2}^{\left( 2p-1\right) /2} =\left( \mu _{2}+\frac{\mathbb{G}%
_{n}(A(2))}{\sqrt{n}}\right) ^{\frac{2p-1}{2}}+ o_{p}(n^{-1/2}).
\end{equation*}

\begin{equation*}
=\mu _{2}^ {\frac{2p-1}{2}}+\frac{2p-1}{2}\mu_{2}^{\frac{2p-3}{2}}\frac{%
\mathbb{G}_{n}(A(2))}{\sqrt(n)}+o_{p}(n^{-1/2}).
\end{equation*}

\noindent and then

\begin{equation*}
\sqrt{n}\left( \mu_{n,2}^{\left( 2p-1\right) /2}-\mu _{2}^{\left(
2p-1\right) /2}\right) =\left( \frac{2p-1}{2}\right) \mu_{2}^{\frac{2p-3}{2}}%
\mathbb{G}_{n}(A(2))+o_{p}(1),
\end{equation*}

\bigskip \noindent and next, by noticing, by the Weak law of Large numbers, that  $\mu _{n,\ell}\rightarrow
\mu_{\ell}$, for all $\ell\leq 2k$, whenever the $(2k)^{th}$ moment of the $X_i$'s are finite, we have

\begin{equation*}
\sqrt{n}\left( b_{n,p}-b_{p}\right)
\end{equation*}

\begin{equation*}
=\mathbb{G}_{n}\left( \sigma ^{-(2p-1)}A(2p-1)-\frac{1}{2}(2p-1)\sigma
^{-(2p+1)}\mu_{2p-1}A(2)\right) +o_{p}(1).
\end{equation*}

\begin{equation*}
\mathbb{G}_{n}\left( B(p)\right) +o_{p}(1)\rightarrow \mathbb{G}\left(
B(p)\right) ,
\end{equation*}

\bigskip \noindent where $B(p)$ is given in (\ref{hfep_part1_moEstimno2}). By the very same methods, we
have

\begin{equation*}
\sqrt{n}\left( a_{n,p}-a_{p}\right) =\mathbb{G}_{n}\left( C(p)\right)
+o_{p}(1),
\end{equation*}

\bigskip \noindent Important applications of these laws concern extension of the Jarque-Berra test for normality to almost any distribution function provided that the moment exist at the dimension we want to work on. Such an extension has been done first in \cite{loJB01}. It will be further developed in \cite{loJB02}.

\chapter{The General Poverty Index} \label{hfep_welfare_gpi}

\subsection{Representation of the GPI}
In this paper, we use the GPI in a unified approach that leads to an asymptotic representation for a large class of indices classified in three kinds. We are entering into the details of the poverty theory nor in the general description of the poverty indexes (See 
\cite{loGPI2013} for details on those questions). We are just giving the general description of the indexes and provide their a unified \textit{GRI}.\\

\noindent Here the observed random variable $X$ is non-negative and represents an income or an expense. $Z>0$ is a fixed number and considered as a threshold and $Q_{n}=n\mathbb{F}_{n,(1)}(Z)$ is the number of individual in the sample whose value $X$ is below the $Z$. $\mu _{1},\mu _{2},\mu _{3},\mu _{4}$ are constants.\\

\noindent Let us suppose given measurable mapping $A(p,q,z)$, $w(t)$, and $d(t)$\ of $p,q\in N,$\ and $z,t\in R$ and

$$
B(Q_{n},n)=\sum_{i=1}^{q}w(i).
$$ 

\noindent The General Poverty Index proposed by \cite{loGPI} and \cite{loGPI2013} is of the form

\begin{equation}
GPI_{n}=\frac{A(Q_{n},n,,Z)}{nB(Q_{n},n)}\sum_{j=1}^{Q_{n}}w(\mu _{1}n+\mu
_{2}Q_{n}-\mu _{3}j+\mu _{4})\text{\ }d\left( \frac{Z-X_{j,n}}{Z}\right), \ n\geq 1.
\label{gpi01}
\end{equation}

\bigskip \noindent This class of indices contains among others :\\

\noindent (1) The Foster-Greer-Thorbecke (FGT) index of parameter \cite{fgt84} defined for $\alpha \geq 0,$%

\begin{equation}
FGT_{n}(\alpha )=\frac{1}{n}\overset{Q_{n}}{\underset{j=1}{\sum }}\left( 
\frac{Z-X_{j,n}}{Z}\right) ^{\alpha },  \ n\geq 1.  \label{sall2}
\end{equation}

\bigskip \noindent (2) The Sen poverty measure (\cite{sen}) :  

\begin{equation}
P_{Sen}=\frac{2}{n(Q_{n}+1)}\overset{Q_{n}}{\underset{j=1}{\sum }}
(Q_{n}-j+1)\left( \frac{Z-X_{j,n}}{Z}\right), \ n\geq 1.  \label{sall3}
\end{equation}

\bigskip \noindent (3) The Kakwani (\cite{kakwani}) class of poverty measures : 

\begin{equation}
P_{KAK,n}(k)=\frac{Q_{n}}{n\Phi _{k}(Q_{n})}\overset{Q_{n}}{\underset{j=1}{%
\sum }}(Q_{n}-j+1)^{k}\left( \frac{Z-X_{j,n}}{Z}\right), \ n\geq 1.  \label{sall4}
\end{equation}

\bigskip \noindent where  

\begin{equation*}
\Phi _{k}(Q_{n})=\sum_{j=1}^{j=Q_{n}}j^{k\text{ \ }}=B(Q_{n},n)
\end{equation*}

\bigskip \noindent (4) The Shorrocks (\cite{shorrocks}) index  

\begin{equation}
P_{SH,n}=\frac{1}{n^{2}}\overset{Q_{n}}{\underset{j=1}{\sum }}(2n-2j+1)\left( \frac{Z-X_{j,n}}{Z}\right) ,  \label{sall5}
\end{equation}

\bigskip \noindent (5) The Thon (\cite{thon}) proposed\bigskip\ the following measure  

\begin{equation*}
P_{Th}=\frac{2}{n(n+1)}\overset{Q_{n}}{\underset{j=1}{\sum }}(n-j+1)\left( 
\frac{Z-X_{j,n}}{Z}\right), \ n\geq 1. \label{Thon} 
\end{equation*}

\bigskip \noindent In \cite{loGPI} and \cite{loGPI2013}, a \textit{GRI} Formula has been given under the following conditions.\\

\noindent First we consider the threshold condition:

\bigskip

\noindent (H1) There exist $\beta >0$ and $0<\xi <1$ such that, 
\begin{equation*}
0<\beta <F_{(1)}(Z)<\xi <1.
\end{equation*}

\bigskip

\noindent Next we have form conditions (on the indices):

\bigskip

\noindent (H2a) There exist a function $h(p,q)$ where $(p,q)\in \mathbb{N}%
^{2}$ and a function $c(s,t)$ where $(s,t)\in (0,1)^{2}$ such that,
when $n\rightarrow +\infty ,$ 
$$
\max_{1\leq j\leq Q}\left\vert A(n,Q)h^{-1}(n,Q)w(\mu _{1}n+\mu _{2}Q-\mu_{3}j+\mu_{4})-c(Q/n,j/n)\right\vert$$ $$=o_{\mathbb{P}}(n^{-1/2});
$$

\bigskip

\noindent (H2b) There exists a function $\pi (s,t)$ with $(s,t)\in \mathbb{R%
}^{2}$ such that, when $n\rightarrow +\infty ,$ 
\begin{equation*}
\max_{1\leq j\leq Q}\left\vert w(j)h^{-1}(n,Q)-\frac{1}{n}\pi
(Q/n,j/n)\right\vert =o_{\mathbb{P}}(n^{-3/2}).
\end{equation*}%

\bigskip

\noindent Further we need regularity conditions on $c$ and $\pi$:

\bigskip

\noindent (H3) The functions  $c(\cdot )$ and $\pi (\cdot )$ have uniformly continuous partial derivatives, that is
\begin{equation*}
\lim_{(k,l)\rightarrow (0,0)}\sup_{(x,y)\in (0,1)^{2}}\left\vert \frac{%
\partial c}{\partial y}(x+l,y+k)-\frac{\partial c}{\partial y}%
(x,y)\right\vert =0
\end{equation*}%
and 
\begin{equation*}
\lim_{(k,l)\rightarrow (0,0)}\sup_{\beta \leq x\leq \xi ,\text{{}}y\in
(0,1)}\left\vert \frac{\partial c}{\partial x}(x+l,y+k)-\frac{\partial c}{%
\partial x}(x,y)\right\vert =0;
\end{equation*}%

\bigskip

\noindent (H4) The functions $y\rightarrow \frac{\partial c}{\partial y}%
(x,y)$ and $y\rightarrow \frac{\partial \pi }{\partial y}(x,y)$  are monotonous.

\bigskip

\noindent (H5) The distribution function $F_{(1)}$ is increasing.

\bigskip

\noindent (H6) There exist $H_{0}>0$ and $H_{\infty }<+\infty $ such that
$$
H_{0}<H_{c}(F_{(1)})=\int_{0}^{+\infty }c(F_{(1)}(Z),F_{(1)}(y))\gamma (y)dF_{(1)}(y)<H_{\infty },
$$
\noindent and 
$$
H_{0}<H_{\pi }(F_{(1)})=\int_{0}^{+\infty }\pi (F_{(1)}(Z),F_{(1)}(y))e(y)dF_{(1)}(y)<H_{\infty }
$$
\noindent where 
$$
\gamma (x)=d\left( \frac{Z-x}{Z}\right) \mathbb{I}_{(x\leq Z)}\text{ and }%
e(x)=\mathbb{I}_{(x\leq Z)}\text{ for }x\in \mathbb{R}.
$$

\bigskip 

\noindent Based on these hypotheses, we put

$$
J(F_{(1)})=H_{c}(F_{(1)})/H_{\pi }(F_{(1)}),
$$

$$
h(\cdot )=H_{\pi }^{-1}(F_{(1)})h_{c}(\cdot )-H_{c}(F_{(1)})H_{\pi }^{-2}(F_{(1)})h_{\pi
}(\cdot )+K(F_{(1)})e(\cdot ),  
$$

\noindent *\textit{ }with 
$$
h_{c}(\cdot )=c(F_{(1)}(Z),F_{(1)}(\cdot ))\gamma (\cdot ),\text{ }h_{\pi }(\cdot )=\pi
(F_{(1)}(Z),F_{(1)}(\cdot ))e(\cdot ), 
$$

$$
K(F_{(1)})=H_{\pi }^{-1}(F_{(1)})K_{c}(F_{(1)})-H_{c}(F_{(1)})H_{\pi }^{-2}(F_{(1)})K_{\pi }(F_{(1)})
$$%
\noindent where 

$$
K_{c}(F_{(1)})=\int_{0}^{1}\frac{\partial c}{\partial x}(F_{(1)}(Z),s)\gamma
(F_{(1)}^{-1}(s))ds,$$
$$K_{\pi }(F_{(1)})=\int_{0}^{1}\frac{\partial \pi }{\partial x}%
(F_{(1)}(Z),s)e(F_{(1)}^{-1}(s))ds, 
$$

$$
q(\cdot )=H_{\pi }^{-1}(F_{(1)})q_{c}(\cdot )-H_{c}(F_{(1)})H_{\pi }^{-2}(F_{(1)}) q_{\pi }(\cdot ), 
$$
\bigskip \noindent and 
$$
q_{c}(\cdot )=\frac{\partial c}{\partial y}(F_{(1)}(Z),F_{(1)}(\cdot ))\gamma (\cdot
),q_{\pi }(\cdot )=\frac{\partial \pi }{\partial y}(F_{(1)}(Z),F_{(1)}(\cdot
))e(\cdot ). 
$$

\bigskip \noindent and $\ell(s)=q(F_{(1)}^{-1}(s)$, $s \in (0,1)$.\\

\bigskip We have the following \textit{GRI} Formulas.

\begin{theorem} \label{theolosall}
Suppose that (H1)-(H6) are true, then we have the following representation
\begin{equation}
\sqrt{n}(J_{n}(F_{(1)})-J(F_{(1)}))=\mathbb{G}_{n,(1)}(h)+\beta _{n,(1)}(\ell )+o_{\mathbb{P}}(1). 
\tag{R}
\end{equation}
\end{theorem}

\bigskip 

\begin{table}[h!]
\caption{Specific functions of the poverty measures }\label{spfGPI}
\begin{tabular}{l|cc}
\hline\hline
\textbf{Mesure} & \textbf{$h$} & \textbf{$q$} \\ \hline
&  &  \\ 
\textbf{Shorrocks} & $2\left( 1-F_{(1)}(y)\right) \left( \frac{Z-y}{Z}\right)
\mathbb{I}_{(y\leq Z)}$ & $-2\left( \frac{Z-y}{Z}\right) \mathbb{I}_{(y\leq Z)}$ \\ 
&  &  \\ 
\textbf{Thon} & $2\left( 1-F_{(1)}(y)\right) \left( \frac{Z-y}{Z}\right) \mathbb{I}_{(y\leq
Z)}$ & $-2\left( \frac{Z-y}{Z}\right) \mathbb{I}_{(y\leq Z)}$ \\ 
&  &  \\ 
\textbf{Sen} & $h_s$ & $q_s$ \\ 
&  &  \\ 
\textbf{Kakwani} & $h_k$ & $q_k$ \\ \hline\hline
\end{tabular}
\end{table}

\bigskip \noindent where%
$$
h_{s}(y) =\left\{ 2\left[ \left( 1-\frac{F_{(1)}(y)}{F_{(1)}(Z)}\right) \left( \frac{%
Z-y}{Z}\right) \right. \right.$$ $$ 
-\left. \left. \left( \frac{F_{(1)}(y)}{F_{(1)}(Z)}\right) \left( \frac{J_{s}(F_{(1)})}{F_{(1)}(Z)}%
\right) \right] +K_{s}(F_{(1)})\right\} \mathbb{I}_{(y\leq Z)},
$$

\bigskip \noindent and
$$
q_{s}(y)=-\frac{2}{F_{(1)}(Z)}\left[ \left( \frac{Z-y}{Z}\right) +\frac{J_{s}(F_{(1)})%
}{F_{(1)}(Z)}\right] \mathbb{I}_{(y\leq Z)}.
$$

\bigskip \noindent with
$$
J_{s}(F_{(1)})=2\int_{0}^{F_{(1)}(Z)}\left( 1-\frac{s}{F_{(1)}(Z)}\right) \left( \frac{%
Z-F_{(1)}^{-1}(s)}{Z}\right) ds,
$$

$$
K_{s}(F_{(1)})=2\left( 1-\frac{1}{ZF_{(1)}(Z)}\int_{0}^{F_{(1)}(Z)}F_{(1)}^{-1}(s)ds\right) +\frac{%
J_{s}(F_{(1)})}{F_{(1)}(Z)}.
$$

\bigskip \noindent And

$$
h_{k}(y) =\left\{ (k+1)\left[ \left( 1-\frac{F_{(1)}(y)}{F_{(1)}(Z)}\right) ^{k}\left( 
\frac{Z-y}{Z}\right) \right. \right. $$
$$-\left. \left. \frac{J_{k}(F_{(1)})}{F_{(1)}(Z)}\left( \frac{F_{(1)}(y)}{F_{(1)}(Z)}\right) ^{k}%
\right] +K_{k}(F_{(1)})\right\} \mathbb{I}_{(y\leq Z)},
$$

\noindent and
$$
q_{k}(y) =-\frac{k(k+1)}{F_{(1)}(Z)}\left[ \left( 1-\frac{F_{(1)}(y)}{F_{(1)}(Z)}\right)
^{k-1}\left( \frac{Z-y}{Z}\right) \right.
$$ 
$$ 
+\left. \frac{J_{k}(F_{(1)})}{F_{(1)}(Z)}\left( \frac{F_{(1)}(y)}{F_{(1)}(Z)}\right) ^{k-1}\right]
\mathbb{I}_{(y\leq Z)}
$$
\noindent where%
$$
J_{k}(F_{(1)})=(k+1)\int_{0}^{F_{(1)}(Z)}\left( 1-\frac{s}{F_{(1)}(Z)}\right) ^{k}\left( \frac{%
Z-F_{(1)}^{-1}(s)}{Z}\right) ds,
$$

\bigskip \noindent and
$$
K_{k}(F_{(1)}) =\frac{k(k+1)}{F_{(1)}(Z)}\int_{0}^{F_{(1)}(Z)}\left( 1-\frac{s}{F_{(1)}(Z)}\right)
^{k-1}\left( \frac{Z-F_{(1)}^{-1}(s)}{Z}\right) ds$$
$$ +\frac{J_{k}(F_{(1)})}{F_{(1)}(Z)}.$$

\bigskip\noindent Notice that the functions are indexed by $k$ for the Kakwani measure. For the FGT measure of index $\alpha$, we have that $q=0$ and 

$$
h(x)=\max(0, (Z-x)/Z)^{\alpha}.
$$

\chapter{Asymptotic Representation of Takayama's statistics} \label{chepRepTakayama}

The one-dimensional Takayama statistic (\cite{takayama}) is originally defined for a non-negative random variable $X$. Here, the not defined notation are supposed to be already done in Chapter \ref{hfep_gateway_intro} (page \pageref{hfep_gateway_intro}). The Takayama welfare measure is given, for $n\geq 1$, by

$$
T_n=1+\frac{1}{n} + \frac{1}{n^2 \mu_n(1)} \sum_{1\leq j \leq n\mathbb{F}_{n,(1)}(Z)} (n_j+1) d(X_{n-j+1,n}),
$$

\bigskip \noindent where $\mu_n(1)$ is the empirical mean for a sample of size $n\geq 1$, $d(x)$ is some measurable function of $x \in \mathbb{R}_{+}$. Originally $d$ is the identity function. But we will treat the general case. We have that $T_n$ is composed of the statistics $\mu_n(1)$ with

$$
C_n=\frac{1}{n^2} \sum_{1\leq j \leq n\mathbb{F}_{n,(1)}(Z)} (n_j+1) d(X_{n-j+1,n})
$$

$$
\mu=\mathbb{E}X \in \mathbb{R} and  
$$

\noindent We will need the following conditions :\\

\noindent (C1) $0 < \mathbb{E}d(X) \in \mathbb{R}$.\\

\noindent (C2) $0<F_{(1)}(Z)<1$.\\

\noindent (C3) For all $0<H<uep(F_{(1)})$, the measurable function $q$ is continuous on $[0,H]$.\\

The \textit{GRI} of the Takayama is given as follows.\\

\begin{theorem} \label{repTakayama} Under conditions (C1), (C2) and (C3), we have :\\

\noindent (1) For $Id(x)=x$ for $x\in \mathbb{R}$,
$$
\mu_n(1)=\mu + n^{-1/2} \mathbb{G}_{n,(1)}(Id)+o_{\mathbb{P}}, \ n\geq 1,
$$

\noindent (2) For 
$$
C=\int_{0}^{Z} (1-F_{(1)}(x)) \ dF_{(1)}(x),
$$

$$
h_c(x)=\left(1-\mathbb{F}_{(1)}(x)\right) d(x) 1_{\left(x \leq Z\right)}, \ \ x\in \mathbb{R}_+
$$

\bigskip \noindent and 

$$
q(x)=-d(x) 1_{\left(x \leq Z\right)}, \ x\in \mathbb{R}_+ \ and \ \ell(s)=q\left(F_{(1)}^{-1}(s)\right), \ s\in (0,1),
$$

\bigskip \noindent we have

$$
\sqrt{n}(C_n-C)=\mathbb{G}_{n,(1)}(h_c) + \int_{0}^{1}  \mathbb{G}_{n,(1)}(\tilde{f})\,\ell(s)\,ds+ o_{\mathbb{P}}(1).
$$

\bigskip \noindent (3) For 

$$
T=\frac{1}{\mu} \int_{0}^{Z} (1-F_{(1)}(x)) \ dF_{(1)}(x),
$$

\noindent and 

$$
h(x)=\mu^{-1}(h_c-C\mu^{-1})Id), \ x\in \mathbb{R},
$$

\noindent We have

$$
\sqrt{n}(T_n-T)=\mathbb{G}_{n,(1)}(h) + \int_{0}^{1}  \mathbb{G}_{n,(1)}(\tilde{f}_s)\,\ell(s)\,ds+ o_{\mathbb{P}}(1).
$$
\end{theorem}

\bigskip \noindent We already know (see Chapter \ref{hfep_part1_moEstim}, page \pageref{hfep_part1_moEstim}) that $\mu_n(1)$ has the \textit{GRI} given in Point (1) of the Theorem.\\

\noindent Before we come to establishing the \textit{GRI} of $C_n$, we remark that condition (C3) implies that for any $0<H<uep(F_{(1)})$,

$$
\varpi(q,\delta,H)=\sup{(x,y)\in [0,H]^2: \ |x-y|<\delta} |q(x)-q(y)|\rightarrow 0, \ as \delta \searrow 0,
$$

\noindent where $\varpi(q,\delta,H)$ is the $\delta$-uniform continuity modulus of $q$ on $[0,H]$ and for $0<h<1$,

$$
\varsigma(d,h)=\sup_{0\leq s \leq h} |d\left(F_{(1)}^{-1}(s)\right)|<+\infty.
$$

\bigskip \noindent Let us establish \textit{GRI} for 

$$
C_n=\frac{1}{n^2} \sum_{1\leq j \leq n\mathbb{F}_{n,(1)}(Z)} (n-j+1) d(X_{n-j+1,n}), \ n\geq 1.
$$

\bigskip \noindent We suppose that the underlying \textit{cdf} $F_{(1)}$ is continuous. Hence, by using the rank statistics in the lines in 
Section \ref{hfep_gateway_sec4} in Chapter \ref{hfep_gateway_intro}, we have, for $n\geq 1$,

\begin{eqnarray*}
C_n&=&\frac{1}{n^2} \sum_{1\leq j \leq n} (n-R_{j,n}+1) d(X_{j})\\
&=&\frac{1}{n} \sum_{1\leq j \leq n} \left(1-\mathbb{F}_{n,(1)}(X_j)+\frac{1}{n}\right) d(X_{j}) 1_{\left(X_j\leq Z\right)}.
\end{eqnarray*}

\noindent Based on the finiteness of the mathematical expectation of $d(X)$ and the boundedness of the $\mathbb{F}_{n,(1)}(\circ)$'s, and by using the law of large numbers, we easily see that

$$
C_n\rightarrow C=\int (1-F_{(1)}(x)) \noindent (a) \ dF_{(1)}(x) , \ as \ n\rightarrow +\infty. 
$$

\bigskip \noindent and 

$$
C_n=\frac{1}{n} \sum_{1\leq j \leq n} \left(1-\mathbb{F}_{n,(1)}(X_j)\right) d(X_{j} 1_{\left(X_j\leq Z\right)} +o_{\mathbb{P}}(n^-1).
$$

\noindent We get for $n\geq 1$,

\begin{eqnarray*}
C_n&=&\frac{1}{n} \sum_{1\leq j \leq n} \left(1-\mathbb{F}_{(1)}(X_j)\right) d(X_{j} 1_{\left(X_j\leq Z\right)}\\
&-&\frac{1}{n} \sum_{1\leq j \leq n} \left(\mathbb{F}_{(1)}(X_j)-\mathbb{F}_{(1)}(X_j)\right) d(X_{j}) 1_{\left(X_j\leq Z\right)}
+o_{\mathbb{P}}(n^-1).
\end{eqnarray*}

\bigskip  \noindent By denoting

$$
h(x)=\left(1-\mathbb{F}_{(1)}(x)\right) d(x) 1_{\left(x \leq Z\right)}, \ \ x\in \mathbb{R}_+
$$

\bigskip \noindent and

$$
q(x)=-d(x) 1_{\left(x \leq Z\right)}, \ x\in \mathbb{R}_+ \ and \ \ell(s)=q\left(F_{(1)}^{-1}(s)\right), \ s\in (0,1),
$$

\bigskip \noindent and hence, we reach half on the way, that is

\begin{eqnarray}
n^{1/2} \left(A_n-A\right)= \mathbb{G}_{n,(1)}(h)+Re_n(\ell)+o_{\mathbb{P}}(1). \ (RT01)
\end{eqnarray}

\bigskip  \noindent To do the other half way, we have to check  Conditions (CRe1) and (CRe2) in page \pageref{cre1}. As to Condition (CRe1), we have
$$
\mathbb{E}q(X)=\int_{0}^{Z} d(x) F_{1}(x) \ dF_{1}(x). 
$$

\noindent Condition (CRe2) is checked by showing that 
$$
A_n=\int_{0}^{1} \sqrt{n} \left( s - \mathbb{V}_{n,(1)}(s)\right) \Delta_n(s)\, ds \rightarrow 0 \ \ (CRe2)
$$ 

\noindent where

$$
\Delta_n(s)=\biggr(\ell\left(\mathbb{V}_{n,(1)}(s)\right)-\ell(s)\biggr), \ s\in (0,1). 
$$

\noindent We have, for $n\geq 1$ and $s \in (0,1)$,

\begin{eqnarray*}
\Delta_n(s)&=&d\left(\mathbb{F}_{n,(1)}^{-1}(s)\right) - d\left(\mathbb{F}_{(1)}^{-1}(s)\right) 1_{\left(s \leq \mathbb{F}_{n,(1)}(Z)\right)}\\
&+& 
d\left(\mathbb{F}_{(1)}^{-1}(s)\right) \biggr(1_{(s \leq \mathbb{F}_{(n),1}(Z))}-1_{(s \leq \mathbb{F}_{(1)}(Z))}\biggr)\\
&=:&\Delta_n(1,s)+\Delta_n(2,s)
\end{eqnarray*}

\bigskip \noindent But, since $F_{(1)}(Z)<1$, there exists $\eta>0$, such that $F_{(1)}(Z)+\eta<1$. By the uniform convergence of the uniform quantile process, for any $\varepsilon>0$ there exists $n_0$ such that for any $\neq n_0$

$$
\mathbb{P}( (\Delta_n \leq \eta) \cup (F_{n,(1)}(Z) \leq F_{(1)}(Z)+\eta)  )\leq \varepsilon.
$$

\bigskip \noindent where

$$
D_n=\sup_{s \in (0,1)} \biggr|\left(\mathbb{V}_{n,(1)}(s)-s\right)\biggr|, \ n\geq 1.
$$

\bigskip \noindent Let us split $A_n$ into

\begin{eqnarray*}
A_n&=&\int_{0}^{1} \sqrt{n} \left( s - \mathbb{V}_{n,(1)}(s)\right) \Delta_n(1,s)
+\int_{0}^{1} \sqrt{n} \left( s - \mathbb{V}_{n,(1)}(s)\right) \Delta_n(2,s)\\
&+&A_n(1)+A_n(2)
\end{eqnarray*}

\noindent Denote $\Omega_n=(\Delta_n \leq \eta) \cup (\mathcal{F}_{n,(1)}(Z) \leq F_{(1)}(Z)+\eta)$, $n\geq 1$. On $(\Delta_n \leq \eta)$, $n\geq n_0$, we have

\begin{eqnarray*}
\left|A_n(1)\right|&=&\int_{s \in (0,1), \ s \leq \mathbb{F}_{n,(1)}(Z)} \left|\sqrt{n} \left( s - \mathbb{V}_{n,(1)}(s)\right) \Delta_n(1,s)\right| \ ds\\
&+&\int_{s \in (0,1), \ s > \mathbb{F}_{n,(1)}(Z)} \left|\sqrt{n} \left( s - \mathbb{V}_{n,(1)}(s)\right) \Delta_n(1,s)\right| \ ds\\
&\leq& \varpi(d,D_n,Z) \int_{(0,1)} \left|\sqrt{n} \left( s - \mathbb{V}_{n,(1)}(s)\right)\right| \ ds\\
&+& \varsigma(q,G(Z)+\eta) \left(\mathbb{F}_{(1)}(Z)-\mathbb{F}_{n,(1)}(Z)\right)^{+}\\
&=:&B_n(1,1)+B_n(1,2).
\end{eqnarray*}

\bigskip \noindent where  $x^+=\max(0,x)$ for any $x \in \mathbb{R}$. By classical results on uniform empirical processes, we have
$$
\int_{0}^{1}  \biggr|\sqrt{n} \left( s - \mathbb{V}_{n,(1)}(s)\right)\biggr| \ ds \rightarrow \int_{0}^{1} |B(s)| \ ds\equiv Y,
$$

\bigskip \noindent where $(B(s), \ s\in (0,1))$ is a standard Brownian Bridge so that $Y$ has a finite expectation and by then is finite \textit{a.e} and next,

$$
\int_{0}^{1}  \biggr|\sqrt{n} \left( s - \mathbb{V}_{n,(1)}(s)\right)\biggr| \ ds
$$

\bigskip \noindent is bounded in probability. As a result, we have $B_n(1,1) \rightarrow_{\mathbb{P}} 0$ since
$\varpi(d,D_n,Z) \rightarrow 0$, as $n\rightarrow 0$. As well, $B_n(1,2) \rightarrow_{\mathbb{P}} 0$ since 
$b(q,G(Z)$ is bounded. We also have

$$
\left|A_n(2)\right| \leq \left|\left(\mathbb{F}_{(1)}(Z)-\mathbb{F}_{n,(1)}(Z)\right)\right| \int_{(0,1)} \left|\sqrt{n} \left( s - \mathbb{V}_{n,(1)}(s)\right)\right| \ ds 
$$

\bigskip \noindent which goes to zero in probability for the same reasons given before. In Total, for 
$B_n=A_n(1,1)+A_n(1,2)+A_n(2)$, we have for $n\geq n_0$

$$
\mathbb{P}( \left|A_n\right|> B_n) \leq \varepsilon, 
$$

\bigskip \noindent with $0\leq B_n \rightarrow_{\mathbb{P}} 0$ as $n\rightarrow +\infty$. Thus, we may and do apply Lemma \ref{lemTech01} in Chapter \ref{introPart2} (See page \pageref{lemTech01}) to conclude that 

$$
A_n(\eta) \rightarrow_{\mathbb{P}} 0, \ as n\rightarrow 0.
$$

\bigskip \noindent which closes the proof of Point (b).\\

\bigskip \noindent As to point (c), it is enough to use the techniques provided in the proof of Lemma \ref{lemma.tool.2} in Chapter \ref{hfep_gateway_intro}, page \pageref{lemma.tool.2}. The proof of the \textit{GRI} is done. $\square$.

\chapter{Conclusion} \label{hfep_conclusion}

\noindent We have set a frame has been set up for establishing General (Asymptotic) Representations for Indices \textit{GRI} for a large class of statistics. In the Gaussian field we have described, we are able to study asymptotic joint distributions of different statistics including temporal (longitudinal) and spatial configurations. As well, in the spatial case, the statistical estimation of the default of decomposability is handle based on the \textit{GRI} formula.\\

\noindent Based on these results, we are going to open two important project :\\

\noindent (a) The handbook of \textit{GRI}'s is open. Any contributor will present a specific or class of statistics and establish the \textit{GRI} along with the full proof. The contribution has to respect the notation given in this portal in order to be coherently included. The contribution in for a chapter will be assigned a digital object identifier and cite as an independent publication. The authors of such contribution will be allowed to use and adapt the packages described below.\\

\noindent (b) Since all the results described above which also will be extended to new \textit{GRI} depend only on functions $h$ and $\ell$, a package of computer programs has to be done in different languages. Actually, this package exists. It should be done again in a detailed writing and extended to other language. An R package is schedule.

\end{document}